\documentclass[twocolumn]{aastex63}
\usepackage{lineno}

\usepackage{graphicx}
\usepackage{wrapfig}
\usepackage{natbib}
\usepackage{color}
\graphicspath{{./figures/}} 
\usepackage{mathtools}
\usepackage{epstopdf}
\usepackage[autostyle]{csquotes}
\usepackage{hyperref} 
\def    \apjl  		{\rm {ApJL}}

\def    \apjl  		{\rm {ApJL}}

\def	\cm		{\,{\rm {cm}}}
\def	\K		{\,{\rm K}}
\def	\g		{\,{\rm {g}}}
\def	\mum	{\,{\mu \rm{m}}}

\def	\s		{\,{\rm {s}}}
\def	\H		{\,{\rm {H}}}

\def    \P     {\,\boldsymbol{P}\,}
\def    \B     {\,\boldsymbol{B}\,}
\def    \k     {\,\boldsymbol{k}\,}
\def    \J     {\,\boldsymbol{J}\,}

\def \bea {\begin{eqnarray}}
\def \ena {\end{eqnarray}}

\usepackage{amsmath}	
\usepackage{amssymb}	

\begin{document}

\shorttitle{Depolarization from Clouds to Starless cores.}
\title{Multiscale Synthetic observations of Polarized dust emission: On the origin of Depolarization effect from Molecular clouds to Starless cores and Constraints on Dust Physics}

\author{Nguyen Chau Giang}
\affil{Korea Astronomy and Space Science Institute, Daejeon 34055, Republic of Korea}
\email{chaugiang@kasi.re.kr}
\affil{Department of Astronomy and Space Science, University of Science and Technology, 217 Gajeong-ro, Yuseong-gu, Daejeon, 34113, Republic of Korea}

\author{Thiem Hoang}
\affil{Korea Astronomy and Space Science Institute, Daejeon 34055, Republic of Korea}
\affil{Department of Astronomy and Space Science, University of Science and Technology, 217 Gajeong-ro, Yuseong-gu, Daejeon, 34113, Republic of Korea}

\author{Blakesley Burkhart}
\affil{Department of Physics and Astronomy, Rutgers, The State University of New Jersey, 136 Frelinghuysen Rd, Piscataway, NJ 08854, USA}
\affil{Center for Computational Astrophysics, Flatiron Institute, 162 Fifth Avenue, New York, NY 10010, USA}

\author{Doris Arzoumanian}
\affil{Institute for Advanced Study, Kyushu University, Japan}
\affil{Department of Earth and Planetary Sciences, Faculty of Science, Kyushu University, Nishi-ku, Fukuoka 819-0395, Japan}

\author{Nguyen Bich Ngoc}
\affil{Department of Astrophysics, Vietnam National Space Center, Vietnam Academy of Science and Technology, 18, Hoang Quoc Viet, Nghia Do, Cau Giay, Ha Noi,
Vietnam} 

\author{Jihye Hwang}
\affil{Institute for Advanced Study, Kyushu University, Japan}
\affil{Department of Earth and Planetary Sciences, Faculty of Science, Kyushu University, Nishi-ku, Fukuoka 819-0395, Japan}

\author{Aran Lyo}
\affil{Korea Astronomy and Space Science Institute, Daejeon 34055, Republic of Korea}
\affil{Department of Astronomy and Space Science, University of Science and Technology, 217 Gajeong-ro, Yuseong-gu, Daejeon, 34113, Republic of Korea}
 
\begin{abstract}
Submillimeter observations of polarized dust emission from molecular clouds to starless cores frequently report a decrease in polarization fraction (p) with increasing dust emission intensity (I). This feature is commonly attributed to the alignment loss of dust grains or the geometrical effects of magnetic fields, yet, the detailed contributions remain unclear. To investigate the mechanism responsible for the depolarization, we use POLARIS to perform the multiscale synthetic dust polarization observations at $850\mum$ from magnetically aligned dust grains by RAdiative Torques (RATs) mechanism. We adopt three collapsing cloud models with different magnetic energy levels and explore the effects of grain magnetic properties and grain growth on dust polarization. The field geometrical effect is the dominant depolarization mechanism at $N_{\rm H} < 10^{21}-10^{22} \cm^{-2}$. We find that if grains grow beyond $>0.5\mum$ and are superparamagnetic (SPM) with large iron clusters, RATs remain effective at high column densities, and grain alignment loss contributes to depolarization only at $N_{\rm H} > 10^{23}\cm^{-2}$. If neither of these conditions is satisfied, the alignment loss (in cases of insufficient grain growth); or the reduced grain alignment efficiency by gaseous damping (for paramagnetic grains or SPM grains with small iron cluster sizes) can become the dominant depolarization mechanism at $\rm N_{\rm H} > 10^{22} \cm^{-2}$, regardless of how tangled the magnetic field lines in our simulation. Finally, we show that the depolarization mechanism and the underlying dust physical properties inside starless cores may be identified through the $p-I$ slope and the mean p at the core center.

\end{abstract}



\section{Introduction}\label{sec:intro}
Magnetic fields ($\B$) are one of the crucial components shaping the density structure of the interstellar medium (ISM), molecular clouds (MCs), filaments, and regulate the formation of cores, protostars, disks, and outflows (see recent reviews by, e.g., \citealt{Crutcher_2012_review}, \citealt{Hennebelle_Inutsuka_2019_review}, \citealt{Wurster_Li_2018_review}, \citealt{Pudritz_Ray_2019_review}, \citealt{Tsukamoto_2022_review}, \citealt{Kate_2023_review}). However, the accurate understanding of the dynamical importance of magnetic fields during the star formation is still unclear. Several observational techniques are used to study magnetic fields, including Zeeman splitting measurements \citep{Crutcher_2012_review, Pillai_2016_zeeman, Thompson_2019_zeeman, ching_2022_zeeman}, Faraday rotation \citep{Tahani_2018, Tahani_2022_review}, polarized molecular-line emission \citep{Gigart_1999_line_polarization, Ching_2016_line_polarization, Cortes_2021_line_polarization}, starlight polarization \citep{Anderson_2015, Panapoulou_2019} and thermal dust emission polarization from magnetically aligned dust grains (see review in, i.e., \citealt{Andersson_2015}, \citealt{Hull_2019_review}, \citealt{Kate_2023_review}). Among these methods, the last technique is widely applied for studying magnetic fields in star-forming regions owing to their ability to probe the wide spatial scales range. After the first detection of polarized dust emission in the Dust Ring at the Center of the Galaxy by \cite{Hildebrand_1993}, polarized dust emission has been extensively observed in whole Galactic sky map by Planck (i.e., \citealt{Planck_2015_XIX}), to MCs of several tens to hundreds pc in size (see, e.g., the review by \citealt{Kate_Fissel_2019_review}), down to sub-pc scale of dense cores (see, e.g., \citealt{Kate_2023_review}) using single-dish facilities such as the Stratospheric Observatory for Infrared Astronomy (SOFIA) and James Clerk Maxwell Telescope (JCMT). Observations toward protostellar envelopes of thousands au scale, down to protostellar and protoplanetary disks with few tens of au in size (see reviews by \citealt{Hull_2019_review}, \citealt{Maury_2022_review}, \citealt{Tsukamoto_2022_review}) have been carried out with interferometers such as the Submillimeter Array (SMA), the Very Large Array (VLA), and the Atacama Large Millimeter/submillimeter Array (ALMA). These rich observations reveal a complicated picture of magnetic fields during the early stages formation and evolution of protostars.

To establish thermal dust polarization from magnetically aligned dust grains as a reliable tracer of magnetic fields in star forming regions, significant effort in theoretical and simulation side has recently been made to better understanding the alignment dynamics of dust grains in dense environments (\citealt{Hoang_2021_polarization_hole}, \citealt{Hoang+2022}, see also the review by \citealt{Tram_Hoang_2022_review}). The current leading alignment mechanism for paramagnetic (PM) grains is based on radiative torques (RATs), originally proposed by \cite{Dolginov_1976}, numerically demonstrated by \cite{Draine_Weingartner_1996}, and analytically developed by \cite{Lazarian_Hoang_2007a, Hoang_Lazarian_2008}. They demonstrate that irregular dust grains exposed to an anisotropic radiation field can be spun up to suprathermal rotation and aligned with the ambient magnetic field through the alignment component of radiative torques. The suprathermal rotation stabilizes their alignment direction, allowing them to produce polarized dust emission with polarization fractions exceeding $\sim 1\%$. This feature makes RATs become more efficient than the classical paramagnetic relaxation mechanism (or Davis–Greenstein mechanism; \citealt{David_1951}) in driving the magnetic alignment of PM grains. \cite{Hoang_Lazarian_2016_mrat} then expanded the works to superparamagnetic (SPM) grains with embedded iron clusters and found that the paramagnetic relaxation mechanism can be strengthened due to the high magnetic susceptibility of SPM grains. Consequently, SPM grains can efficiently be aligned with $\B$ due to the joint action of RATs and the enhanced paramagnetic relaxation, known as the Magnetically enhanced RAdiative Torque (MRAT) mechanism. \cite{Hoang_Lazarian_2016_mrat} found that the MRAT mechanism can lead SPM grains including high amount of iron inclusions to achieve perfect magnetic alignment, regardless of dust properties (size, shape, composition) and the relative orientation between the grain spinning axis and the local magnetic field. Numerical tools for modeling the grains alignment and polarized emission have been also developed recently. The release of the public POLArized RadIative Simulator (POLARIS) originated by \cite{Reissl_2014, Reissl_2016} and improved by \cite{Giang_2023a, Giang_et_al_2024} with RATs/MRAT physics; and the DustPol-py code (\citealt{Lee_2020}, \citealt{Tram_2021}, \citealt{Tram_2025})  \footnote{POLARIS solves the full radiative transfer of the Stokes parameters within the given analytical three-dimensional model/magneto-hydrodynamic simulation, while DustPol-py models polarized emission in the optically thin regime without explicitly accounting for magnetic-field geometry.} provide powerful platforms for testing the grain alignment theory and studying the evolution of grain properties from polarized dust emission across a diverse astrophysical environments. 

One of the most prominent features of thermal dust polarization observed at sub-millimeter wavelengths toward star-forming regions is the systematic decrease of the polarization degree $p(\%)$ toward higher column densities $N_{\rm H}$ and thermal dust emission intensities $I$, commonly referred to as the depolarization effect. The depolarization is widely detected in MCs \citep{Planck_2015_XX, Fissel_2016, Sullivan_2021}, dense cores \citep{Alves_2014, Kwon_2018, Kate_2021}, and protostellar envelopes \citep{Cox_2015, Maury_2018, Sadavoy_2018, Kwon_2019}. Observationally, depolarization at the pc-scale MCs is commonly characterized by a power-law relation between $p(\%)$ and column density $N_{\rm H}$. The corresponding power-law index, denoted by $\alpha_{\rm p-N_{\rm H}}$, quantifies the rate at which $p(\%)$ decreases toward regions of higher column density \citep{Planck_2015_XX}. Similarly, depolarization at the sub-pc scale of dense cores and the thousand au scale of protostellar envelopes \citep{Whittet_2008} is characterized by a power-law relation between $p(\%)$ and the dust emission intensity, $I$, with the corresponding power-law index denoted by $\alpha_{\rm p-I}$. On MC scales, $\alpha_{\rm p-N_{\rm H}}$ varies from $\sim 0.1$ in the Planck cloud sample \citep{Planck_2015_XX} to $\alpha_{\rm p-N_{\rm H}} \sim 0.46$–$1.6$ in \cite{Sullivan_2021} sample. In starless cores, $\alpha_{\rm p-I}$ spans from $\sim 0.5–0.6$ in L1689N, SMM-16, and L1689B \citep{Kate_2021}, \citep{Kalory_2023}, to $\alpha_{\rm p-I} \sim 1$ in Oph C \citep{Kwon_2018, Kate_2019, Liu_2019}. Yet, the reason behind the diverse slope of the depolarization seen from clouds to starless cores is not yet quantified in detailed \footnote{This paper focuses only on the depolarization from MCs to starless cores. The discussion of the depolarization at thousands au scale of Young Stellar Objects can be found in \cite{Giang_et_al_2024}}.

There are two main categories that are usually proposed to explain the origin of the depolarization effect. The first one is the dust origin, such as the alignment loss of dust grains due to the inefficient RAT alignment in dense environments  (\citealt{Cho_Lazarian_2005}, \citealt{Bethell_2007}, \citealt{Hoang_2021_polarization_hole}), or the change of grain properties, i.e., shape, composition. The second one is the geometrical effect of magnetic fields, including the disorganization of the plane-of-sky (POS) magnetic fields along the line-of-sight (LOS); the bending of magnetic fields toward the LOS, and beam averaging effects (see, e.g., \citealt{Falceta-Gonçalves_2008},  \citealt{Planck_XII_2020}). On MC scales where $N_{\rm H} \sim 10^{20}-10^{22}\cm^{-2}$, \cite{Planck_2015_XX} found an anticorrelation between polarization fraction and polarization angle dispersion $S$, and an approximately constant $p \times S$ quantity across the wide observed density range. They thus suggest the increasing angle dispersion of the POS magnetic fields along the LOS and the change in the field inclination angle to the major depolarization mechanisms at the pc-cloud scale. This finding is then confirmed by subsequent numerical studies of dust polarization in MCs by \cite{Chen_2016} and synthetic dust polarization modeling with POLARIS by \cite{Seifried_2019}. \cite{King_2019}, in contrast, suggested the depolarization seen in Vela C to be the product of the combination of the field geometrical effect and the reduced grain alignment efficiency. Yet, the physics behind the weak grain alignment degree at low densities $n_{\rm H} \sim 10^{2}-10^{4}\cm^{-2}$ remains unclear. Toward sub-pc core scale where $N_{\rm H} > 10^{22}\cm^{-2}$, \cite{Chen_2016} showed that the increasing tangledness of magnetic fields induced by gravitational contraction can further lower $p(\%)$ at the core center. However, this study assumes perfect magnetic alignment of dust grains, that may overestimate the impact of magnetic fields on the depolarization seen inside starless cores. 

\cite{Padoan_2001}, \cite{Cho_Lazarian_2005}, \cite{Pelkonen_2007}, \cite{Bethell_2007}, \cite{Hoang_2021_polarization_hole} showed that the alignment loss can reproduce the decrease of $p(\%)$ inside this object if grains grow insignificantly. But if grains can grow to micron-sized inside dense cores (as suggested by theoretical studies by \citealt{Hirashita_2013,Fukihara_2026} and observational studies by \citealt{Foster_2006,Steinacker_2014,Wang_2015,Schnee_2014}), they can still be magnetically aligned by interacting with infrared photons of ISRF \citep{Cho_Lazarian_2005,Pelkonen_2007,Bethell_2007}. If this is the case, the magnetic field tangledness may also contribute to the depolarization at the core scale, as suggested by \cite{Chen_2016}. However, how strongly this field geometrical effect interferes with the alignment efficiency in producing the depolarization from clouds to starless cores is not quantified in detail in previous studies. Up to date, no multi-scale synthetic dust polarization observations from pc scales clouds to sub-pc scales starless cores that consider the effects of magnetic fields and aligned dust grains simultaneously have been performed. The exact relative contribution of each mechanism to the depolarization at these scales thus remains unclear.  
 
Besides the lack of multi-scale synthetic observations of dust polarization; the modeling of grain alignment inside MCs and starless cores is also not quantified in detail. As shown in recent theoretical study by \cite{Hoang+2022} and synthetic dust polarization modeling by \cite{Giang_2023a}, \cite{Giang_et_al_2024}, the strong gas-grain collision in protostellar environments where $n_{\rm H} \sim 10^{7}-10^{11}\cm^{-3}$ can significantly suppress the internal and external alignment of dust grains with magnetic fields. This issue can cause the depolarization seen at thousands au scale of Class 0/I Young Stellar Objects (YSOs, \citealt{Giang_et_al_2024}). Although starless cores are typically less dense ($n_{\rm H} \sim 10^{6}-10^{7}\cm^{-3}$) than protostellar environments, it remains unclear whether sub-micron and micron-sized grains can maintain their efficient magnetic alignment as assumed in \cite{Cho_Lazarian_2005}, \cite{Bethell_2007}, \cite{Hoang_2021_polarization_hole}. In addition, given the high sensitivity of grain alignment efficiency on the grain magnetic susceptibility (\citealt{Hoang_Lazarian_2016_mrat}) in star-forming regions (\citealt{Hoang_2023}, \citealt{Giang_2023a}), how this factor affects dust polarization detected inside MCs and starless cores also has not yet been explored.

Therefore, the first goal of this study is to perform multiscale synthetic observation of polarized dust emission, including different magnetized core models and dust models to consistently understand the mechanism behind the depolarization produced from MCs to starless cores. The second goal is to investigate the connection between the physical properties of dust grains (i.e., grain magnetic properties, grain growth) and the observed quantities of dust polarization (i.e., polarization fraction, the power-index of $p-N_{\rm H}$ or $p-I$ relation). We expect that these correlations may provide some hints for probing again the dust physics via dust emission polarization observations. The structure of our paper is as follows: we first describe the simulations used for the study and the POLARIS post-processing framework in Section \ref{sec:arepo_polaris}. The analysis of effects of magnetic fields and grain alignment efficiency on dust polarization will be shown in Section \ref{sec:Slos_gamma_align_p}, followed by the effect of grain magnetic properties and grain growth on the depolarization (Section \ref{sec:grain_depolarization}). Further discussion and summary on our findings are in Sections \ref{sec:discuss} and \ref{sec:summary}, respectively.

\section{Data description and synthetic polarized dust emission modeling}\label{sec:arepo_polaris}

\subsection{AREPO simulations}\label{sec:arepo}
We adopt the magnetohydrodynamic (MHD) datasets of \cite{Mocz_2017} and \cite{Burkhart_2020}, which simulate the prestellar cores formation inside turbulent, magnetized MCs using the moving mesh, quasi-Lagrangian AREPO code \citep{Springel_2010}. The code can dynamically adapt the computational grid to the evolving physical system, allowing for accurate capture dynamics inside new forming substructure. The simulations are scale-free, with the box size $L_{0}$ (in unit of pc), the initial magnetic-field strength $B_{0}$ (in unit of $\mu G$), and the total cloud mass $M$ (in unit of $M_{\odot}$) to be scaled as:
\bea 
L_{\rm 0} = 5.2\Bigg(\frac{c_{\rm s}}{0.2\rm ~km ~s^{-1}}\Bigg) \Bigg(\frac{n_{\rm H}}{1000 \cm^{-3}}\Bigg)^{-1/2} \Bigg(\frac{M_{\rm s}}{10}\Bigg) ~~~~\\ 
B_{\rm 0} = 1.2, 12, 36, 120\Bigg(\frac{c_{\rm s}}{0.2\rm ~km~ s^{-1}}\Bigg) \Bigg(\frac{n_{\rm H}}{1000 \cm^{-3}}\Bigg)^{1/2} ~~~\\
M = 4860 \Bigg(\frac{c_{\rm s}}{0.2\rm ~km~ s^{-1}}\Bigg)^{3} \Bigg(\frac{n_{\rm H}}{1000 \cm^{-3}}\Bigg)^{-1/2} \Bigg(\frac{M_{\rm s}}{10}\Bigg)^{3}, ~~~
\ena
where $c_{\rm s}$ is the sound speed, $n_{\rm H}$ is the gas volume density, and $M_{\rm s}$ is the Sonic Mach number. At the starting time, solenoidal turbulence is injected into the computational domain using an Ornstein–Uhlenbeck process \citep{Federrath_2010, Bauer_2012, Federrath_2015} to drive the gas density and magnetic field structure. Gravity is subsequently turned on after the gas density distribution reaches the steady state of several eddy turnover times. The cloud collapses isothermally, and the simulations will end when a prestellar core is resolved at a few au with the peak density of $n_{\rm H} \sim 10^{7}$–$10^{8}~\mathrm{cm^{-3}}$ (see \citealt{Mocz_2017} for a detailed description of the stopping criteria). The initial magnetic field is uniform and oriented along the $x$-direction. \cite{Mocz_2017} considers four sets of the plasma beta parameter $\beta = 0.0025, 0.028, 0.25, 25$ (which characterizes the ratio of thermal pressure to magnetic pressure) and Alfvénic Mach number $M_{\rm A} = 0.35, 1.2, 3.5, 35$ (which characterizes the relative importance of turbulence and magnetic fields) to simulate the prestellar formation inside the strong-field cloud (with an initial magnetic-field strength of $\rm B_{\rm 0} = 120\mu G$, model B100), moderate-field cloud ($\rm B_{\rm 0} = 36\mu G$, model B30), weak-field cloud ($\rm B_{\rm 0} = 12\mu G$, model B10), and very weak-field cloud ($\rm B_{\rm 0} = 1.2\mu G$, model B1). The corresponding initial mass-to-flux ratios (ratio of gravitational to magnetic energy) in each cloud will be $\mu_{\Phi,0} \simeq \sqrt{5\pi/3} M_{\rm A} = 0.8, 2.7, 8, 80$, respectively. In the rest of our study, we only use the cloud models B30, B10, and B1 because the initial magnetic field strength of model B100 is far beyond a few tens of $\mu G$ of $\B$ detected in MCs by Zeeman measurements (\citealt{Crutcher_2010}, \citealt{Thompson_2019_zeeman}). A summary of three cloud models used in the study is provided in Table~\ref{tab:model_cloud}.

 \begin{table*}
  \centering
         \caption{Initial setups of the four AREPO simulations used in \cite{Mocz_2017}. Only models B30, B10, and B1 are used in this study.}
  \begin{tabular} {cccccc}
  \hline 
\textbf{Cloud model} & \textbf{Initial magnetic} & \textbf{Plasma beta} & \textbf{$\text{Alfv\'{e}nic}$ Mach number} & \textbf{Mass-to-flux ratio} & \textbf{Model description} \\
  & \textbf{field strength} & $\beta$ &   $M_{\text{A}}$ & $\mu_{\Phi,0}$ &  \\

  \hline 
B30 & $36  \mu G$ & 0.028 & 1.2 & 2.7 & Moderate magnetic field \\ 
B10 & $12  \mu G$ & 0.25 & 3.5 & 8 & Weak magnetic field \\ 
B1 & $1.2  \mu G$ & 25 & 35 & 80 & Very weak magnetic field \\
 
   \hline 
    \label{tab:model_cloud}
    \end{tabular}\\
\end{table*}

\begin{figure*}[h]
\centering
\includegraphics[width=\textwidth,height=\textheight,keepaspectratio]{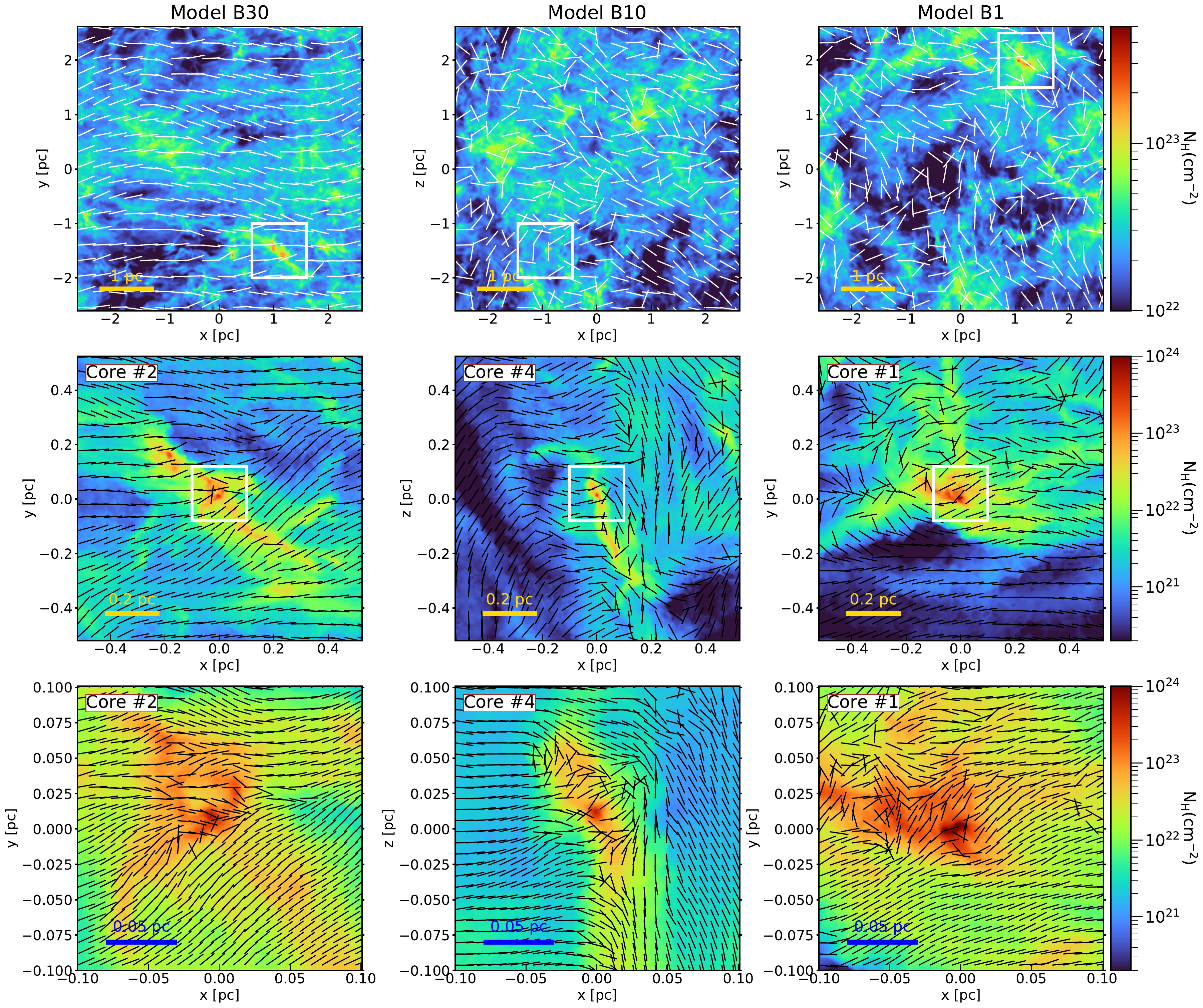}
    \caption{Upper row: column gas density maps overplotted with the density-weighted projected magnetic field orientation on the POS (black vectors), shown for the cloud scale of 5 pc. Left column shows results of intermediately magnetized cloud obtained on x$-$y plane (model B30), middle column shows results of weakly magnetized cloud seen on x$-$z plane (model B10), and right column shows results for very weaky magnetized cloud observed on x$-$y plane (model B1). Middle and lower rows: zoom-in view of the white box shown in the upper row, at the scale of 1 pc and 0.2 pc, respectively. Magnetic fields are rather well preserved from $\sim 5$ pc to $\sim 0.1$ pc of starless cores formed inside intermediately magnetized cloud, but they are highly distorted in weakly magnetized clouds due to large scale turbulent motions.}
     \label{fig:gas_density}
\end{figure*} 

\begin{table*}
\centering
\caption{Protostar information}
\begin{tabular}{llll}
\hline
Quantity & Symbol & Value & References \\   
\hline
\multicolumn{4}{c}{\textbf{Radiation source}}\\
\hline 
Interstellar radiation field strength  & $G_{\rm 0}$      &   1 \\ 

\hline
\multicolumn{4}{c}{\textbf{Dust model}}\\
\hline
Grain axial ratio           & $s$                    & 0.5, 0.625, 0.7143 & \cite{Draine_Hensley_2021a, Draine_Henseley_2021b, Draine_Henseley_2021c}\\
Grain porosity  & P & 0.0, 0.2, 0.4 & \cite{Draine_Hensley_2021a, Draine_Henseley_2021b, Draine_Henseley_2021c}\\
Dust-to-gas mass ratio      & $\eta$               & 0.01 & \cite{Bohlin_1978}\\
Initial size distribution   & $\rm dn/da$              & $\sim a^{-3.5}$ & \cite{Mathis_1977}\\
Minimum grain size          & $a_{\rm min}$        & 5nm   & \\ 
Maximum grain size          & $a_{\rm max}$        & $0.25, 0.5, 1, 2\mum$ & \cite{Hirashita_2013, Pagani_2010, Schnee_2014} \\
Iron atoms/cluster (for SPM) & $N_{\rm cl}$  & 50,500,1000 & \cite{Hoang_Lazarian_2016_mrat, Yang_2021}\\
Volume filling factor  & $\phi_{\rm sp}$ & 0.03 & \cite{Bradley_1994, Martin_1995}\\
of iron clusters & & \\
Iron fraction inside PM & $f_{\rm p}$ & 0.1 & \cite{Draine_Weingartner_1996} \\
  \hline
    \label{tab:parameter}
    \end{tabular}   
\end{table*}

From each cloud model, we identify regions with the peak density within 1000 au greater than $n_{\rm H} > 5\times 10^{6}\cm^{-3}$ to be the dense core, and regions with the peak density greater than $n_{\rm H} \geq 10^{8}\cm^{-3}$ to be the prestellar core. Following these criteria, we obtain 5 dense cores and 1 prestellar core in the cloud model B30; 4 dense cores and 1 prestellar core in model B10; and 2 dense cores and 3 prestellar cores in model B1. In model B1, gas can efficiently mix and accrete toward denser regions by turbulence due to the weak magnetic support, which explains why more prestellar cores can be formed inside weakly magnetized clouds.

We show in Figure \ref{fig:gas_density} the gas column density distribution within 5 pc cloud scale of models B30, B10, and B1. The zoom-in maps to one of their cores is shown in the middle and lower rows. The density and magnetic field morphology of all cores identified in these cloud are presented in Appendix \ref{sec:appen_core_illustration}. In the moderate-field cloud (model B30; left column), gas forms filamentary structures perpendicular to the well-ordered magnetic field along the $x$-direction. This well-ordered magnetic field is preserved from $\sim 5$ pc down to $\sim 0.2$ pc core scale (Figures \ref{fig:gas_density}$\rm a_{\rm 1}, ~b_{\rm 1}, ~c_{\rm 1}$), but it is then bent by gravitational contraction at $\sim 0.1$ pc scale (Figure \ref{fig:gas_density}$\rm c_{\rm 1}$). In contrast, in weak- and very weak-field clouds (models B10 and B1), gas and magnetic fields are governed by large-scale turbulent motions, producing vortical, distorted magnetic fields following the clumpy density structures on the $\sim 5$ pc cloud scale (Figures~\ref{fig:gas_density}$\rm a_{\rm 2}$, $\rm a_{\rm 3}$). The field then becomes more distorted inside starless cores by gravity effects (Figures \ref{fig:gas_density}$\rm b_{\rm 2}, ~\rm b_{\rm 3}, ~\rm c_{\rm 2}, ~\rm c_{\rm 3}$). 

\subsection{POLARIS post-processing}\label{sec:polaris}
\subsubsection{Radiative transfer simulation}\label{sec:MCRT}
We post-process the three above datasets with POLARIS to simulate polarized thermal dust emission inside MCs. We first perform the 3D Monte-Carlo Radiative Transfer (RT) simulation to get the radiation field distribution required for modeling dust temperature. We consider interstellar radiation field (ISRF) to be the major heating source of dust grains inside MCs, assuming $G_{0} = 1$. Their spectral energy distribution will follow the Mathis ISRF \citep{Mathis_1983, Camps_2015}, with wavelengths spanning from $\lambda_{\rm min} = 0.1\mum $ to $\lambda_{\rm max} = 3$ mm. We assume the uniform dust-to-gas mass ratio of $\eta = 0.01$ (typical values found in ISM of the Milky Way) from clouds to starless cores \footnote{Indeed, $\eta$ is found to vary among star-forming regions, i.e., $\eta \sim 0.0105-0.0125$ in OphA (\citealt{Liseau_2015}), $\eta \sim 0.0085-0.012$ in $A_{\rm V} < 10$ mag and $\eta \sim 0.0033-0.0025$ in $A_{\rm V} < 10$ mag of M17 (\citealt{Zhao_2025}). Theoretical studies also predict the increase of $\eta$ inside dense filamentary structures of MCs if large grains beyond $>0.25\mum$ exist there (\citealt{Hopkins_2016}, \citealt{Tricco_2017}, \citealt{Commercon_2023}, \citealt{Fukihara_2026}). In this case, the grain alignment efficiency inside cores will worsen due to stronger attenuation of the ISRF by stronger dust reddening effect. However, since large grains only occupy a small amount inside the entire dust size distribution, we do not expect the significant variation of $\eta$ from clouds to starless cores (\citealt{Tricco_2017}). Therefore, it may be still safe to adopt the constant $\eta = 0.01$ across clouds to cores in our study.
 }. We choose the Astrodust model \citep{Draine_Hensley_2021a, Draine_Henseley_2021b, Draine_Henseley_2021c}, and assume the grain size distribution to follow a standard MRN size distribution with $\rm dn/da = C n_{\rm H} a^{-3.5}$ \citep{Mathis_1977}, where $C$ is the normalization constant determined by the dust-to-gas mass ratio $\eta$. The minimum grain size is set to $a_{\rm min} = 5\rm nm$ and the maximum grain size is varied from $0.25\mum$ to $2\mum$ to understand the impact of grain growth on dust polarization (see Section \ref{sec:synthetic_dust_polarization}).
  
Interstellar photons are injected inward from the boundaries of the computational domain. \textsc{Polaris} keeps tracking all scattering and absorption events between each photon and dust grains along their trajectory, storing their deposited energy in each passing cell until the photon is absorbed by dust grains \citep{Lucy_1999}. Dust emission is considered right after each absorption event to guarantee the energy conservation. We use $5 \times 10^{7}$ photon packages per wavelength and 35 wavelengths spanning $0.1\mu$m to 3 mm for the RT \footnote{We adopt 35 wavelengths because of computational limitations. We did a test using a larger number of wavelengths and a higher number of photon packages per wavelength, and received almost similar radiation field and dust temperature distributions as the simulation done with 35 wavelengths. The resulting dust temperature and radiation field maps with our choice of MCRT setup also exhibit negligible Monte Carlo noise, which is sufficient to accurately compute the following Stokes $I$, $Q$, and $U$ maps.}.At the end of RT, the equilibrium temperature of each grain size, $T_{\rm d}(a)$, is determined again from the local radiation energy density stored inside the cell. We assume the coupling between gas and grains inside MCs and starless/prestellar cores where $n_{\rm H} > 10^{4}\cm^{-3}$ (\citealt{Goldsmith_2001}, \citealt{Galli_2002}) and set the gas temperature $T_{\rm g}$ inside each cell to be equal $T_{\rm d}$ the effective dust temperature integrated over the grain size distribution.

\subsubsection{Grain alignment modeling}\label{sec:modelling_grain_alignment}
Knowing the distribution of the radiation field and dust temperature, we then model the magnetic alignment for all grain size inside every cell of the simulation domain following RATs/MRAT alignment theory \citep{Lazarian_Hoang_2007a, Hoang_Lazarian_2008}, using the updated version of POLARIS developed by \cite{Giang_2023a}. POLARIS first determines the range of grain sizes which can be aligned with $\B$ following the suprathermal rotation condition by RATs (which determines the minimum alignment size $a_{\rm align}$) and the fast Larmor precession condition (which determines the maximum alignment size). Then, it determines the external alignment mechanism (RATs or MRAT) for each grain size within the alignment size range based on their magnetic relaxation parameter $\delta_{\rm m}$ ($\delta_{\rm m}$ characterizes how fast the magnetic relaxation timescale compared to gas damping timescale, \citealt{Hoang_Lazarian_2016_mrat}, \citealt{Hoang+2022}, \citealt{Giang_2023a}). We consider RAT alignment for grains having $\delta_{\rm m} < 1$, and MRAT alignment for grains having higher $\delta_{\rm m}$ (\citealt{Hoang_Lazarian_2016_mrat}). Following the setup from \cite{Giang_2023a}, we choose the fraction of grains aligning with $\B$ at high-\textit{J} attractors, $f_{\rm high-J} = 0.25$ for RAT alignment, and $f_{\rm high-J} = 0.5$ for grains having $1 \leq \delta_{\rm m} \leq 10$ and $f_{\rm high-J} = 1$ for grains having $\delta_{\rm m} > 10$. Knowing the fraction of grains aligned with $\B$ at high$-$ and low$-$J attractors (i.e., grains having the magnetic alignment at thermal rotation), we finally determine the internal alignment efficiency for all grain sizes inside each group using the Barnett relaxation mechanism (see detailed in \citealt{Giang_2023a}). The resulting density-weighted grain alignment states along the LOS inside all cores identified in the three cloud models are presented in Appendix \ref{sec:appen_grain_alignment}.

\begin{figure*}
\centering
    \includegraphics[width=\textwidth,height=\textheight,keepaspectratio]{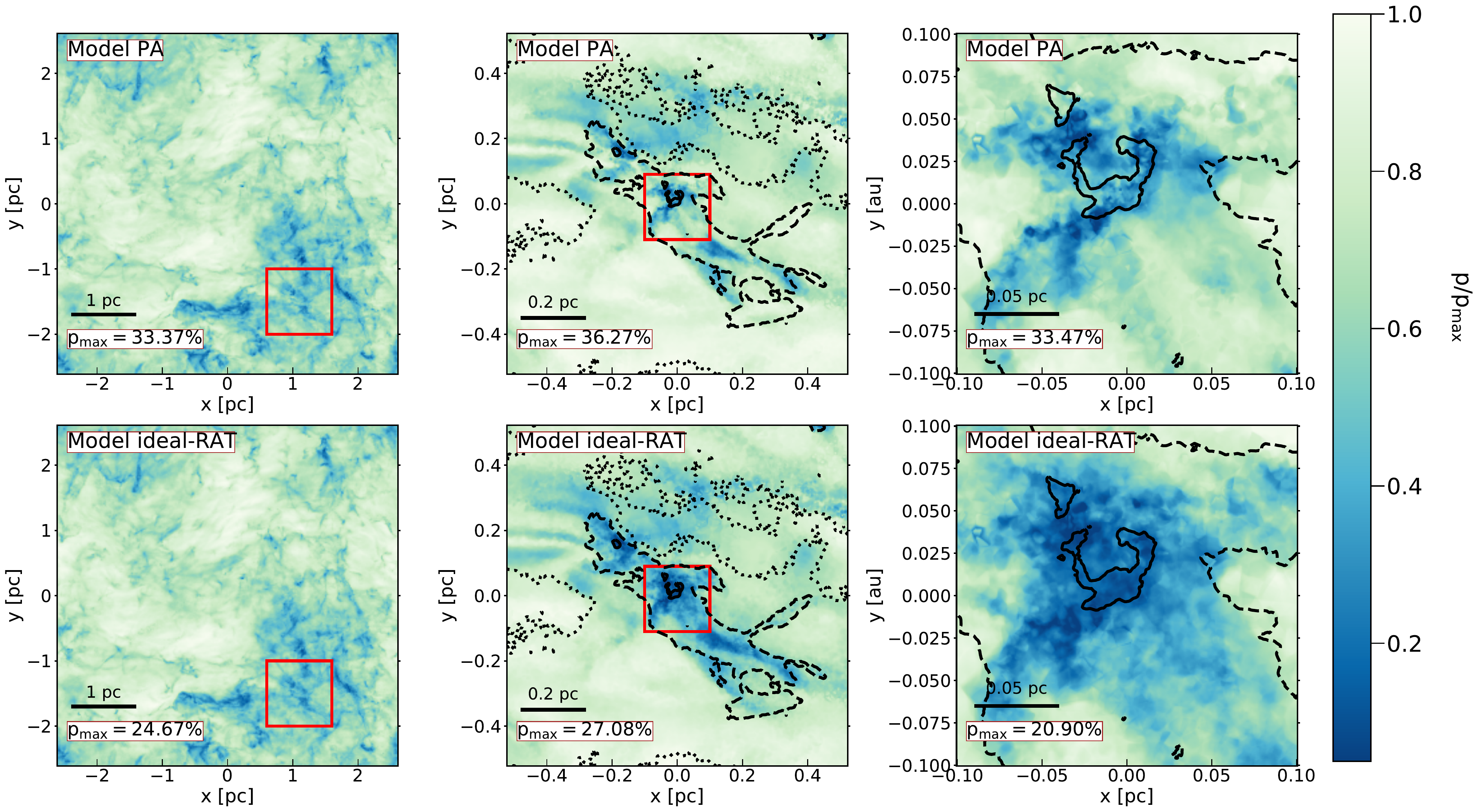}
    \caption{Normalized polarization degree map obtained at 5 pc cloud scale (left column), 1 pc (middle column) and 0.2 pc (right column) core scale of core $\#2$ in model B30 (left column of Figure \ref{fig:gas_density}). Upper row shows results obtained from model PA, while lower row shows results from model ideal$-$RAT, assuming $a_{\rm max} = 1\mum$. The normalized maximum polarization degree $p_{\rm max}$ at each scale is denoted in the lower left corner of each panel. Dotted, dashed, and solid black contours mark the positions of $N_{\rm H} = 10^{21}, 10^{22}, 10^{23}\cm^{-2}$. Model ideal$-$RAT shares similar reduction factors $p/p_{\rm max}$ inside the cloud (left column) and in the outer part of 1 pc core scale (middle column) to model PA, but it shows stronger depolarization in the inner 0.2 pc region (right column).}
     \label{fig:polarization_degree_map}
\end{figure*} 

Knowing the alignment size range, the internal alignment state, and the external alignment mechanism, the net magnetic alignment degree of each grain size in the RAT framework will be quantified by the Rayleigh reduction factor \citep{Reissl_2014, Reissl_2016}, which is defined as:
 
\bea 
R = f_{\rm high-J}Q_{\rm J}^{\rm highJ} Q_{\rm X}^{\rm highJ}   + (1-f_{\rm highJ}) Q_{\rm J}^{\rm lowJ} Q_{\rm X}^{\rm lowJ},
\ena 

where $Q_{\rm J}$ and $Q_{\rm X}$ are the internal and external alignment degrees of dust grains, with the superscripts $^{\rm highJ}$ and $^{\rm lowJ}$ referring to alignment state at the high- and low-\textit{J} attractor. We assume the perfect external alignment for all alignment states ($Q_{\rm J}^{\rm highJ} = Q_{\rm J}^{\rm lowJ} = 1$), while adjust the internal alignment degree depending on the competitive of Barnett relaxation timescale and gas damping timescale (see detailed in \citealt{Giang_2023a}). Values for $Q_{\rm X}$ and $Q_{\rm J}$ considered to each alignment case are summarized in Table \ref{tab:parameter}. Generally, grains can achieve perfect magnetic alignment if they have $f_{\rm high-J} = 1$ by the MRAT mechanism and experience fast internal relaxation at high$-$\textit{J} attractors, or $Q_{\rm X}^{\rm high-J} = 1$. The alignment efficiency is lower if grains have smaller $f_{\rm high-J}$, and it is further reduced when grains experience slow internal relaxation at high- and low-\textit{J} attractors, i.e., smaller $Q_{\rm X}^{\rm low-J}$ and $Q_{\rm X}^{\rm high-J}$\footnote{The internal alignment efficiency of grains undergoing slow internal relaxation remains uncertain. For example, if the grain angular momentum is initially close to a principal axis of the grain inertia, the perfect internal alignment may still occur regardless of their long internal relaxation timescale compared to gas damping timescale (\citealt{Hoang_Lazarian_2009}). However, how popular this state is unclear. In our study, we assume low $Q_{\rm X}$ because we think the strong gas randomization will prevent grains from returning to their lowest rotational energy states despite their thermal or suprathermal rotation.}.

   \begin{table*}
  \centering
         \caption{\textsc{Polaris} run initial setup}
  \begin{tabular} {ccccc}
  \hline 
\textbf{Model name} & \textbf{Grain alignment effect} & \textbf{Alignment size range} & \textbf{Slow internal relaxation $^{a}$} & $f_{\rm high-J}$ \\
  \hline 
PA & No & -- & -- & -- \\
ideal-RAT & Yes & $a_{\rm align} - a_{\rm max}$ & No & 1 \\
MRAT & Yes & $a_{\rm align} - a_{\rm max,JB}^{\rm Lar}$ & Yes & $\delta_{\rm m}$$^{b}$ \\
   \hline 
    \label{tab:model_POLARIS}
    \end{tabular}\\
    \footnotesize{($^{a}$): consider the slow internal relaxation if grain size have the Barnett relaxation timescale being larger than the gas damping timescale} \\
    \footnotesize{($^b$): $f_{\rm high-J}$ is determined by the magnetic relaxation ratio $\delta_{\rm m}$ which depends on the grain size, grain magnetic properties, and gas damping timescale.}\\
\end{table*} 
\subsubsection{Synthetic polarized dust emission simulation}\label{sec:synthetic_dust_polarization}
Given the distribution of dust temperature and alignment degree of dust grains over the simulation domain, we finally solve the polarized radiative transfer of the Stokes parameters to generate synthetic maps of Stokes $I$, $Q$, and $U$. We place a detector including $256 \times 256$ pixels at a distance of 100 pc to the source to observe $5\rm pc \times 5\rm pc \times 5\rm pc$ cloud-scale region at $850\mum$. For zoom-in observations, we extract subvolumes of $1\rm pc \times 1\rm pc \times 0.2\rm pc$ and $0.2\rm pc \times 0.2\rm pc \times 0.2\rm pc$, centered on the peak volume density of each starless/prestellar core, from the cloud to minimize the contamination from foreground and background dust emission. We use the same detector plane to observe their producing polarized dust emission. The resulting spatial resolution changes from 0.02 pc for observations toward 5 pc cloud scale, to 805 au for zoom-in observations toward 1 pc core scale, and 161 au for observations at 0.2 pc core scale. The polarization fraction in unit of percentage is calculated as:
\bea 
p = \frac{\sqrt{Q^{2} + U^{2}}}{I}\times 100\%.
\ena 

\subsubsection{Model setup and Synthetic polarization degree map}\label{sec:model_setup}
To disentangle the effects of magnetic field morphology and grain alignment efficiency on polarized dust emission, we consider model Perfect Alignment (PA) and ideal Radiative Torque alignment (ideal$-$RAT). We assume perfect magnetic alignment (or $f_{\rm high-J} = 1$) for all dust grains from $a_{\rm min}$ to $a_{\rm max}$ in model PA \footnote{Note that model PA is quite unrealistic inside starless/prestellar cores because very small grains are expected to not magnetically align due to the strong depletion of UV-optical ISRF photons there. The perfect alignment of all dust grains may approximately happen if small grains are quickly depleted by grain growth process, but how efficient it is is still unclear. Regardless this isuee, we still adopt model PA as a reference to clearly indicate the contribution of aligned dust grains on the depolarization seen from clouds to starless cores.}, while only allowing grains larger than the minimum alignment size $a_{\rm align}$ (Section \ref{sec:modelling_grain_alignment}) to be perfectly aligned with $\B$ in model ideal$-$RAT. We use the compact (porosity $P=0$) Astrodust model with the axial ratio of $s = 0.7143$ (a typical value for interstellar grains, \citealt{Draine_Henseley_2021b, Draine_Henseley_2021c}), and the maximum grain size of $a_{\rm max} = 1\mum$ as a fiducial dust model. Finally, we capture the complicated behavior of aligned dust grains inside MCs and starless cores by considering model MRAT, which accounts for both internal and external alignment efficiencies of grains determined in Section \ref{sec:modelling_grain_alignment}. Within the MRAT framework, we consider one model of PM grains containing $10\%$ of iron atoms (characterized by an iron fraction $f_{\rm p} = 0.1$), and three models of SPM grains to understand the impact of grain magnetic properties on dust polarization. For SPM grains, we assume iron clusters to occupy $\sim 9\%$ inside the grain volume (parameterized by a volume filling factor $\phi_{\rm sp} = 0.03$), and choose to vary the number of iron atoms per cluster (or adjust iron cluster size) to change the grain magnetic susceptibility. A summary of all parameters adopted for the \textsc{Polaris} post-processing is provided in Tables~\ref{tab:parameter} and \ref{tab:model_POLARIS}.

We show in Figure \ref{fig:polarization_degree_map} the synthetic polarization degree map of the cloud model B30 and the zoom-in maps toward the embedded starless cores determined in the left column of Figure \ref{fig:gas_density}. The systematic decrease of the polarization degree toward high density region is clearly seen in both models PA (upper row) and ideal$-$RAT (lower row), from cloud to core scale. The normalized polarization fraction $p/p_{\rm max}$ obtained in model ideal$-$RAT shares the same values with model PA in 5 pc cloud and 1 pc core scale where $N_{\rm H} < 10^{22}\cm^{-2}$ (black dashed contour). But in the inner 0.2 pc region (right column), the depolarization obtained in model ideal$-$RAT appears prominent than the one from model PA. Specifically, while model PA shows $p/p_{\rm max}$ to reduce from $\sim 0.8$ at $N_{\rm H} \sim 10^{22}\cm^{-2}$ to $\sim 0.4$ at $N_{\rm H} \sim 10^{23}\cm^{-2}$ (upper right panel), model ideal$-$RAT yields $p/p_{\rm max} \sim 0.6$ at $N_{\rm H} \sim 10^{22}\cm^{-2}$ and smaller $p/p_{\rm max} \sim 0.1$ at $N_{\rm H} \sim 10^{23}\cm^{-2}$ (lower right panel). In addition to producing stronger depolarization, model ideal-RAT also results in lower maximum polarization degrees due to their narrower alignment size range compared to the assumption in model PA\footnote{Zoom-in synthetic observations at 1 pc and 0.2 pc scale yields higher maximum polarization degree than observations at 5 pc scale due to the removal of foreground/background contaimination (see description in Section \ref{sec:synthetic_dust_polarization}).}.

\section{Effect of Magnetic field and Aligned dust grains on Dust polarization from Clouds to Starless cores}\label{sec:Slos_gamma_align_p}
This section focuses on quantifying the impact of the magnetic field morphology and grain alignment efficiency on dust polarization properties shown in Figure \ref{fig:polarization_degree_map}. Following \cite{Chen_2016}, both variations in the inclination angle of the magnetic field relative to the LOS and the dispersion of the POS magnetic field along the LOS can reduce the net observed dust polarization. Let $\gamma$ be the angle between the local magnetic field direction from the MHD simulation and the LOS, the mean field inclination angle $\langle \gamma\rangle$ of the study domain will be calculated from the quantity:
\bea 
 \langle \cos^{2}\gamma \rangle  = \frac{\int_{\rm los} \cos^{2} \gamma n_{\rm H} dl}{\int_{\rm los} n_{\rm H} dl}, 
\ena
that gives:
\bea 
\langle \gamma \rangle = \arccos \sqrt{ \langle \cos^{2}\gamma \rangle }.\label{eq:<gamma>}
\ena

Let $\theta$ be the angle of the local POS magnetic field to the North direction, the tangledness of the POS magnetic fields along the LOS can be quantified via parameter $S_{\rm los}$ (\citealt{Seifried_2019}), which follows:
\bea 
S_{\rm los} = \sqrt{\frac{\int_{\rm los} (\theta - \langle \theta \rangle) n_{\rm H} dl}{\int_{\rm los} n_{\rm H} dl}}, \label{eq:Slos}
\ena
where $\langle \theta \rangle$ is the mean POS magnetic field orientation. Following \cite{Valdivia_2022}, we first weight $\sin 2\theta$ and $\cos 2\theta$ with density $n_{\rm H}$ and integrate them along the LOS to obtain $\langle \sin 2\theta \rangle$ and $\langle \cos 2\theta \rangle$. The mean POS field orientation is then calculated by:
\bea 
\langle \theta \rangle = \frac{1}{2}\arctan \frac{\langle \sin 2{\theta} \rangle}{\langle \cos 2{\theta} \rangle}.\label{eq:<theta>}
\ena
 
Finally, we calculate the density-weighted minimum alignment size $\langle a_{\rm align} \rangle$ to quantify the impact of alignment size range on dust polarization, which is given as:
\bea 
\langle a_{\rm align} \rangle =\frac{\int_{\rm los} a_{\rm align} n_{\rm H} dl}{\int_{\rm los} n_{\rm H} dl}.\label{eq:<align>}
\ena

All quantities $\langle \gamma \rangle$, $S_{\rm los}$, and $\langle a_{\rm align} \rangle$ are weighted by $n_{\rm H}$ in order to emphasize the contribution of magnetic fields and aligned dust grains in areas where almost all polarized dust emission is produced \footnote{For each observed scale, three above quantities are calculated from the uniform 3D datacube of $n_{\rm H}$, $B$, and $a_{\rm align}$ which shares the similar resolution with the synthetic polarization map to account for the averaging of $\B$ fields inside large pixel size. With 161 au resolution within 0.2 pc core scale, we may underestimate the disorganization and bending of $\B$ in the core center because 161 au is larger than the smallest resolution of all chosen cores. However, since almost no grain alignment is found at this scale, we do not expect a significant change in our conclusion about the relative contribution of aligned dust grains and magnetic field morphology on the depolarization detected there.}. The spatial distribution of $S_{\rm los}$, $\langle \gamma \rangle$, and $\langle a_{\rm align} \rangle$ for all dense cores identified in three cloud models is presented in Appendix \ref{sec:appen_core_illustration}.

\begin{figure*}
\centering      \includegraphics[width=\textwidth,height=0.2\textheight,keepaspectratio]{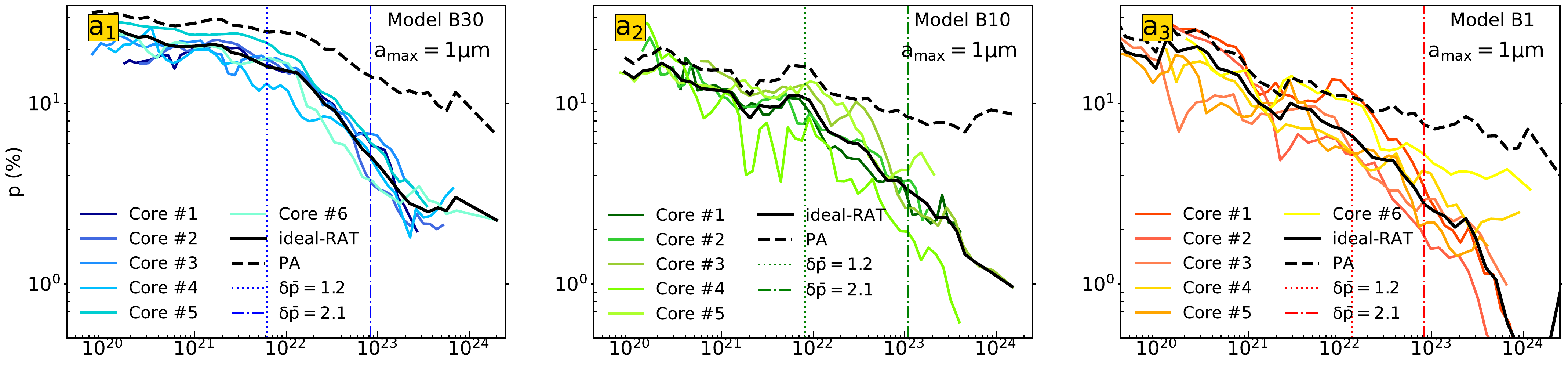}
\includegraphics[width=\textwidth,height=0.2\textheight,keepaspectratio]{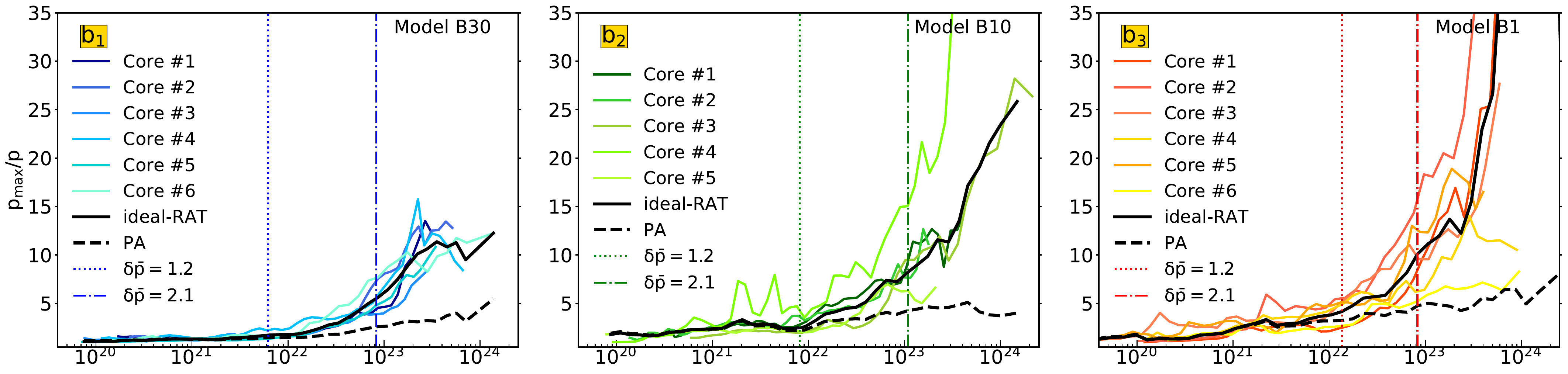}
\includegraphics[width=\textwidth,height=0.2\textheight,keepaspectratio]{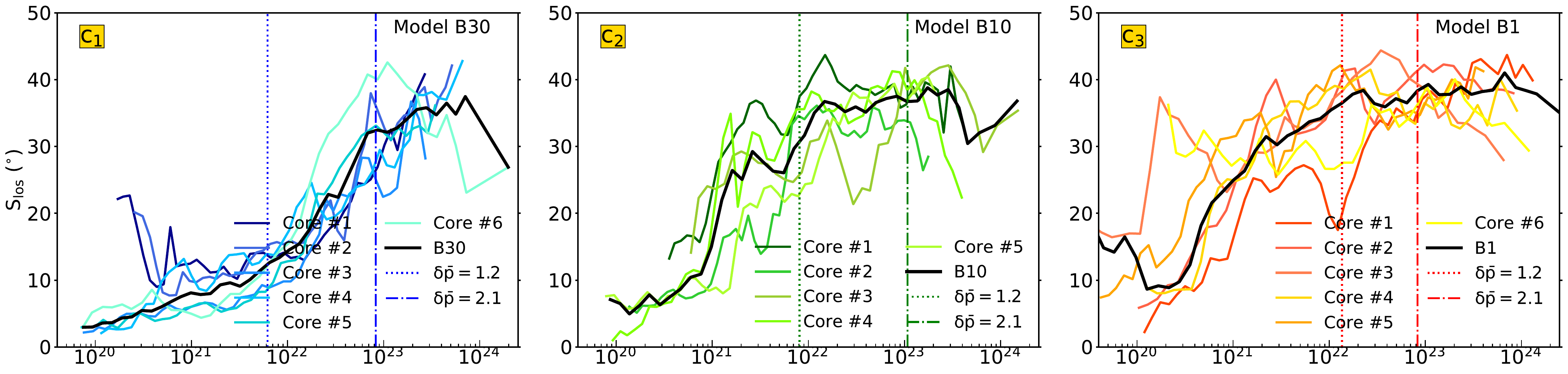}
\includegraphics[width=\textwidth,height=0.2\textheight,keepaspectratio]{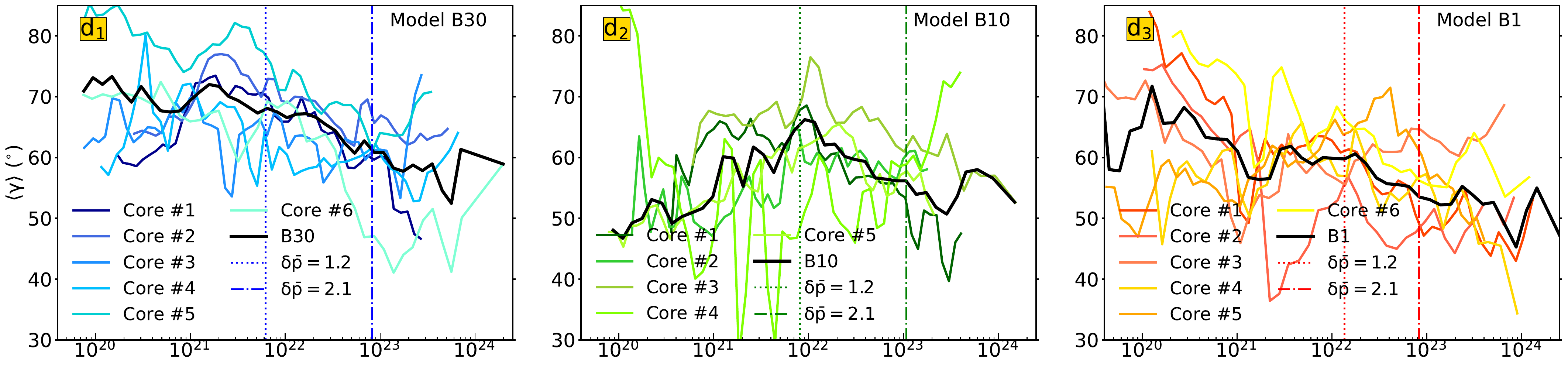}
\includegraphics[width=\textwidth,height=0.2\textheight,keepaspectratio]{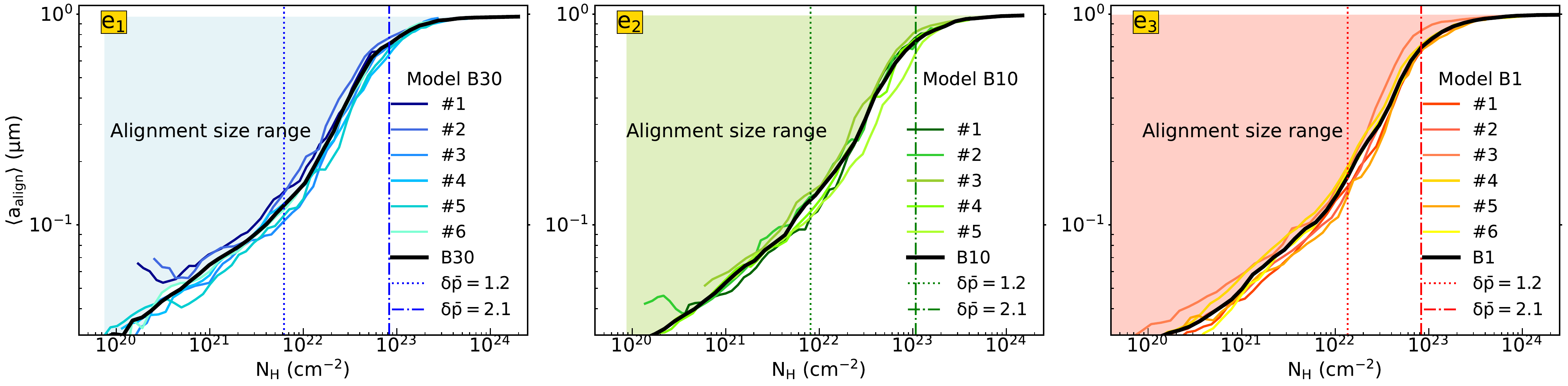}
\caption{Columns from left to right show the results for model B30, B10, and B1 corresponding to intermediate to weak magnetized cloud models. \textbf{Rows a and b}: variation of $p(\%)$ and normalized polarization degree $p_{\rm max}/p$ with column densities $N_{\rm H}$ within 1 pc core scale. \textbf{Rows c, d, e}: variation of the angle dispersion of the POS magnetic field along the LOS $S_{\rm los}$, the field inclination angle $\langle \gamma \rangle$, and the density$-$weighted minimum alignment size $\langle a_{\rm align} \rangle$ with $N_{\rm H}$. Color lines show the mean tendency of each quantity found in each individual core with model ideal$-$RAT, with the summarized tendency in the black solid lines. We include the mean tendency summarized from all cores in model PA in black dashed line for comparison. Two vertical lines mark the column densities at which the ratio $\delta \bar{p} = [p_{\rm max}/p]^{\rm ideal-RAT}/[p_{\rm max}/p]^{\rm PA}=  1.2$ and 2.1. \ textbf {The colored area in row e marks the alignment size range that contributes to producing dust emission polarization}. The depolarization is present in all cores in both models PA and ideal$-$RAT, with stronger depolarization at higher $N_{\rm H}$ in model ideal$-$RAT.}
     \label{fig:p_Slos_gamma_align_NH}
\end{figure*}

\subsection{Depolarization inside moderatelly magnetized clouds}\label{sec:depolarization_B30}
We present in the left column of Figure \ref{fig:p_Slos_gamma_align_NH} the variation of polarization degree, magnetic field morphology, and alignment size range as a function of column density of all starless/prestellar cores \footnote{We use all polarization and column density data found at both 1 pc scale map (with resolution of 805 au) and 0.2 pc scale map (with resolution of 161 au) to find the mean correlation with $N_{\rm H}$.} found in model B30, assuming $a_{\rm max} = 1\mu m$. In Figure \ref{fig:p_Slos_gamma_align_NH}$\rm b_{\rm 1}$, we show the variation of the normalized polarization degree $p_{\rm max}/p$ with $N_{\rm H}$, with $p_{\rm max}$ the maximum polarization fraction measured at the 1 pc core scale. Then, we calculate the quantity $\delta \bar{p}$:
\bea 
\delta \bar{p} = \frac{[p_{\rm max}/p]^{\rm ideal-RAT}}{[p_{\rm max}/p]^{\rm PA}},
\label{eq:Diff}
\ena
to quantify the relative contribution of aligned dust grains to the depolarization compared to the magnetic field geometrical effect. We consider the density range where $\delta \bar{p} < 1.2$ (models ideal$-$RAT and PA share similar $p_{\rm max}/p$) is the region where the magnetic field morphology is the major factor producing the reduction of $p(\%)$ with $N_{\rm H}$. Meanwhile, we consider the density range where $\delta \bar{p} > 2.1$ to be the area where the poor magnetic alignment of dust grains plays a more significant role than the field geometrical effect in producing the depolarization. The position of $\delta \bar{p} = 1.2$ and $\delta \bar{p} = 2.1$ is marked by vertical dashed and dotted lines in each panel. 
  
The depolarization is detected in all cores found in model B30, for both models PA and ideal$-$RAT (Figure \ref{fig:p_Slos_gamma_align_NH}$\rm a_{\rm 1}$). The decline of $p(\%)$ with increasing $N_{\rm H}$ correlates well with the enhanced disorganization of the POS magnetic field $S_{\rm los}$ (Figure \ref{fig:p_Slos_gamma_align_NH}$\rm c_{\rm 1}$), the decreasing mean field inclination angle $\langle \gamma \rangle$ (Figure \ref{fig:p_Slos_gamma_align_NH}$\rm d_{\rm 1}$), and the increasing minimum alignment size $\langle a_{\rm align} \rangle$ (Figure \ref{fig:p_Slos_gamma_align_NH}$\rm e_{\rm 1}$) toward the core center. In the outer core where $N_{\rm H} \lesssim 7\times10^{21}\cm^{-2}$, the depolarization predicted by models PA and ideal-RAT is comparable ($\delta \bar{p} \sim 1$, dashed$-$dot vertical line) due to the wide alignment size range with $\langle a_{\rm align} \rangle \lesssim 0.1\mum$ there (Figure \ref{fig:p_Slos_gamma_align_NH}$\rm e_{\rm 1}$. Toward the inner core where $N_{\rm H} \gtrsim 10^{22}\cm^{-2}$, model ideal$-$RAT produces stronger depolarization than model PA (Figures \ref{fig:p_Slos_gamma_align_NH}$\rm a_{\rm 1}$ and \ref{fig:p_Slos_gamma_align_NH}$\rm b_{\rm 1}$) as a result of the rapid reduction of the alignment size range from $\langle a_{\rm align} \rangle \sim 0.1\mu m$ at $N_{\rm H} \sim 10^{22}\cm^{-2}$ to $\langle a_{\rm align} \rangle \sim 1\mum$ at $N_{\rm H} \gtrsim 10^{23}\cm^{-2}$ (Figure \ref{fig:p_Slos_gamma_align_NH}$\rm e_{\rm 1}$). It seems that for starless/prestellar cores formed inside moderately magnetized clouds, the narrowing alignment size range will overcome the magnetic field morphology effect in producing the depolarization at density $N_{\rm H} > 10^{23}\cm^{-2}$ (i.e., $\delta \bar{p} = 2.1$, dashed vertical line).

In addition to enhancing the depolarization feature, the reduction of the alignment size range also lowers the polarization degree at the core center, yielding typical values of $p \sim 2-3\%$ there. By contrast, our model shows $p \sim 8$–$10\%$ in the core center if the depolarization is driven by the magnetic field effect only (Figure \ref{fig:p_Slos_gamma_align_NH}$\rm a_{\rm 1}$, model PA).

\subsection{Depolarization inside weakly magnetized clouds}\label{sec:depolarization_B10_B1}
We perform the same analysis for all cores of models B10 and B1 to understand the nature behind the depolarization observed inside weakly magnetized clouds. Results are shown in the middle and right columns of Figure \ref{fig:p_Slos_gamma_align_NH}. Opposite to the well-ordered magnetic field in model B30, the high distortion of magnetic fields in models B10 and B1 is observed even at low column densities $N_{\rm H} \lesssim 10^{22}\cm^{-2}$. As shown in Figures \ref{fig:p_Slos_gamma_align_NH}$\rm c_{\rm 2}$ and \ref{fig:p_Slos_gamma_align_NH}$\rm c_{\rm 3}$, one gets $S_{\rm los}$ increases rapidly from $\sim 10–20^{\circ}$ at $N_{\rm H} \sim 10^{20}\cm^{-2}$ to $\sim 30–40^{\circ}$ at $N_{\rm H} \sim 10^{22}\cm^{-2}$, substantially larger than $S_{\rm los} \lesssim 15^{\circ}$ measured at similar column densities in the cloud model B30 (Figure \ref{fig:p_Slos_gamma_align_NH}$\rm c_{\rm 1}$). The field inclination angle with the LOS at $N_{\rm H} \lesssim 10^{22}\cm^{-2}$ is also diverse and more complicated, with $\langle \gamma \rangle \sim 40-60^{\circ}$ in model B10 and $\langle \gamma \rangle \sim 50-70^{\circ}$ in model B1 (Figures \ref{fig:p_Slos_gamma_align_NH}$\rm d_{\rm 2}$ and \ref{fig:p_Slos_gamma_align_NH}$\rm d_{\rm 3}$), difference from the large, stable $\langle \gamma \rangle \sim 70^{\circ}$ in model B30 (Figure \ref{fig:p_Slos_gamma_align_NH}$\rm d_{\rm 1}$). In the inner core where $N_{\rm H} \sim 10^{22}-10^{23}\cm^{-2}$, gravitational contraction further enhances the field disorganization along the LOS and reduces the field inclination angle, yielding higher $S_{\rm los}$ and smaller $\langle \gamma \rangle$ in models B10 and B1 than values found in model B30 at the same density range. However, in the innermost region where $N_{\rm H} > 10^{23}\cm^{-2}$, the field morphology found in models B30, B10, and B1 nearly share similar behaviors, with high $S_{\rm los} \sim 40^{\circ}$ and $\langle \gamma \rangle \sim 50-60^{\circ}$. The impact of the viewing angle on the tangling and bending of magnetic fields will be discussed in Section \ref{sec:view_angle}.

The highly distorted magnetic fields in models B10 and B1 strengthen the reduction of $p(\%)$ with $N_{\rm H}$ predicted by model PA, reaching typical values of $p \sim 5$-$8\%$ at the core center (Figures~\ref{fig:p_Slos_gamma_align_NH}$\rm a_{\rm 2}$ and \ref{fig:p_Slos_gamma_align_NH}$\rm a_{\rm 3}$; black dashed lines). The dominant role of magnetic field geometry in producing the depolarization also slightly extends to higher column densities, increasing from $N_{\rm H} \sim 7 \times 10^{21}\,\cm^{-2}$ in model B30 to $N_{\rm H} \sim 10^{22}\,\cm^{-2}$ in models B10 and B1 (black dash-dotted vertical lines).

 \begin{figure*}
\centering
    \includegraphics[width=\textwidth,height=\textheight,keepaspectratio]{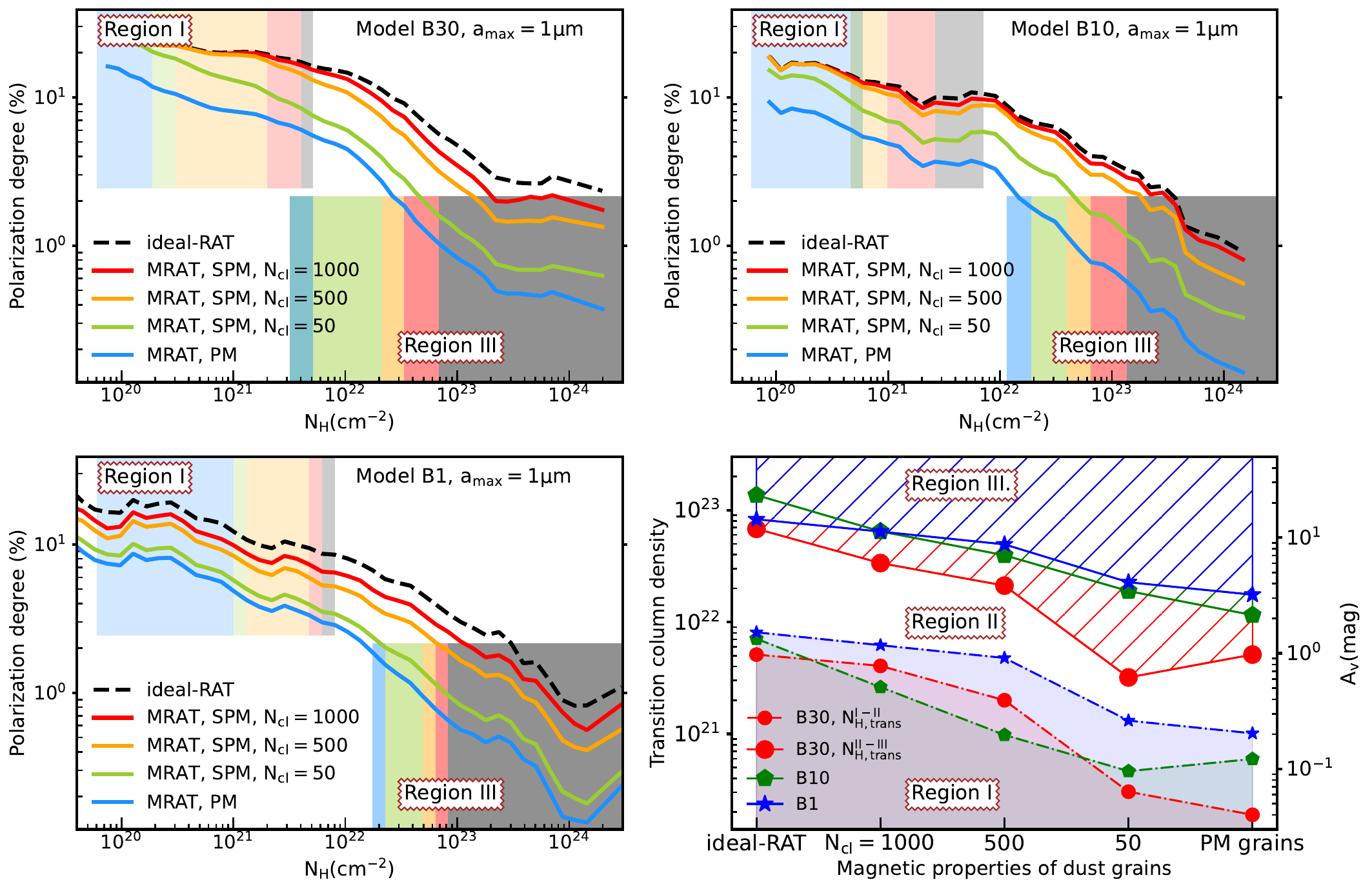}
    \caption{Effect of grain magnetic properties on the $p–N_{\rm H}$ relation found inside starless and prestellar cores of models B30 (upper left panel), B10 (upper right panel), and B1 (lower left panel). Black dashed lines show the mean $p-N_{\rm H}$ variation obtained from all cores with model ideal$-$RAT, while colored lines illustrate model MRAT of SPM grains with $N_{\rm cl} = 1000$ (red), $500$ (orange), $50$ (green), and PM grains (blue), assuming $a_{\rm max} = 1\mum$. Light colored block in the upper left part of each panel marks region I, where depolarization is dominated by the geometrical effect of magnetic fields. Dark colored block in the lower right part marks region III, where the reduced grain alignment efficiency dominates the depolarization. The intermediate area corresponds to region II, where both mechanisms contribute to lower $p(\%)$ with $N_{\rm H}$\footnote{The transition column density $N_{\rm H,trans}^{\rm I-II}$ and $N_{\rm H,trans}^{\rm II-III}$ of each dust model shown in the panels is found using the summarized polarization data from all cores. We check and confirm the similarity of transition column densities for each individual core while using the same dust model.}. Lower right panel: predicted transition column densities $N_{\rm H,trans}^{\rm I-II}$ and $N_{\rm H,trans}^{\rm II-III}$ for different magnetic properties of dust grains. The corresponding visual extinctions $A_{\rm V}$ are indicated on the right y-axis. If large grains are PM or SPM with small iron clusters size (blue, purple), the reduced internal and external alignment efficiency by gaseous damping can start contributing to the depolarization at $N_{\rm H} > 10^{21}\cm^{-2}$ and totally replace the field geometrical effect in lowering $p(\%)$ with $N_{\rm H}$ at $N_{\rm H} > 10^{22}\cm^{-2}$.}
     \label{fig:p_NH_Ncl}
\end{figure*}

However, despite the highly distorted magnetic fields, the narrowing of the grain alignment size range caused by inefficient RATs remains the dominant depolarization mechanism over magnetic field geometry on the $\sim 0.1$ pc scale where $N_{\rm H} > 10^{23}\,\cm^{-2}$ (black dashed vertical lines; Figures~\ref{fig:p_Slos_gamma_align_NH}$\rm b_{\rm 2}$ and \ref{fig:p_Slos_gamma_align_NH}$\rm b_{\rm 3}$), even when grains grow significantly to $\sim1\,\mum$. As illustrated in Figures \ref{fig:p_Slos_gamma_align_NH}$\rm a_{\rm 2}$ and \ref{fig:p_Slos_gamma_align_NH}$\rm a_{\rm 3}$, ideal-RAT model predicts $p \sim 1\%$ at the center of starless/prestellar cores identified in models B10 and B1, much lower than $p \sim 5$-$8\%$ predicted by model PA.
 
 \begin{figure*}
\centering
    \includegraphics[width=\textwidth,height=\textheight,keepaspectratio]{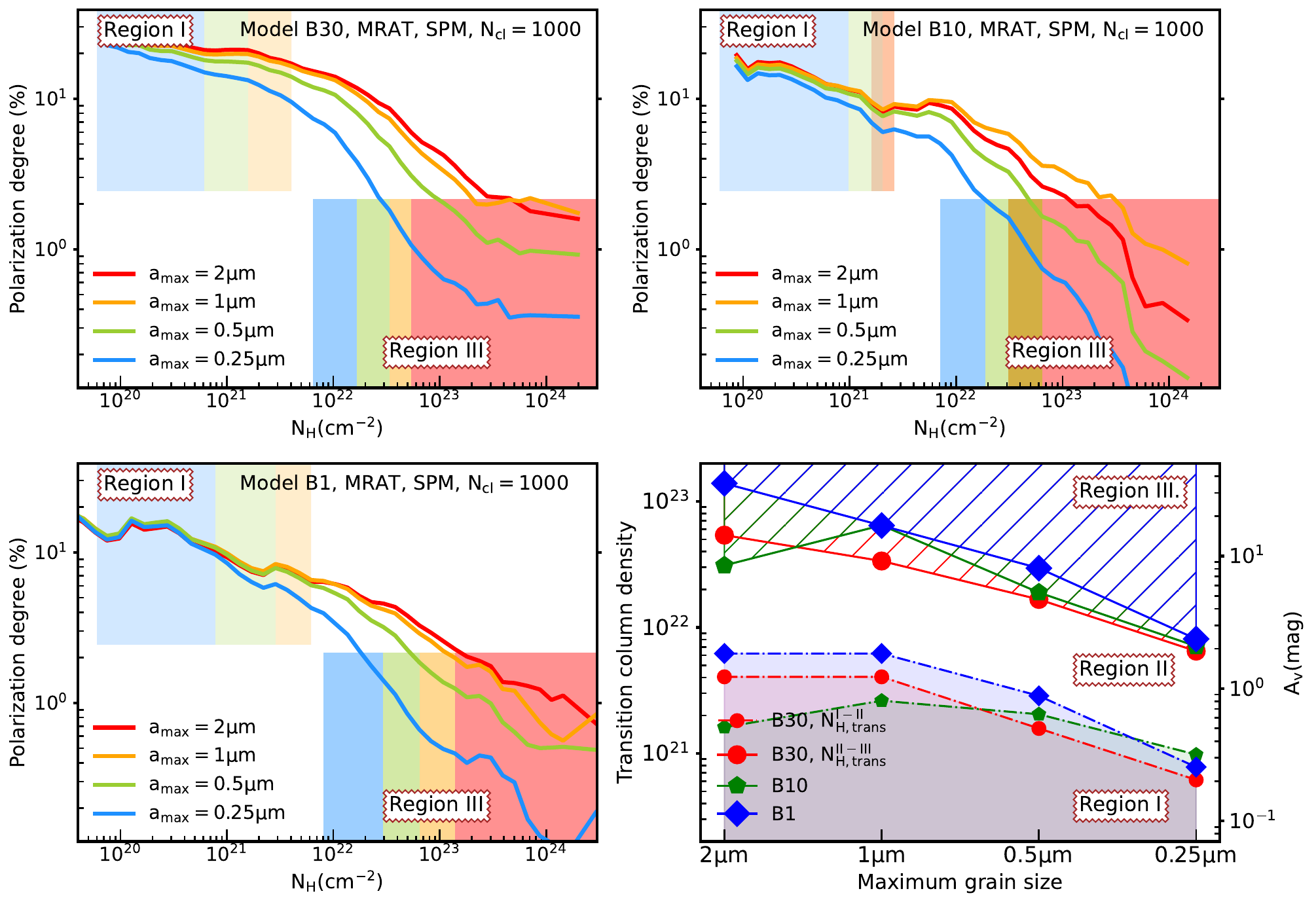}
    \caption{Similar results to Figure \ref{fig:p_NH_Ncl} but for the effect of grain growth on the $p–N_{\rm H}$ relation and the transition column densities $N_{\rm H,trans}^{\rm I-II}$ and $N_{\rm H,trans}^{\rm II-III}$, assuming model MRAT for SPM grains with $N_{\rm cl} = 1000$. When grains grow insignificantly inside starless cores, the alignment loss can join with the field geometrical effect to produce the depolarization at $N_{\rm H} > 10^{21}\cm^{-2}$. It will become the dominant depolarization mechanism at $N_{\rm H} > 10^{22}\cm^{-2}$.}
     \label{fig:p_NH_amax}
\end{figure*}
 
\section{Effects of Grain Physical Properties on the Depolarization within Starless cores}\label{sec:grain_depolarization}

One disadvantage of model ideal$-$RAT is that it does not yet account for the complex alignment of dust grains with magnetic fields inside starless/prestellar cores. As discussed in Section \ref{sec:intro}, assuming all aligned grains to achieve perfect magnetic alignment may underestimate the contribution of grain alignment to the depolarization detected at $<1$ pc core scale, especially if grains grow to micron-sized. To address this limitation, this section focuses on analysing the effect of dust properties on the depolarization predicted by model MRAT. We follow the definition of $\delta \bar{p}$ (Equation~\ref{eq:Diff}) introduced in Section \ref{sec:depolarization_B30} to quantify the relative contributions of magnetic field geometry and grain alignment efficiency to the depolarization. We define the density regime with $\delta \bar{p} < 1.2$ as region I, where magnetic field geometry is the dominant depolarization mechanism. The density regime with $\delta \bar{p} > 2.1$ is defined as region III, where inefficient grain alignment becomes the primary depolarization mechanism. The intermediate density regime between region I and II is referred to as region II, where both effects contribute to reduce $p(\%)$ with $N_{\rm H}$. We define the column density separating regions I and II as $N_{\rm H,trans}^{\rm I-II}$, and that separating regions II and III as $N_{\rm H,trans}^{\rm II-III}$. We investigate the impact of grain magnetic properties and grain growth on dust polarization in Sections \ref{sec:Iron_depolarization} and \ref{sec:amax_depolarization}. The effects of the grain axial ratio and grain porosity will be discussed in Appendix \ref{sec:s_P_depolarization}.

\subsection{Effect of grain magnetic properties}\label{sec:Iron_depolarization}
We present in Figure \ref{fig:p_NH_Ncl} the effect of grain magnetic properties on the $p–N_{\rm H}$ relation obtained within 1 pc scale of all cores identified in models B30, B10, and B1, assuming $a_{\rm max} = 1\mum$. The summary of the transition densities $N_{\rm H,trans}^{\rm I-II}$ and $N_{\rm H,trans}^{\rm II-III}$ obtained from each dust model is illustrated in the lower right panel. In all three cloud models, the perfect magnetic alignment assumed for all dust grains beyond $a_{\rm align}$ in model ideal-RAT can be reproduced only if dust grains are SPM with $N_{\rm cl} \gtrsim 1000$ iron atoms/cluster. Under this condition, grains can achieve efficient internal alignment and align with $\B$ through MRAT mechanism with $f_{\rm high-J} \simeq 1$ (Figure \ref{fig:grain_alignment_all_core}). Therefore, similar to model ideal-RAT, the narrower of the alignment size range will start lowering $p(\%)$ with $N_{\rm H}$ at $N_{\rm H} > 10^{22}\cm^{-2}$ and become the leading depolarization mechanism at $N_{\rm H} > 10^{23}\cm^{-2}$ (lower right panel). By reducing the size of iron cluster sizes (reducing $N_{\rm cl}$) inside SPM grains, the polarization fraction is systematically reduced, and the transition boundaries between regions I-II, and II-III ($N_{\rm H,trans}^{\rm I-II}$ and $N_{\rm H,trans}^{\rm II-III}$) will shift significantly toward lower column densities (Figure \ref{fig:p_NH_Ncl}, lower right panel) due to the increasing abundance of grains aligning with $\B$ at low-\textit{J} attractors with slow internal relaxation (see detailed explanation in Appendix \ref{sec:appen_grain_alignment}).
 
In particular, for PM and SPM with $N_{\rm cl} = 50$ (blue and green), the reduction of internal and external alignment efficiencies of dust grains by gaseous damping (Appendix \ref{sec:appen_grain_alignment}) can partially contribute to depolarization in model B30 at $N_{\rm H,trans}^{\rm I-II} \sim 2–3 \times 10^{20}\cm^{-2}$ (corresponding to $A_{\rm V} \lesssim 0.1$ mag); then become the dominated depolarization mechanism at $N_{\rm H,trans}^{\rm II-III} \sim 6 \times 10^{21}-2\times10^{22}\cm^{-2}$ (or $A_{\rm V} \sim 1$ mag), even with $a_{\rm max} \sim 1\mum$ in the core center (lower right panel, red lines). In weakly magnetized clouds (models B10 and B1), the transition densities obtained in this dust model slightly shift toward higher column density owing to the stronger distortion of magnetic fields, reaching $N_{\rm H,trans}^{\rm I-II} \sim 5 \times 10^{20}$–$10^{21}\cm^{-2}$ (or $A_{\rm V} \sim 0.1–0.2$ mag) and $N_{\rm H,trans}^{\rm II-III} \sim 1-2 \times 10^{22}\cm^{-2}$ (or $A_{\rm V} \sim 3$ mag, lower right panel, green and blue lines).

\subsection{Effect of grain growth}\label{sec:amax_depolarization}
We next examine in Figure \ref{fig:p_NH_amax} the effect of grain growth on the $p–N_{\rm H}$ relation and the locations of the transition column densities $N_{\rm H,trans}^{\rm I-II}$ and $N_{\rm H,trans}^{\rm II-III}$, assuming model MRAT for SPM grains with $N_{\rm cl} = 1000$. As the maximum grain size decreases from $a_{\rm max} = 2 \mum$ to $0.25\mum$, $p(\%)$ will decline with $N_{\rm H}$ faster, and $N_{\rm H,trans}^{\rm I-II}$ and $N_{\rm H,trans}^{\rm II-III}$ will shift toward lower column densities owing to the faster reduction of the alignment size range \footnote{As discussed in Section \ref{sec:Iron_depolarization}, SPM grains with $N_{\rm cl} = 10^{4}$ can maintain their efficient magnetic alignment by MRAT mechanism. Therefore, the reduction of the alignment size range is the dominant factor limiting grain alignment efficiency inside starless cores here.}. In cloud model B30, the reduction of the alignment size range will partly contribute, then fully replace the geometrical effect of magnetic field to become the dominant depolarization mechanism at $N_{\rm H,trans}^{\rm II-III} \sim 1-2\times 10^{21}\cm^{-2}$ and $N_{\rm H,trans}^{\rm II-III} \sim 1-2\times 10^{22}\cm^{-2}$ (corresponding to $A_{\rm V} \sim 0.3-3$ mag, lower right panel, red lines) if $a_{\rm max} \sim 0.25–0.5\mum$ inside starless cores. Similar feature will happen at $N_{\rm H,trans}^{\rm I-II} \sim 1-4 \times 10^{21}\cm^{-2}$ and $N_{\rm H,trans}^{\rm II-III} \sim 1-3 \times 10^{22}\cm^{-2}$ for models B10 and B1, regardless of the high distortion of magnetic fields inside starless cores formed inside weakly magnetized clouds.

\section{Discussion}\label{sec:discuss}
\subsection{Grain alignment state inside starless cores}\label{sec:discuss_grain_alignment}
It is well known that sub-micron grains inside starless cores can be misaligned with the ambient magnetic field owing to the strong shielding of UV and optical photons of ISRF (\citealt{Cho_Lazarian_2005}, \citealt{Pelkonen_2007}, \citealt{Hoang_2021_polarization_hole}). Meanwhile, micron-sized grains, which can be formed via the grain growth process in dense environments, can be magnetically aligned and produce polarized dust emission inside starless cores by interactions with infrared photons of ISRF (\citealt{Cho_Lazarian_2005,Bethell_2007,Pelkonen_2007,Pelkonen_2009}). However, our detailed modeling of grain alignment shows that this picture is not as simple as this.

By analysing the grain alignment state inside 14 starless cores identified from our simulated clouds (Appendix \ref{sec:appen_grain_alignment}), we find that if micron-sized grains are not SPM containing big iron clusters, i.e., $N_{\rm cl} < 1000$ iron atoms/cluster; they will tend to be aligned with $\B$ by RATs only, with $\sim 75\%$ of grains being aligned with $\B$ at low-\textit{J} attractors and all of them will experience slow internal relaxation due to inefficient Barnett relaxation mechanism (Figures \ref{fig:grain_alignment} and \ref{fig:grain_alignment_all_core}). The increasing abundance of grains at low-\textit{J} with inefficient internal alignment is similar to the scenario of alignment loss, which happens when grains grow insignificantly with $a_{\rm max} \sim 0.25-0.5\mum$ (\citealt{Cho_Lazarian_2005}, \citealt{Pelkonen_2007}, \citealt{Hoang_2021_polarization_hole}). The slow internal relaxation not only induce weak polarized thermal dust emission, but also can make the alignment direction of dust grains to the local $\B$ undetermined (see \citealt{Giang_2023b} for protostellar disks), which further complicates the interpretation of polarization angles observed inside sub-pc scale of starless/prestellar cores. 

\begin{figure*}
\centering
    \includegraphics[width=\textwidth,height=\textheight,keepaspectratio]{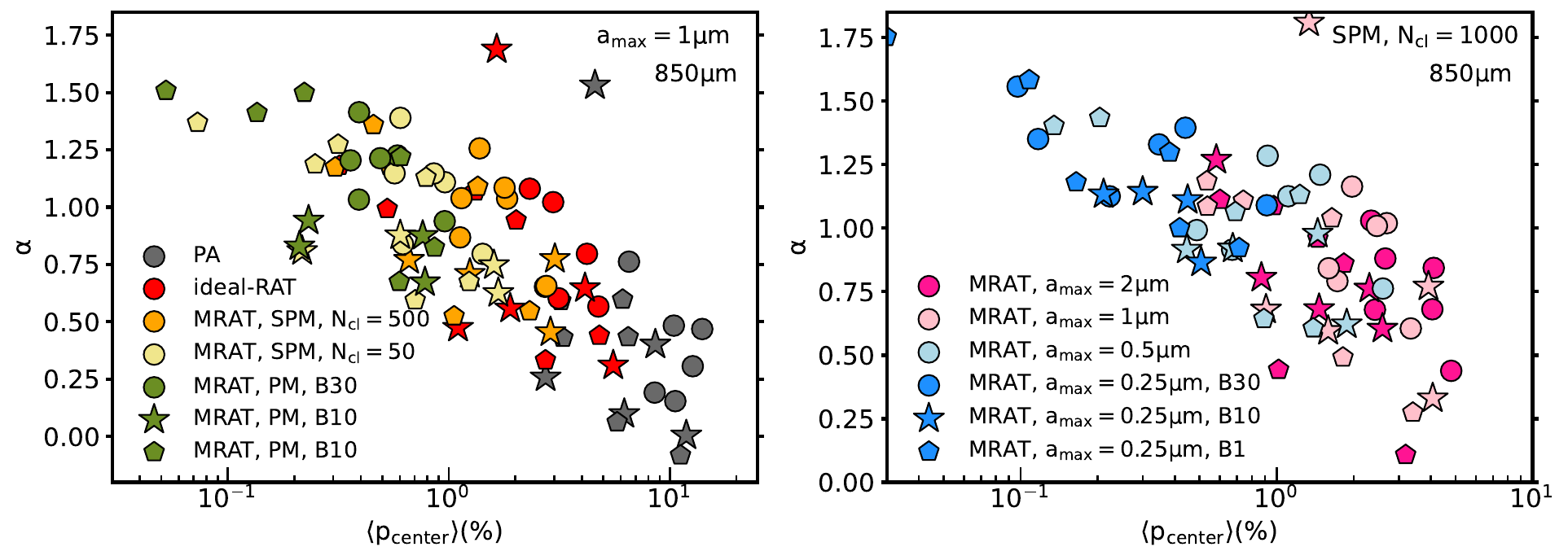}
    \caption{Variation of the $p-I$ slope, $\alpha$ obtained from synthetic polarization data, as a function of the mean polarization degree found at the location of maximum dust emission intensity, $\langle p_{\rm center} \rangle$. Left panel: effect of grain alignment model and grain magnetic properties on the relation $\langle p_{\rm center} \rangle-\alpha$, assuming $a_{\rm max} = 1\mum$. Right panel: effect of grain growth on the correlation, considering model MRAT for SPM grains with $N_{\rm cl} = 1000$. Circles, stars, and hexagons represent starless cores formed in models B30, B10, and B1, respectively. One expects to obtain lower $\langle p_{\rm center} \rangle$ and higher $\alpha$ by decreasing the size of iron clusters embedded inside dust grains and reducing grain growth efficiency. }
     \label{fig:pcenter_alpha_intrinsic}
\end{figure*}

We note that dust grains can also be internally aligned by Inelastic relaxation mechanism. However, this mechanism prefers micron-sized grains than sub-micron grains. Furthermore, it is more sensitive to grain rotational velocity, making it less efficient than Barnett relaxation inside starless/prestellar cores. We test the inelastic relaxation mechanism for sub-micron and micron-sized grains with small iron clusters and confirm the inefficiency of this mechanism in helping grains to have fast internal relaxation inside starless cores. Besides this scenario, increasing the volume filling factor of iron clusters inside dust grains (increasing $\phi_{\rm sp}$ to beyond $>0.03$ assumed in our model) may help SPM grains with small iron cluster size to experience stronger Barnett relaxation and MRAT mechanism, thus better alignment with magnetic fields.

\subsection{Depolarization mechanism - a view from Multiscale Synthetic Dust polarization}\label{sec:discuss_depolarization_origin}
The reduction of polarization degrees of polarized dust emission at submillimeter wavelengths toward denser regions is widely observed from MCs to starless cores. Yet, the physical origin of this phenomenon remains debated. By performing the multiscale synthetic observation of dust polarization from clouds to starless cores, we confirm the dominant role of magnetic field geometry in driving depolarization on the pc-scale clouds where $N_{\rm H} < 10^{21}\cm^{-2}$, as conclusions by \cite{Planck_2015_XX}, \cite{Chen_2016}, and \cite{Seifried_2019} (see Figures \ref{fig:polarization_degree_map} and \ref{fig:p_Slos_gamma_align_NH}). Within $\sim 1$ pc scale of starless cores where $N_{\rm H} > 10^{21}\cm^{-2}$, the depolarization can be classified into two groups. In the first group, the geometrical effect of magnetic fields will continue to produce the depolarization up to $N_{\rm H} \sim 10^{22}\cm^{-2}$ (similar as findings of \citealt{Chen_2016}), then it will combine with the reduction of the alignment size range to lower $p(\%)$ at thousands au scale where $N_{\rm H} < 10^{23}\cm^{-2}$. Beyond this column density, the depolarization mechanism switches to the effect of aligned dust grains, regardless of how complex the magnetic field from our MHD simulation can be (Figure \ref{fig:p_Slos_gamma_align_NH}). As shown in Section \ref{sec:Iron_depolarization}, this picture can only happen if SPM grains contain big iron cluster sizes and are growing to micron-sized inside starless/prestellar cores (Figures \ref{fig:p_NH_Ncl} and \ref{fig:p_NH_amax}, lower right panel).
 
For the inefficient grain growth (i.e., $a_{\rm max} < 0.5\mum$) and for dust grains with small iron inclusion sizes (Section \ref{sec:discuss_grain_alignment}), the grain alignment effect can start contributing to the depolarization inside starless/prestellar cores at lower densities of $N_{\rm H} > 10^{21}\cm^{-2}$. In case of inefficient grain growth (with $a_{\rm max} \sim 0.25-0.5\mum$), the alignment loss can dominate the field geometrical effect to produce the depolarization at $N_{\rm H} > 10^{22}\cm^{-2}$ (Figure \ref{fig:p_NH_amax}, lower right panel), consistent with earlier theoretical expectations \citep{Cho_Lazarian_2005,Bethell_2007,Pelkonen_2007,Hoang_2021_polarization_hole}. Similarly, if micron-sized grains can form inside starless cores, but having small iron clusters size, the reduction of the internal and external alignment degree by efficient gaseous damping will become the major depolarization mechanism at the same density range $N_{\rm H} > 10^{22}\cm^{-2}$ (Figure \ref{fig:p_NH_Ncl}, lower right panel). We found consistent results among 14 cores formed inside intermediately and weekly magnetized clouds. As the grain alignment dynamics are sensitive to the gas volume density and the magnetic field strength rather than density structure and magnetic field morphology, we expect the results not to vary much between starless/prestellar cores formed inside different star formation scenarios, i.e., collapsing clouds, cloud-cloud collision. However, the density and magnetic field strength may be altered under the radiation and mechanical feedback from nearby stars. Further studies are needed to understand the nature behind the depolarization detected inside starless/prestellar cores formed inside strong feedback environments. However, we note that the identification of grain alignment as the dominant depolarization mechanism depends on our choice of $\delta \bar{p} = 2.1$ (Section \ref{sec:depolarization_B30}). Adopting higher thresholds will narrow the area where grain alignment efficiency is considered as the dominant depolarization mechanism. Additionally, if the core has stronger foreground contamination, it may enhance the contribution of magnetic field geometry to the observed depolarization at higher column densities.
 
Interestingly, \cite{King_2019} already mentioned a joint contribution of magnetic field geometry and grain alignment efficiency to the depolarization observed in Vela C at column densities $N_{\rm H} \sim 10^{21}$–$10^{23}\cm^{-2}$. In that study, they adopt a broken power-law to describe the variation of the  polarization efficiency with column densities. They justify their assumption with the variation $a_{\rm align}$ with $A_{\rm v}$ and volume density. However, the physical mechanism behind the reduction of the grain alignment efficiency at low densities of $n_{\rm H} \sim 10^{2}-10^{4}\cm^{-3}$ remains unclear. Further studies that consider the detailed grain alignment processes and their dependence on the grain magnetic properties and grain sizes may help to reveal the origin of the depolarization and the underlying dust physics inside Vela C better.

\subsection{From Depolarization to Dust properties - Synthetic Dust polarization results}\label{sec:discuss_relation_intrinsic}
Observationally, the depolarization observed in sub-pc scale of dense cores is commonly characterized by a power-law relation, $p \propto I^{-\alpha}$, where $I$ is the thermal dust emission intensity (e.g., \citealt{Kate_2019,Kate_2021,Kalory_2023,Choi_2024}). The power-law index, $\alpha$, is often used to characterize the grain alignment efficiency (e.g., \citealt{Whittet_2008,Kate_2019}). Within this framework, $\alpha \sim 0$ suggests that grains remain efficiently aligned throughout the core, while $\alpha \sim 1$ implies grains to significantly lose their magnetic alignment toward high intensity areas. However, these interpretations have not yet been tested using synthetic dust polarization data. Furthermore, as discussed in Sections \ref{sec:grain_depolarization} and \ref{sec:discuss_depolarization_origin}, the depolarization mechanism also depends closely on the grain magnetic properties and the maximum grain size. In this section, we will investigate the relationship between the dust model and the power-law index $\alpha$, which may help improve the interpretation of dust physics underlying dust polarization signal observed toward starless and prestellar cores.

We use synthetic polarization data obtained at 0.2 pc scale, which corresponds to the $\sim 3^{\prime}$ radius field of view of JCMT observations at $\sim 140$ pc (the typical distance to starless core sample in nearby regions measured by JCMT). We rebin synthetic Stokes $I$, $Q$, $U$ maps from 161 au (Section~\ref{sec:synthetic_dust_polarization}) to 560 au, which corresponds to $4^{\prime \prime}$ pixel size of JCMT. Then, we convert fluxes from mJy/px to mJy/beam, using a JCMT beam size at $850\mum$ with full-width-half-maximum $\rm FWHM = 14.1^{\prime \prime}$. As discussed in Sections \ref{sec:Iron_depolarization}, \ref{sec:amax_depolarization}, and \ref{sec:discuss_depolarization_origin}, the depolarization mechanism at $0.2$ pc scale (corresponding to $N_{\rm H} > 10^{21}\cm^{-2}$, Figures \ref{fig:gas_density}) can change depending on the grain alignment efficiency with $N_{\rm H}$. Therefore, the $p-N_{\rm H}$ relation, and then $p-I$ relation \footnote{Note that $850\mum$ is sensitive to cold grains in the core center. Therefore, the peak intensity of thermal dust emission will overlap with the peak of column density, inducing the similar behavior of $p-I$ and $p-N_{\rm H}$.} can be described by the broken power-law function. Since the outermost part of the studied region can be significantly affected by noise (Section \ref{sec:discuss_relation_noise}), we will focus only on the slope $\alpha$ derived from the variation of $p(\%)$ with $I$ near the core center. Besides, we compute the mean polarization degree at the peak intensity and denote it as $\langle p_{\rm center} \rangle$. The relation of $\alpha$ and $\langle p_{\rm center} \rangle$ for each core calculated by each dust model used in Section \ref{sec:grain_depolarization} is presented in Figure \ref{fig:pcenter_alpha_intrinsic}.  

We find that if the depolarization inside starless cores is produced only by the disorganization and bending of magnetic fields (model PA), one would expect relatively high central polarization fractions, $\langle p_{\rm center} \rangle \sim 3$–$15\%$, accompanied by shallow $p–I$ slopes, $\alpha \sim 0$–$0.6$. Cores forming in weakly and very weakly magnetized clouds tend to show slightly lower $\langle p_{\rm center} \rangle$ and steeper $\alpha$ than those forming in moderately (or strongly) magnetized clouds. However, the assumption of perfect alignment for all grain sizes inside starless cores is unrealistic (Section \ref{sec:model_setup}). When the grain alignment conditions are taken into account (model ideal-RAT), $\langle p_{\rm center} \rangle$ systematically decreases to $\sim 2$–$6\%$, while $\alpha$ increases to $\sim 0.3$–$1$ owing to the additional contribution from the reduced alignment size range to the depolarization (Figure~\ref{fig:p_Slos_gamma_align_NH}). The $\langle p_{\rm center} \rangle$–$\alpha$ relation further shifts to the upper-left corner of the parameter space when we take the internal and external alignment degrees of dust grains inside starless cores into account. One expects $\langle p_{\rm center} \rangle$ to decrease and $\alpha$ to increase with decreasing iron clusters size embedded inside dust grains (Figure \ref{fig:p_NH_Ncl}). For example, if grains grow to $a_{\rm max} = 1\mum$ but are PM, one expects to obtain $\langle p_{\rm center} \rangle \sim 0.2–1\%$ and $\alpha \sim 0.75–1.5$\footnote{The high power index $\alpha > 1$ arises from the combined effects of the weak polarized emission from the densest regions and the extra tangledness of magnetic fields from foreground. Indeed, such high $\alpha > 1$ never been detected in observational data. But mathematically, it can happen when fitting synthetic results showing the rapid decline of $p(\%)$ with $I$ with the power-law function. This issue will be addressed in Section \ref{sec:discuss_relation_noise}.}.

The shift of $\langle p_{\rm center} \rangle - \alpha$ toward lower polarization degrees and higher power indexes also occurs if the alignment loss happens inside the core center (Figure \ref{fig:p_NH_amax}). For example, one expects $\langle p_{\rm center} \rangle \sim 0.4–2\%$ and $\alpha \sim 0.75-1.5$ (lighblue) if SPM with $N_{\rm cl} = 1000$ can only grow to $a_{\rm max} = 0.5\mum$ in the core center. In the next section, we assess whether the dependence of $\langle p_{\rm center} \rangle$ to grain properties can be recovered in the presence of observational noise.
 
\subsection{From Depolarization Properties to Dust Physical properties - Noise effect}\label{sec:discuss_relation_noise}
\subsubsection{Observed synthetic dust polarization}\label{sec:adding_noise}
To bring synthetic polarization data closer to realistic observations of JCMT, we first convolve the synthetic Stokes $I$, $Q$, and $U$ maps in units of mJy/px generated in Section \ref{sec:discuss_relation_intrinsic} with JCMT beamsize at $850\mum$ \citep{Zamponi_2022}. The process is performed using the \textbf{convolve-fft} package in Python, which convolves the maps with the 2D Gaussian kernel having the standard deviation of $\sigma = \mathrm{FWHM}/\sqrt{8\ln2}$. After convolution, we convert the maps to mJy/beam and add white noise to synthetic data. Following \cite{Seifried_2019}, we consider noise of Stokes I, Q, U to be independent to each other. We assume noise to follow the Gaussian distribution with zero mean and standard deviations $\sigma_I = \sigma_{QU} \simeq 3$ mJy/beam - the rms noise levels measured at $4^{\prime \prime} $ pixel size in Oph C by JCMT \citep{Kate_2019}. We apply the same noise model to all synthetic datasets to avoid introducing additional free parameters related to the noise amplitude. We note that this choice may cause issues on datasets having very low polarized emission intensity (see Section~\ref{sec:recovery}). An example of synthetic polarization maps before and after adding noise and the noise impact of $\alpha$ fitting will be shown in Appendix \ref{sec:append_add_noise}.

\begin{figure*}
\centering \includegraphics[width=\textwidth,height=\textheight,keepaspectratio]{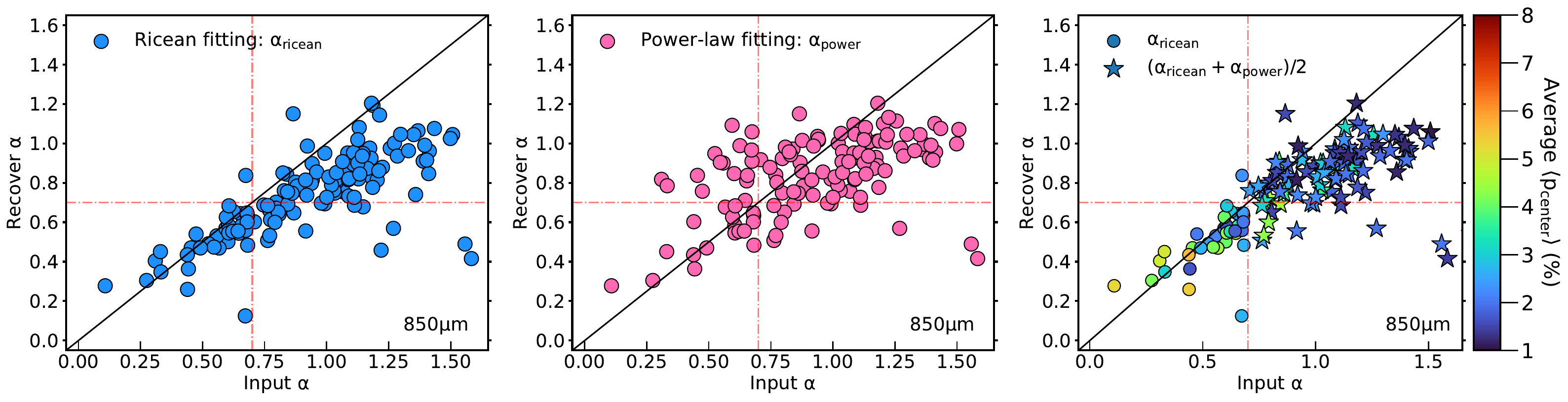}
    \caption{Correlation between the $p–I$ power index recovered from the noisy synthetic polarization data and derived from synthetic data. The slope derived from synthetic data (Figure \ref{fig:pcenter_alpha_intrinsic}) is labeled as Input $\alpha$ on the x-axis. Left panel: correlation of $\alpha_{\rm ricean}$ - the power index recovered from Ricean fitting and $\alpha$. Middle panel: results for $\alpha_{\rm power}$ returned from power-law fitting. Right panel: relation of $\alpha_{\rm ricean}-\alpha$ for $\alpha < 0.7$ (circle), and the median values of $\alpha- (\alpha_{\rm ricean} + \alpha_{\rm power})/2$ for $\alpha > 0.7$ (square). The color bar of the right panel illustrates the average polarization degree found at the location of the maximum dust emission intensity, $\langle p_{\rm center} \rangle$, found from two fitting methods. Black solid line marks the 1:1 relation, while red lines indicate $\alpha = \alpha_{\rm ricean} = \alpha_{\rm power-law} = 0.7$. Ricean fitting works better than Power-law fitting in revealing the power index derived from the synthetic $p-I$ relation in the case $\alpha < 0.7$. For $\alpha \sim 0.7 - 1$, the average of the power index returned from Ricean and power-law fitting can provide values closer to $\alpha$ than each method alone, but it does not work to recover $\alpha > 1$.}
     \label{fig:alpha_fitting}
\end{figure*}

In observational studies, polarization data are usually debiased before fitting the $p$–$I$ relation to the power-law function. However, \cite{Kate_2019} demonstrated that the probability distribution function (PDF) of the observed polarized intensity tends to follow a Ricean distribution, with high signal-to-noise (S/N) data following Gaussian profile, while low S/N data following the skewness of the distribution. In contrast to high S/N data that can show the $p-I$ relation from the emission source, low S/N data always show $p \propto I^{-1}$. As a result, debiased polarization data may still contaminate noise in low intensity regions, that artificially steepens the $p–I$ relation and overestimates the realistic values of $\alpha$. Therefore, \cite{Kate_2019} suggest fitting the non-debiased polarization data with the Ricean distribution to better recover $\alpha$. Using Monte Carlo simulation to generate polarization data, they show that Ricean fitting can recover the $p-I$ slope better when the  slope is $\alpha \lesssim 0.7$. In contrast, when $\alpha > 0.7$, Power-law fitting can return $\alpha$ better than Ricean fitting. However, two fitting methods have not yet been tested using synthetic dust polarization data, that motivates us to re-visit this issue using our samples of dust polarization. The fitting method is described in Appendix \ref{sec:fitting}.

\begin{figure*}
\centering
\includegraphics[width=\textwidth,height=\textheight,keepaspectratio]{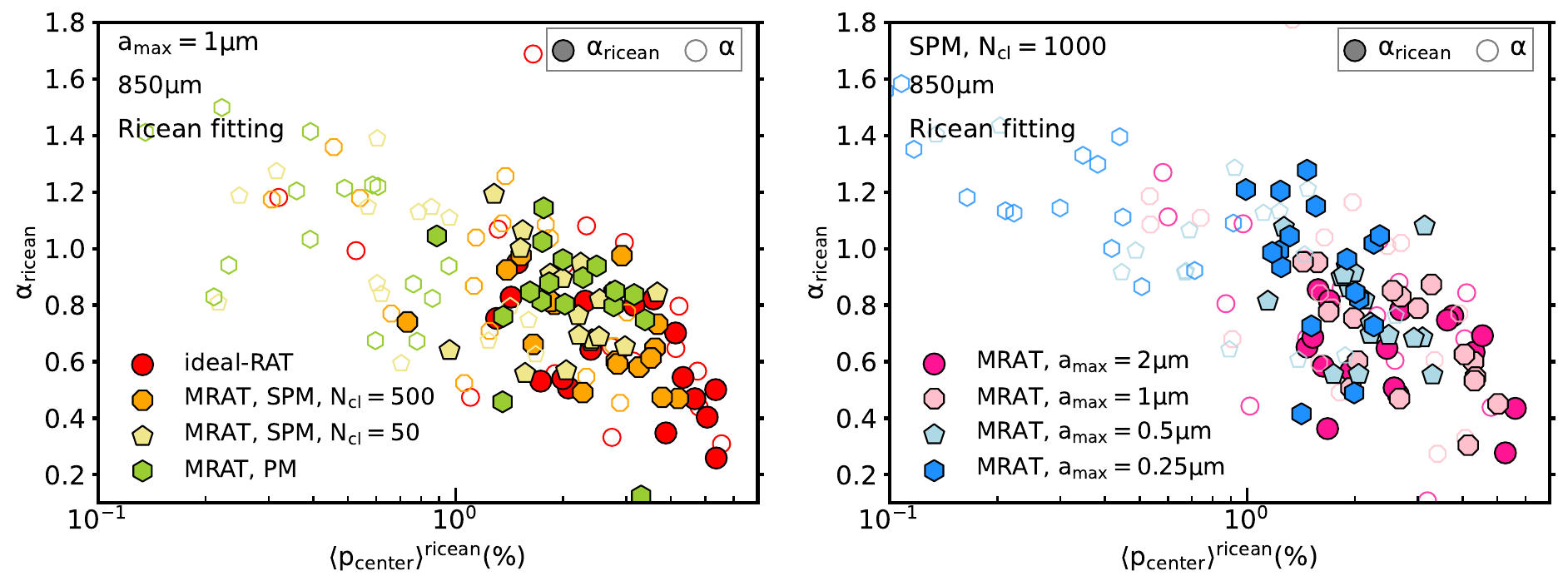}
\includegraphics[width=\textwidth,height=\textheight,keepaspectratio]{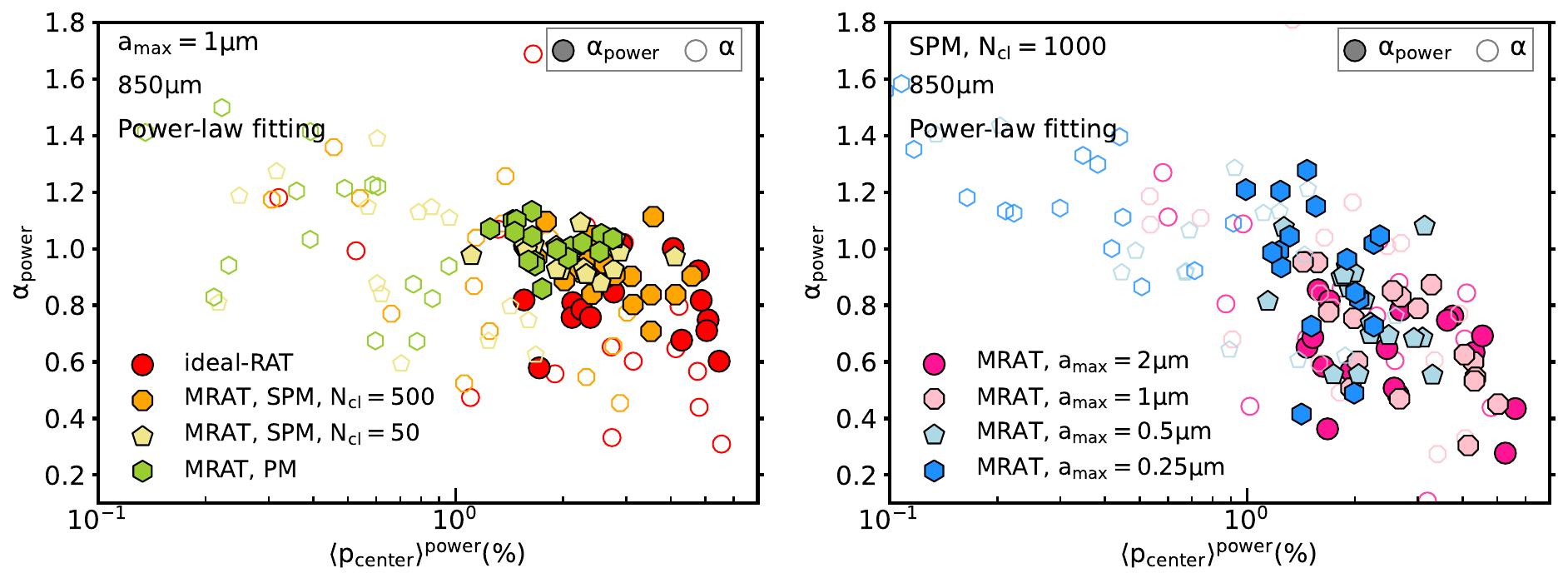}
\caption{Upper row: $\alpha_{\rm ricean} - \langle p_{\rm center} \rangle^{\rm ricean}$ relation recovered from Ricean fitting. Lower row: $\alpha_{\rm power} - \langle p_{\rm center} \rangle^{\rm power}$ relation recoved from the power-law fitting. Left column shows results obtained from different grain alignment models (ideal-RAT and MRAT) and grain magnetic properties, while right column shows results for different grain growth scenarios. Empty circles represent the relation obtained from synthetic polarization data shown in Figure \ref{fig:pcenter_alpha_intrinsic} (except model PA). Ricean fitting can rather well reveal the connection between depolarization properties and dust models in the case $\alpha < 0.7$ (red, orange, pink, hotpink groups). In cases where dust grains produce very low $\langle p_{\rm center} \rangle < 2\%$, the recovered polarization degree at the peak intensity is typically limited to $\sim 1\%$ due to noise contamination.}
     \label{fig:pcenter_alpha_fitting}
\end{figure*}

\subsubsection{Recovery rate} \label{sec:recovery}
We examine the recoverability of the  $p–I$ slope by comparing $\alpha$ to the power index returned from Ricean fitting $\alpha_{\rm ricean}$ and Power-law fitting $\alpha_{\rm power}$ in Figure \ref{fig:alpha_fitting}. In addition, to access whether the $\langle p_{\rm center} \rangle$–$\alpha$ relation shown in Figure \ref{fig:pcenter_alpha_intrinsic} can be recovered from observational data, we show in Figure \ref{fig:pcenter_alpha_fitting} these correlations obtained from the Ricean and Power-law fitting \footnote{$\langle p_{\rm center} \rangle^{\rm ricean}$ and $\langle p_{\rm center} \rangle^{\rm power}$ in Figure \ref{fig:pcenter_alpha_fitting} are the mean polarization fraction at the peak intensity inferred from each fitting method to the observed synthetic $p–I$ relation.}.  
 
From results shown in the left and middle panels of Figure \ref{fig:alpha_fitting}, we statistically confirm the Monte Carlo results of \cite{Kate_2019} that Ricean fitting can recover the slope for $\alpha \lesssim 0.7$ (left panel) better than Power-law fitting, which tends to overestimate $\alpha$, i.e., from $\alpha \sim 0.3-0.7$ to $\alpha_{\rm power} \sim 0.7-1$ (middle panel). Consequently, the $\langle p_{\rm center} \rangle^{\rm ricean}-\alpha_{\rm ricean}$ values returned from Ricean fitting can better reveal again the physical properties of dust grains that produce properties of observed dust polarization (Figure \ref{fig:pcenter_alpha_fitting}, upper row). 

In contrast, when $\alpha > 0.7$, Power-law fitting seems to perform better jobs than Ricean fitting in capturing the steep $p–I$ relation. Although Power-law fitting still overestimates the power index for $\alpha \sim 0.7–1$ and underestimates values for $\alpha > 1$, this method at least returns a consistent, stable recovered range among $\alpha_{\rm power} \sim 0.75–1$. Ricean fitting, by contrast, while preserving a positive correlation of $\alpha$ and $\alpha_{\rm ricean}$, it tends to underestimate the real values of $\alpha$ and exhibits a wide scatter, with recovered values spanning among $\alpha_{\rm ricean} \sim 0.4–1.6$. This issue makes their recovered power index $\alpha_{\rm ricean}$ be highly uncertain in case $\alpha > 0.7$. Beyond the issue of their returned value of $\alpha$, neither Ricean nor Power-law fitting can accurately reproduce again the  $\langle p_{\rm center} \rangle–\alpha$ relation when $\alpha > 0.7$. As shown in Figure \ref{fig:pcenter_alpha_intrinsic}, the high $\alpha > 0.7$ is usually accompanied by $\langle p_{\rm center} \rangle < 1\%$ at the peak intensity. In this case, noise can strongly contaminate into polarization signals at the core center, that explains why both Ricean and Power-law fitting return $\langle p_{\rm center} \rangle \sim 1\%$ \footnote{We note that there are some cases that the extremely low  polarization fractions $\ll 1\%$ are artificially boosted to $\sim 2–3\%$ in the recovered results. This effect happens because we apply the same noise level to all dust models. While this choice is appropriate for sources showing $\langle p_{\rm center} \rangle \gtrsim 1\%$, it may overestimate noise in datasets with weak polarization signals. This issue may not appear in the realistic observations, and such sources would likely be detected at $\langle p_{\rm center} \rangle \sim 1\%$ or remain below detection thresholds.} at $I=I_{\rm max}$ .

\subsubsection{Link to observational results}
Indeed, it is challenging to determine whether Ricean or Power-law fitting is a better method to recover the  $p–I$ slope. Moreover, connecting $\alpha$ to the physical characteristics of dust grains also poses some difficulties owing to the overlapping results produced from different combinations of grain models and magnetic field morphology. But overal, our results can be classified into two main groups:

The first group includes cases showing $\alpha_{\rm ricean} < 0.7$, $\alpha_{\rm power} > 0.7$, and relatively high central polarization fractions $\langle p_{\rm center} \rangle > 2\%$. As discussed in Section \ref{sec:recovery}, these signatures indicate that $\alpha$ likely falls within $\alpha < 0.7$, the regime in which Ricean fitting does better job than Power-law fitting in recovering both $\alpha$ (Figure \ref{fig:pcenter_alpha_fitting}, left panel) and $\langle p_{\rm center} \rangle–\alpha$ relation (Figure \ref{fig:pcenter_alpha_fitting}, upper row). In this group, depolarization is usually produced by the combined effects of the reduced alignment size range and the geometrical effect of magnetic fields. Based on the polarization angle dispersion and the relative importance of magnetic energy over gravitational energy and thermal/non-thermal pressure, we may quantify the contribution of the field geometrical effect to depolarization. But it may be hard to accurately evaluate the role of above factors due to the coupling of $\B$ and dust physics in producing $p(\%)$ (red, orange, hot-pink, and pink groups in Figure~\ref{fig:pcenter_alpha_fitting}). But regardless of this issue, this group can only be produced if dust grains are SPM with big iron cluster size and they are growing significantly to $> 0.5\mum$. With this configuration, SPM grains inside starless cores seems to be aligned with $\B$ by MRAT mechanism (Section \ref{sec:discuss_relation_intrinsic}). Some starless cores observed at submillimeter wavelengths appear to fall into this category. Examples include Pipe$-$109, which exhibits $\langle p_{\rm center} \rangle \sim 8\%$ at the peak intensity \citep{Alves_2014} and $\alpha_{\rm ricean} \sim 0.41$ \citep{Kandori_2020}; L1512 with $\langle p_{\rm center} \rangle \sim 8\%$ and $\alpha_{\rm ricean} \sim 0.48$ \citep{Lin_2024}; and L1495 with $\langle p_{\rm center} \rangle \sim 4\%$ and $\alpha_{\rm ricean} \sim 0.73$. Recently, \cite{Tram_2025}; Quang Anh et al., in prep. suggest dust grains inside Pipe-109 and L1512 to grow significantly and have efficient magnetic alignment in the core center, that matches with our prediction of dust models derived from the observed values of $\alpha$ and $\langle p_{\rm center} \rangle$. If it is a case, our results suggest dust grains inside Pipe-109 and L1512 to be SPM, with iron clusters occupying at least $\sim 9\%$ of the grain volume, and each cluster containing $N_{\rm cl} \gtrsim 500-1000$ iron atoms (Figure \ref{fig:pcenter_alpha_fitting}). Future polarization modeling that explicitly includes grain magnetic properties will be essential to test this prediction for Pipe-109, L1512, and similar objects.

 \begin{figure*}
\centering
\includegraphics[width=\textwidth,height=\textheight,keepaspectratio]{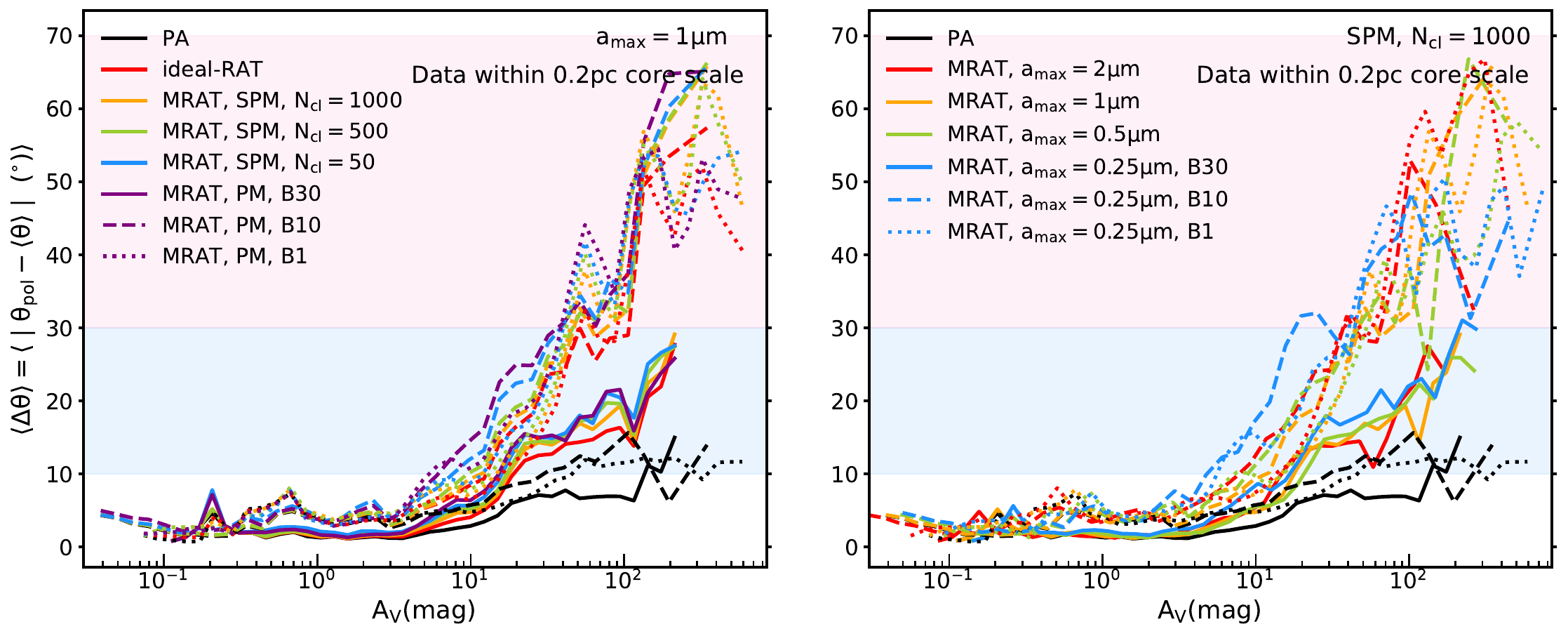}
    \caption{Variation of the mean angle difference $\langle \Delta \theta \rangle$ between magnetic field orientation inferred from dust polarization $\theta_{\rm pol}$ and the mean POS field from MHD simulations $\langle \theta \rangle$ as a function of the visual extinction $A_{\rm V}$. Each colored line represents the tendency averaged over all cores for each dust model. Solid, dashed, and dotted lines illustrate results for cloud models B30, B10, and B1. Left panel: Effect of grain alignment model and grain magnetic properties on $\langle \Delta \theta \rangle - A_{\rm V}$. Right panel: Effect of grain growth on the $\langle \Delta \theta \rangle–A_{\rm V}$ relation. The angle deviation starts earlier, i.e., at lower $A_{\rm V}$, with decreasing magnetization levels inside the parent clouds, decreasing the iron cluster sizes, and decreasing the maximum grain size. For cores formed inside weekly magnetized clouds (models B10 and B1), one can get $\langle \Delta \theta \rangle \sim 40-60^{\circ}$ in the core center due to the grain alignment effect.}  
     \label{fig:delta_phi_Av}
\end{figure*}

The second group includes cases showing low central polarization degree $\langle p_{\rm center} \rangle^{\rm ricean} \approx \langle p_{\rm center} \rangle^{\rm power} < 2\%$. As discussed in Section \ref{sec:recovery}, neither Ricean nor Power-law fitting alone can approximately return the  slope, i.e., Power-law fitting tends to overestimate $\alpha$, while Ricean fitting tends to underestimate them (Figure~\ref{fig:alpha_fitting}). However, taking their average values, $(\alpha_{\rm ricean} + \alpha_{\rm power})/2$, can yield the power index that is closer to their real power indexes of $\alpha \sim 0.7–1$, as demonstrated in the right panel of Figure \ref{fig:alpha_fitting}. This averaging approach does not work when $\alpha > 1$, owing to the large scatter in $\alpha_{\rm ricean}$. Fortunately, such steep slopes $\alpha > 1$ are typically accompanied by extremely low  $\langle p_{\rm center} \rangle < 1\%$ (Figure \ref{fig:pcenter_alpha_intrinsic}), which tends to appear as $\sim 1\%$ of detected polarization fraction in the core center (Figure \ref{fig:pcenter_alpha_fitting}). In these cases, power-law fitting can return more reliable power index than the estimation from Ricean fitting (Figure~\ref{fig:alpha_fitting}). Some starless cores may belong to this second group, including Oph$-$C with $\langle p_{\rm center} \rangle \sim 2\%$, $\alpha_{\rm power} \sim 1$ \citep{Liu_2019} and $\alpha_{\rm ricean} \sim 0.76–0.86$ \citep{Kate_2019}; L43 with $\langle p_{\rm center} \rangle \sim 0.8\%$ and $\alpha_{\rm ricean} \sim 0.83$ \citep{Kalory_2023}; SMM$-$16 with $\langle p_{\rm center} \rangle \sim 1\%$ and $\alpha_{\rm ricean} \sim 0.59$ \citep{Kate_2021}; and L1689B with $\langle p_{\rm center} \rangle \sim 3\%$ and $\alpha_{\rm ricean} \sim 0.56$ \citep{Kate_2021}. Results from Ricean fitting of \cite{Kate_2021} might suggest that SMM$-$16 and L1689B belong to the first group. But because of their low central polarization degrees, further tests with Power-law fitting may be necessary to better classify the depolarization group that two objects belong to.  

As shown in Figure \ref{fig:pcenter_alpha_intrinsic}, the depolarization in the second group is majorly produced by the poor magnetic alignment of dust grains inside starless cores. RAT is found to be the alignment mechanism for dust grains in this case, that recently figured out in L43 by \cite{Scheiter_2025}. However, it remains unclear whether this poor grain alignment arises from the inefficient grain growth, or from large grains with small iron clusters (see the overlap of $\langle p_{\rm center} \rangle - \alpha$ from two scenarios in Figures \ref{fig:pcenter_alpha_intrinsic} and \ref{fig:pcenter_alpha_fitting}). Another observational constraint on the maximum grain size inside starless cores may help to distinguish these possibilities. For example, the detection of Coreshine phenomenons in near- and mid-infrared inside clouds and starless cores (\citealt{Pagani_2010}, \citealt{Steinacker_2014}, \citealt{Lefevre_2014}, \citealt{Kim_2016}) can be interpreted as a sign of micron-sized grains formed by grain-grain coagulation or mantle formation in dense star-forming regions (\citealt{Lefevre_2014}, \citealt{Jones_2016}, \citealt{Ysard_2016}). The dust emissivity index inferred from SED fitting of thermal dust emission can also be used as indirect evidence for grain growth, as theoretical studies predict that the presence of hundred-micron/millimeter-sized grains can produce $\beta < 1$ (\citealt{Testi_2014,Ysard_2019,Carpine_2026}). However, the extremely low $\beta < 1$ is only reported in the inner envelope of Class 0/I Young Stellar Objects (\citealt{Sadavoy_2010}, \citealt{Schnee_2014}, \citealt{Bracco_2017}, \citealt{Cacciapuoti_2023a}) and Class 0/I disks (\citealt{Kwon_2009}, \citealt{Chiang_2012}, \citealt{Miotello_2014}, \citealt{Galametz_2019}). It is unclear to interpret  $\beta \sim 1-2$ as sign of early grain growth due to its complex dependence on the maximum grain size, dust composition, dust model, optical depth effects, LOS temperature variations, and low dust temperatures inside starless/prestellar cores (\citealt{Juvela_2013,Schnee_2014,Carpine_2026}). The high dust opacity $\kappa$ is also used as the sign of early grain growth inside clouds and starless cores (\citealt{Kramer_2003}, \citealt{Suutarinen_2013}, \citealt{Jun_2024}). However, this method relies on the detection of background stars (\citealt{Webb_2017}) and is limited due to the high optical depth at optical-MIR range (\citealt{Juvela_2015}). Regardless of the difficulties of each method, having a first constraint on grain growth will help to better classify the reason behind the depolarization detected in group 2.
 
Another scenario that we do not consider in our study is the evolution of grain shape/grain composition inside starless cores. As discussed in Appendix \ref{sec:s_P_depolarization}, the increase/decrease of the grain axial ratio $s$ and grain porosity $P$ may change the slope of $p-I$ relation due to the strong dependence of $p(\%)$ with $s$ and $P$. \cite{Tram_2025} found that to reproduce the variation of $p(\%)$ with $A_{\rm V}$ from optical to submillimeter wavelengths seen in Pipe-109 \citep{Alves_2014}, dust grains must grow to $\sim $ micron-size and become more elongated, i.e., smaller $s$, toward the core center. This finding supports the anisotropic grain growth scenario proposed by \cite{Hoang_2022}. Future studies that incorporate the evolution of grain shapes and compositions inside starless cores will help to better understanding their impact in controlling the depolarization. 
  
\subsection{Accuracy of magnetic fields inferred from dust polarization inside starless cores} \label{sec:discuss_trace_B} 
 \begin{figure*}
\centering     
\includegraphics[width=\textwidth,height=\textheight,keepaspectratio]{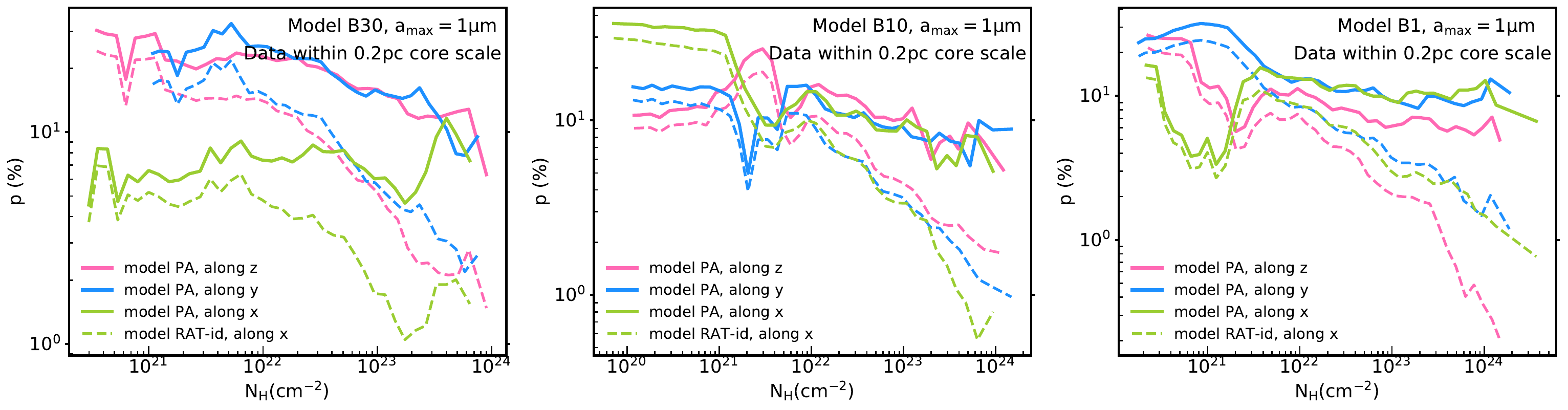}
\includegraphics[width=\textwidth,height=\textheight,keepaspectratio]{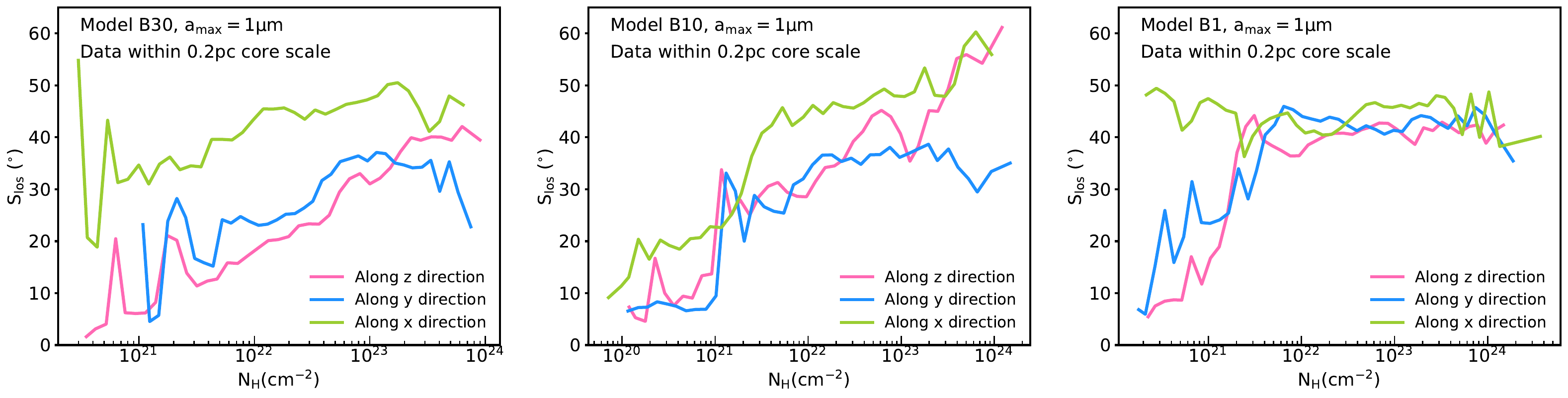}
\includegraphics[width=\textwidth,height=\textheight,keepaspectratio]{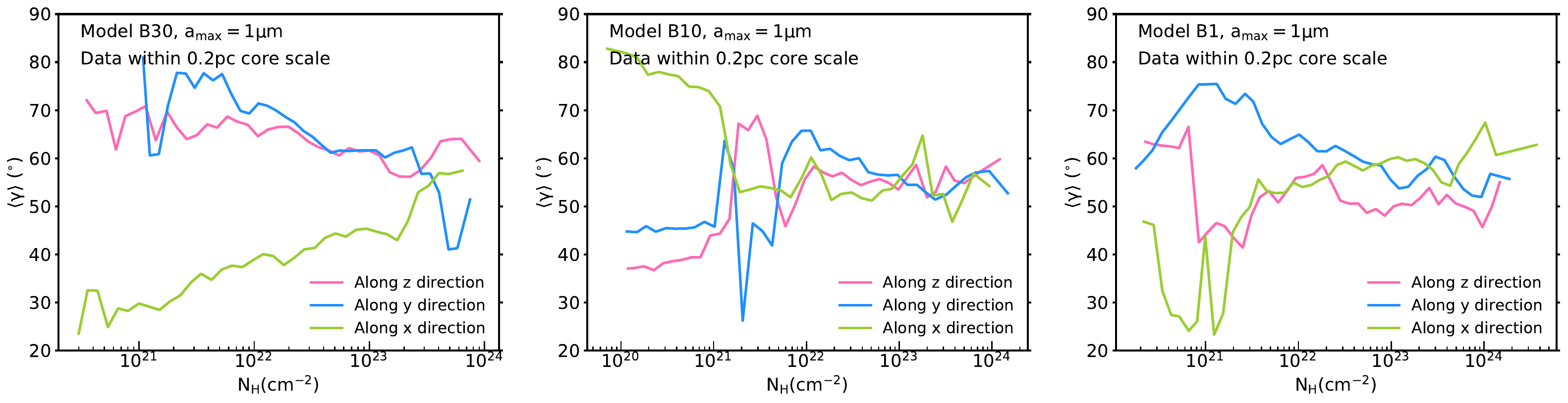}
    \caption{Upper row: Mean variation of $p(\%)$ and $N_{\rm H}$ within 0.2 pc core scale obtained from model PA (solid lines) and ideal$-$RAT (dashed lines) along z$-$, y$-$, and x$-$direction, assuming $a_{\rm max} = 1\mu m$. Results from models B30, B10, B1 are shown from left to right, respectively. Middle and lower row: the corresponding mean variation of $S_{\rm los}$ and $\langle \gamma \rangle$ within 0.2 pc scale as a function of $N_{\rm H}$. The depolarization produced by model RAT$-$id is still much more prominent than model PA despite the complicated morphology of magnetic fields seen along different viewing angles. } 
     \label{fig:viewing_angle_depolarization}
\centering     
\includegraphics[width=\textwidth,height=\textheight,keepaspectratio]{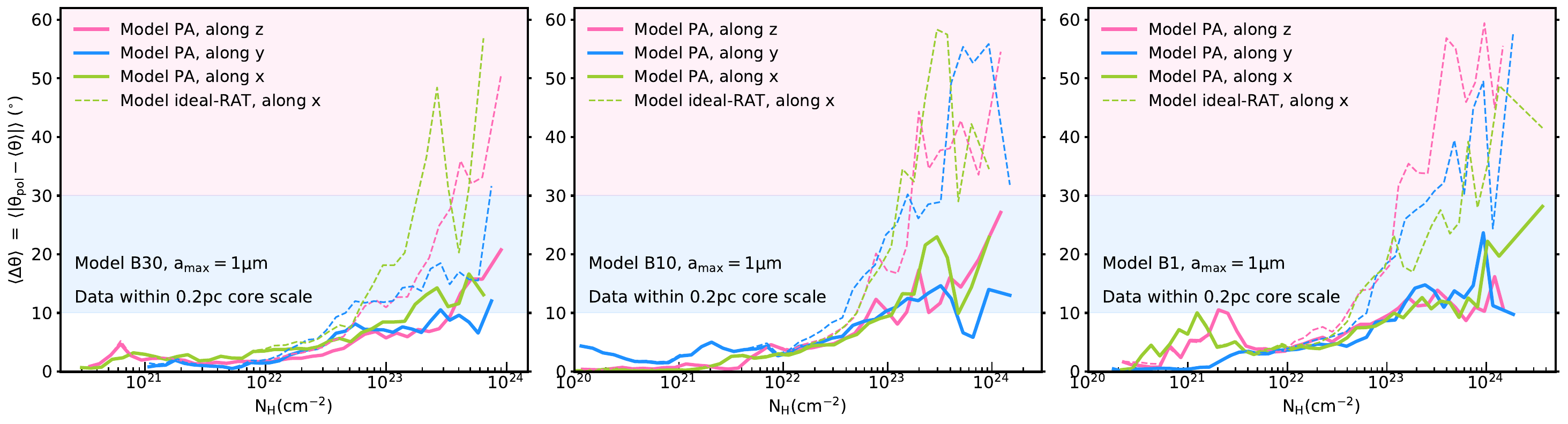}
    \caption{Effect of the viewing angle on the variation of $\langle \Delta \theta \langle$ with $N_{\rm H}$. Solid and dashed lines show the results summarized from all cores with model PA (solid lines) and ideal$-$RAT (dashed lines), assuming $a_{\rm max} = 1\mu m$. The systematically increasing angle difference between $\B$ inferred from dust polarization and the density$-$weighted mean field is seen across different observed directions of different clouds when accounting for the reduced alignment-size range, reaching $\langle \Delta \theta \rangle > 30^{\circ}$ in the peak column density position.}  
     \label{fig:delta_phi_NH_weight_density_direction}
\end{figure*}

Given the complex alignment states of dust grains inside starless cores (Section \ref{sec:discuss_grain_alignment}), whether we can still rely on the field inferred from dust polarization also needs to be re-examined. Previous studies by \cite{Cho_Lazarian_2005}, \cite{Bethell_2007}, and \cite{Pelkonen_2009} showed that dust polarization can continue tracing $\B$ from $A_{\rm V} \sim 3$ mag to $A_{\rm V} \sim 10$ mag if grains grow significantly to $a_{\rm max} \sim 1\mum$ inside starless cores. However, these studies did not model the grain alignment dynamics in detail. As discussed in Section \ref{sec:discuss_relation_intrinsic}, large submicron-sized and micron-sized grains can also be poorly aligned with $\B$ in the core center if they are PM or SPM grains with small iron cluster sizes. Under these conditions, the observed dust polarization carries only limited information about the central magnetic field, similar to the case of insignificant grain growth. To re-examine the reliability of magnetic field orientations inferred from dust polarization toward starless and prestellar cores, we first rotate the polarization angle by $90^{\circ}$\footnote{All starless and prestellar cores are optically thin at $850\,\mum$.} to obtain the inferred POS magnetic field orientation $\theta_{\rm pol}$. We then compare $\theta_{\rm pol}$ with the density-weighted POS mean field orientation $\langle \theta \rangle$ (Equation~\ref{eq:<theta>}) to quantify the discrepancy between two fields $\Delta \theta = \mid \theta_{\rm pol} - \langle \theta \rangle \mid$. The variation of the mean angular difference $\langle \Delta \theta \rangle$ measured within the 0.2 pc core scale as a function of $A_{\rm V}$ is shown in Figure \ref{fig:delta_phi_Av}. We use synthetic dust polarization data without instrumental noise for this analysis. The left and right panels illustrate the effects of grain alignment models, grain magnetic properties, and grain growth on $\langle \Delta \theta \rangle - A_{\rm V}$ relation.

In model PA, the accuracy of the POS magnetic field inferred from dust polarization decreases toward the core center because magnetic fields become disorganized and be bent out of the POS stronger by gravitational contraction (Figure \ref{fig:p_Slos_gamma_align_NH}). The angle discrepancy $\langle \Delta \theta \rangle$ in the center of starless/prestellar cores increases with decreasing the magnetization of their parent clouds. For example, we get $\langle \Delta \theta \rangle \sim 6–7^{\circ}$ in model B30 (black solid line), but $\langle \Delta \theta \rangle \sim 20^{\circ}$ in models B10 and B1 (black dashed and dotted lines). The issue is further worse when considering the alignment states of dust grains into account (models ideal$-$RAT and MRAT). Even in cases that grains grow to micron-size and contain big iron clusters size, we still get $\langle \Delta \theta \rangle \sim 20-30^{\circ}$ in the core center (for model B30) and up to $\sim 50$–$60^{\circ}$ (for models B10 and B1). The angle deviation slightly becomes larger if grains grow inefficiently (right panel) or have small iron cluster sizes (left panel). In the latter case, the inferred field from dust polarization starts showing $\sim 10^{\circ}$ difference from the mean field at $A_{\rm V} > 10$ mag (for model B30) or $A_{\rm V} > 6$ mag (for models B10 and B1), similar to the finding with insignificant grain growth by \cite{Cho_Lazarian_2005} and \cite{Bethell_2007}). Instrumental noise can further increase the angle discrepancy. In our study, we do not consider cores formed in strongly magnetized clouds, which are generally expected to have the well-ordered hourglass-shaped magnetic fields (\citealt{Mocz_2017}, \citealt{Hull_2017a}). Given their ordered magnetic field morphology, we expect smaller angle deviation between the inferred and mean field in the core center, even when the grain alignment effect is taken into account.

Besides magnetic field morphology and grain alignment, another factor that can contribute to the large angular deviation found in core centers is our use of the density-weighted mean magnetic field as the reference. To evaluate this effect, we compare the field inferred from dust polarization with the intensity-weighted mean field in Appendix \ref{sec:appen_wI} but find similar results as Figure \ref{fig:delta_phi_Av}. A systematic increase in $\Delta \theta$ with $N_{\rm H}$ is also found when observing these cores at different viewing angles (Section \ref{sec:view_angle}). If polarized dust emission cannot accurately capture the angle dispersion induced by turbulence in core centers, it may affect the magnetic field strength estimated using the Davis-Chandrasekhar-Fermi method (\citealt{David_1951,Chandrasekhar_1953}), considering the molecular-line tracers used to infer the turbulent velocity dispersion can probe the core center. This issue could lead to the incorrect evaluation of the dynamic importance of magnetic energy in starless and prestellar cores. This question needs to be quantified in future studies.

\subsection{Impact of the viewing angle}\label{sec:view_angle}
Another issue that has not yet been addressed is the impact of the viewing angle on the properties of polarized dust emission. In the previous section, we analyzed the polarization properties of 8 cores viewed along the $z$-axis and 9 cores viewed along the $y$-axis. To generalize the viewing angle effect, we generate synthetic dust polarization observations of the inner 0.2 pc region of each core along the $x$-, $y$-, and $z$-direction, considering models PA and ideal-RAT with $a_{\rm max}=1\mu m$.

The upper row of Figure~\ref{fig:viewing_angle_depolarization} shows the effect of the viewing angle on the $p(\%)$-$N_{\rm H}$ relation. The corresponding variations of $S_{\rm los}$ and $\langle \gamma \rangle$ with $N_{\rm H}$ are shown in the middle and lower rows, respectively. Despite different distortion degrees and bending behaviors of magnetic fields with respect to the LOS when observing the cores at different direction, model ideal$-$RAT still produces stronger depolarization, i.e., faster reduction of $p(\%)$ with $N_{\rm H}$, within the 0.2 pc core scale compared to model PA. However, for cores formed in the moderately magnetized cloud, grain alignment does not completely erase the imprint of magnetic field geometry on the $p$-$N_{\rm H}$ relation. For example, when observing the cores in model B30 along x$-$direction (upper left panel), one can still see the increase of $p(\%)$ with $N_{\rm H}$ at the peak density position from model ideal$-$RAT. This feature originates from the bending of $\B$ from the LOS toward the POS while observing the cores along the mean field direction (lower left panel). This feature is not obvious in cores formed inside weakly magnetized clouds because the initial uniform field is rapidly distorted by large-scale turbulent motions. 

Finally, we examine the effect of viewing angle on the angular difference $\Delta \theta$ discussed in Section \ref{sec:discuss_trace_B}. The results are presented in Figure \ref{fig:delta_phi_NH_weight_density_direction}. Consistent with the results shown in Figure \ref{fig:delta_phi_Av}, the mean angle deviation between the field inferred from dust polarization and the density$-$weighted mean field exceeds $\langle \Delta \theta \rangle \sim 20^{\circ}$ in the centers of cores formed in model B30 when grain alignment is taken into account (model ideal-RAT, dashed lines), regardless of the viewing direction. The accuracy of the inferred magnetic field orientation is further reduced in cores formed inside weakly magnetized clouds, with $\langle \Delta \theta \rangle \sim 30$-$60^{\circ}$ in models B10 and B1, independent of the viewing angle.

\section{Summary}\label{sec:summary}
In this study, we performed multi-scale synthetic modeling of polarized dust emission, from $\sim$5 pc MCs scale down to $\sim$0.2 pc of starless cores. Using three MHD cloud models with different levels of cloud magnetizations, three models of grain alignment, and the broad range of dust physical properties, we investigate the origin behind the depolarization of dust emission from clouds to starless cores and explore the connection between dust polarization properties and the underlying physics of dust grains inside starless/prestellar cores. Our main conclusions are as follows:

\begin{enumerate} 
\item We confirm \cite{Chen_2016} and \cite{Seifried_2019} that the dispersion and bending of magnetic fields by the large-scale turbulence is the major depolarization mechanism inside MCs where $N_{\rm H} < 10^{21}\cm^{-2}$.

\item Adopting the criteria that grain alignment becomes the dominant depolarization mechanism when it produces $p(\%)$ which is smaller than half of those produced by magnetic field geometry alone, we find that the depolarization within $< 1$ pc scale of starless and prestellar cores can be classified into two main scenarios. 

\begin{itemize}
\item In the first case, the field geometrical effect will continue to lead the depolarization up to $N_{\rm H} \sim 10^{22}\cm^{-2}$. Then it will combine with the reduction of the grain alignment size range to lower $p(\%)$ with $N_{\rm H}$ until $N_{\rm H} \sim 10^{23}\cm^{-2}$, and can be totally replaced by grain alignment effect at higher column densities of $N_{\rm H} > 10^{23}\cm^{-2}$. This scenario can only happen if 1) grains grow significantly to micron-size inside starless cores and 2) grains are SPM with embedded big iron clusters.

\item If grains do not grow significantly, or they grow to micron-sized but are not SPM with big iron clusters, grain alignment efficiency will start contributing to the depolarization at $N_{\rm H} \sim 10^{21}$-$10^{22}\cm^{-2}$ regardless of the tangledness and inclination of magnetic fields (to the LOS) in our simulations. At higher column densities ($N_{\rm H} > 10^{22}\cm^{-2}$), depolarization is primarily driven by either (1) the loss of grain alignment or (2) the reduction of the internal and external alignment degrees caused by efficient gaseous damping. The former scenario occurs when grains grow insignificantly with $a_{\rm max} \sim 0.25$--$0.5\,\mum$ in the core center, whereas the latter occurs when dust grains are PM or SPM with small iron cluster sizes (or with low volume filling factor of iron clusters).
 \end{itemize}

\item We found that if SPM grains grow significantly to micron-size inside starless cores and contain big iron clusters, one expects to obtain the depolarization characterized by the considerable central polarization degree $\langle p_{\rm center} \rangle \gtrsim 3\%$ and shallow $p-I$ slope with $\alpha \lesssim 0.7$. Ricean fitting can work well in recovering these coupled values, allowing us to explore the dust physics behind the depolarization. In contrast, if the alignment loss or the reduced internal and external alignment degrees of dust grains is the major depolarization mechanism, one expects to obtain lower $\langle p_{\rm center} \rangle \lesssim 2\%$ and high $\alpha \gtrsim 0.8$. In this range, neither Ricean nor Power-law fitting alone can provide an accurate power index of the $p-I$ relation. We recommend fitting observational data to both methods and adopting their average values to reflect the slope of the depolarization feature better.

\item The accuracy of the magnetic field orientation inferred from dust polarization decreases significantly within sub-pc scale of starless and prestellar cores (at $A_{\rm V} > 5$ mag) as a result of the reduced grain alignment efficiency. The mean angle difference between the magnetic field inferred from dust polarization and the density(intensity)$-$weighted mean field reaches $\sim20$-$30^{\circ}$ in the center of cores formed in intermediately magnetized clouds (model B30) and increases to $\sim 40$--$60^{\circ}$ for cores formed in weakly magnetized clouds (models B10 and B1). The field orientation inferred from dust polarization starts to be different from the mean field at lower $A_{\rm V}$ (or lower $N_{\rm H}$) as the cloud magnetization decreases, the iron cluster size (or volume filling factor) inside dust grains decreases, and the maximum grain size becomes smaller. This feature is independent of the viewing angle, which questions the reliability of magnetic field strengths inside $<0.2$ pc scale of starless and prestellar cores derived from the DCF method using observations of polarized dust emission. 

\end{enumerate}

\section*{Acknowledgements}
N.C.G and T.H. acknowledge the support from the major research project (No. 2026183200)from Korea Astronomy and Space Science Institute (KASI) funded by the Ministry of Science and ICT (MSIT). This work is partially supported by the Vietnam National Foundation for Science and Technology Development (NAFOSTED) under grant number 103.99-2024.36.
 
 
\bibliography{main}
\appendix

\section{Magnetic fields and synthetic dust polarization toward starless cores}\label{sec:appen_core_illustration}

 \begin{figure*}
\centering     
\includegraphics[width=\textwidth,height=\textheight,keepaspectratio]{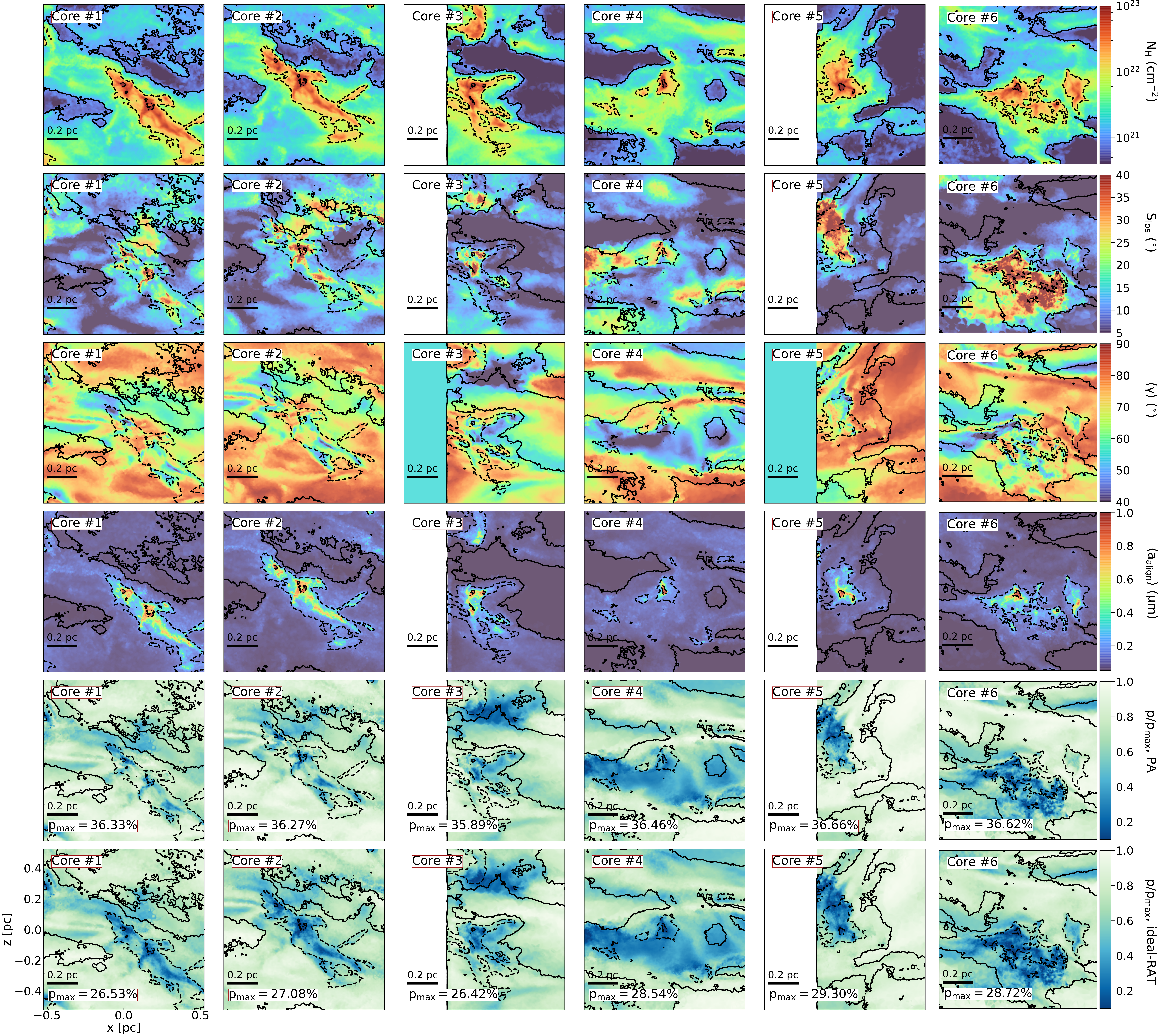}
    \caption{Spatial distribution of the gas column density $N_{\rm H}$ (first row), angle dispersion of the POS magnetic fields along the LOS $S_{\rm los}$ (second row), mean inclination angle of $\B$ toward the LOS $\langle \gamma \rangle$ (third row), and the density$-$weighted minimum alignment size $\langle \rm a_{\rm align} \rangle$ (fourth row) inside 1 pc of all cores identified in model B30. Fifth and sixth rows: Normalized polarization degree map $p/p_{\rm max}$ obtained from model PA and ideal$-$RAT. Solid and dashed contours mark the position of $N_{\rm H} = 10^{21}\cm^{-2}$ and $10^{22}\cm^{-2}$. The well-order magnetic fields is well seen at $N_{\rm H} < 10^{22}\cm^{-2}$ in 6 cores inside moderately magnetized cloud. Models ideal$-$RAT and PA give similar normalized polarization fraction distribution within 1 pc core scale, with lower polarization fractions in areas with higher $S_{\rm los}$ and lower $\langle \gamma \rangle$. It implies the leading contribution of the field geometrical effect on the depolarization observed at $\sim 1$ pc scale of starless and prestellar cores.} 
     \label{fig:density_polarization_B30_clump}
\end{figure*}

\begin{figure*}
\centering      \includegraphics[width=\textwidth,height=\textheight,keepaspectratio]{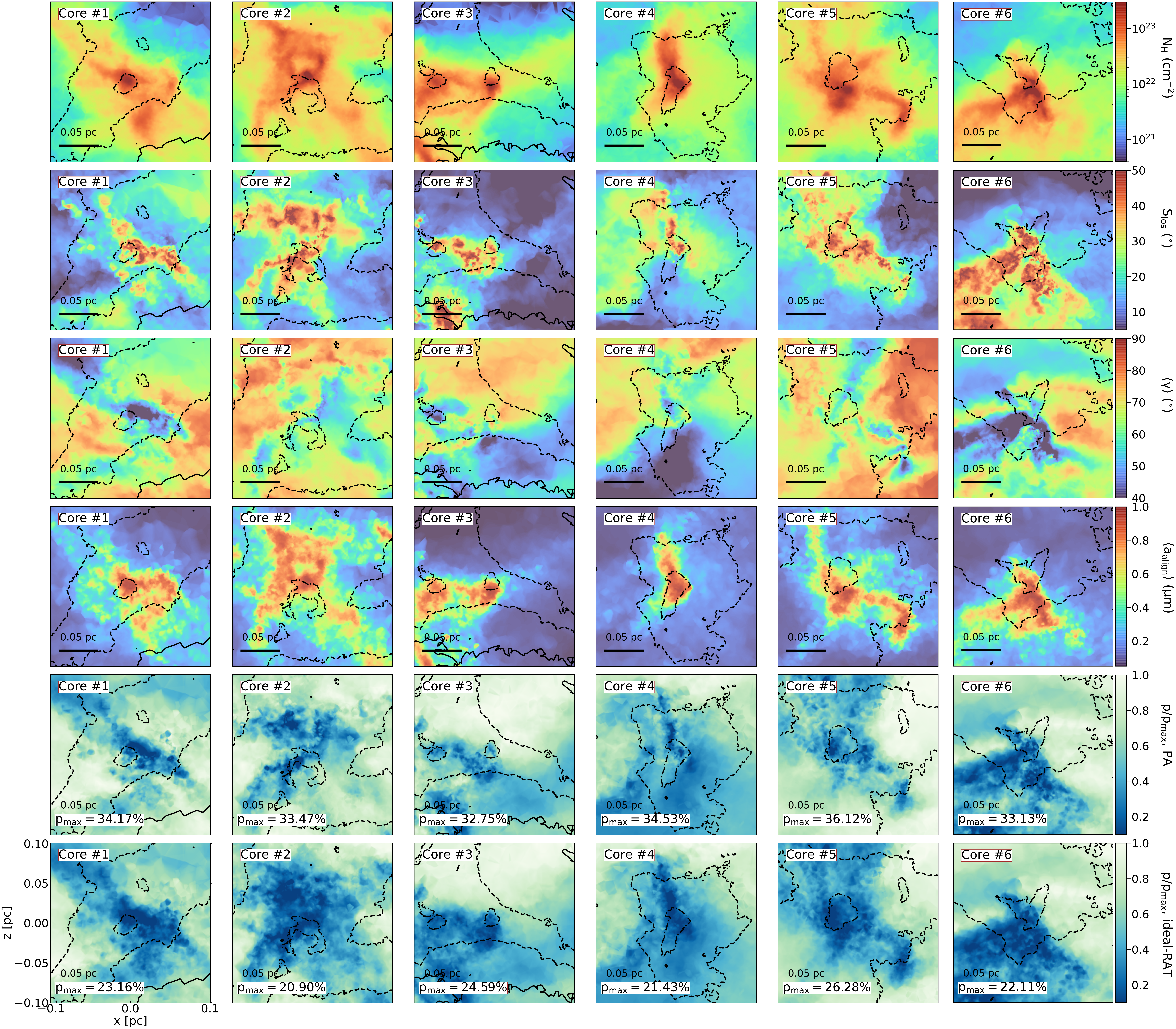}
    \caption{Similar results to Figure \ref{fig:density_polarization_B30_clump} but zooming into the inner 0.2 pc region. Dashed and solid black contours mark the position of $N_{\rm H} = 10^{22}\cm^{-2}$ and $10^{23}\cm^{-2}$. Regardless of the high distortion of magnetic fields in the inner core, model ideal$-$RAT still reveals stronger depolarization effect, i.e., lower $p/p_{\rm max}$, than in model PA in the area where $\langle a_{\rm align} \rangle$ (fourth row) approaches the maximum grain size of $1\mum$.}
     \label{fig:density_polarization_B30_core}
\end{figure*}

We show in the first row of Figure \ref{fig:density_polarization_B30_clump} the gas column density distribution within 1 pc of all cores identified in model B30. The second and third rows show the angle dispersion of the POS magnetic field along the LOS, $S_{\rm los}$, and the mean inclination angle of $\B$ with respect to the observer, $\langle \gamma \rangle$. The fourth row presents the density$-$weighted minimum alignment size $\langle a_{\rm align} \rangle$. The fifth and sixth rows display the normalized polarization fraction $p/p_{\rm max}$ obtained from models PA and ideal-RAT, with the corresponding maximum polarization degree $p_{\rm max}$ in the lower-left corner of each panel. The zoom-in map toward the inner 0.2 pc of each core is provided in Figure \ref{fig:density_polarization_B30_core}. Significant grain growth is assumed in this figure, with $a_{\rm max} = 1\mum$.

Generally, the well-ordered magnetic field along the x$-$direction of the parent clouds is preserved up to $\sim 1$ pc scale of their embedded cores, with the mean field inclination angle of $\langle \gamma \rangle \sim 70$–$80^\circ$ and the small angle dispersions along the LOS $S_{\rm los} < 10^\circ$ at $N_{\rm H} < 10^{22}\cm^{-2}$ (second and third rows of Figure \ref{fig:density_polarization_B30_clump}). At higher column densities, the magnetic field becomes more disorganized and is bent away from the POS due to the combined effects of gravitational contraction and turbulent gas motions inside the collapsing core. Typically, $S_{\rm los}$ rises to $\sim 50^\circ$ while $\langle \gamma \rangle$ decreases to $\sim 40$–$60^\circ$ at $N_{\rm H} > 10^{23}\cm^{-2}$ (Figure \ref{fig:density_polarization_B30_core}). Accompanied by the increasing magnetic field tangling at higher column densities, the grain alignment size range becomes narrower (fourth row) and the polarization fraction obtained in both models PA and ideal$-$RAT reduces toward the core center (fifth and sixth rows). 

At the $\sim1$ pc scale where $N_{\rm H} < 10^{22}\cm^{-2}$ (Figure \ref{fig:density_polarization_B30_clump}), model ideal-RAT produces similar spatial distribution of $p/p_{\rm max}$ to that of model PA, with lower $p/p_{\rm max}$ in regions of higher $S_{\rm los}$ and lower $\langle \gamma \rangle$. Toward the inner $\sim 0.2$ pc, model ideal-RAT systematically produce lower $p/p_{\rm max}$ than model PA (Figure \ref{fig:density_polarization_B30_core}), particularly in the core center where the minimum alignment size $\langle a_{\rm align} \rangle$ approaches the maximum grain size of $1\mum$ (fourth row). The consistent behavior of the 6 cores identified in model B30 suggests the transition in the dominant depolarization mechanism within starless and prestellar cores, from magnetic field geometry on the $\sim1$ pc scale to grain alignment on scales smaller than $\sim0.2$ pc. Similar features are obtained inside starless and prestellar cores identified from weakly magnetized clouds (models B10 and B1) regardless of the highly distortion and the bending of magnetic field lines toward the LOS.

\section{Grain alignment efficiency inside dense cores}\label{sec:appen_grain_alignment}
\begin{figure*}
\centering      
\includegraphics[width=\textwidth,height=\textheight,keepaspectratio]{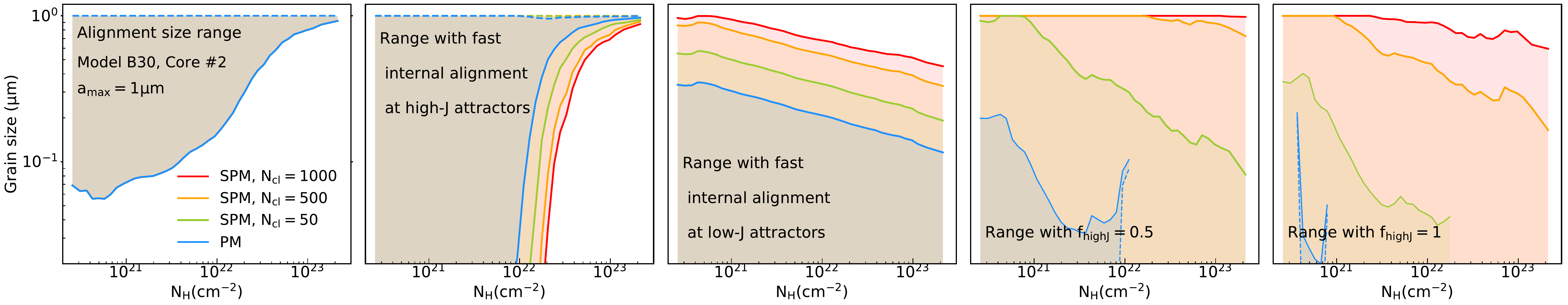}
\includegraphics[width=\textwidth,height=\textheight,keepaspectratio]{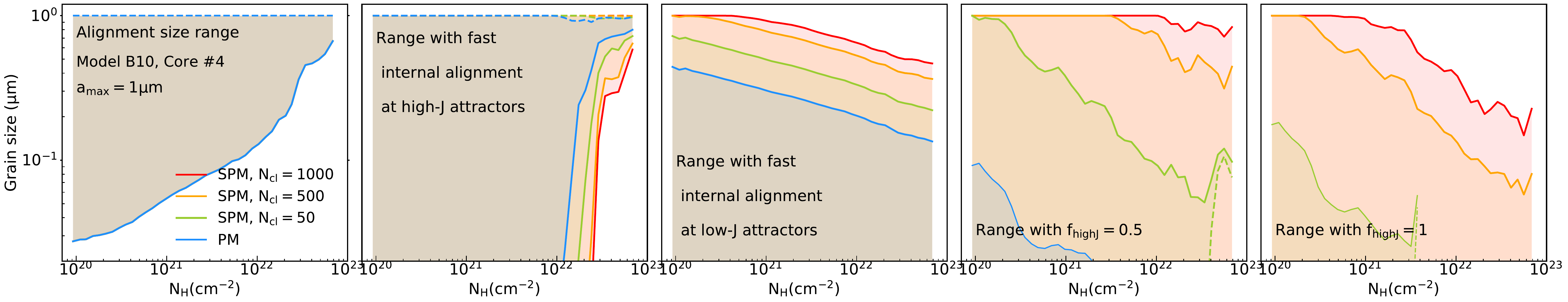}
\includegraphics[width=\textwidth,height=\textheight,keepaspectratio]{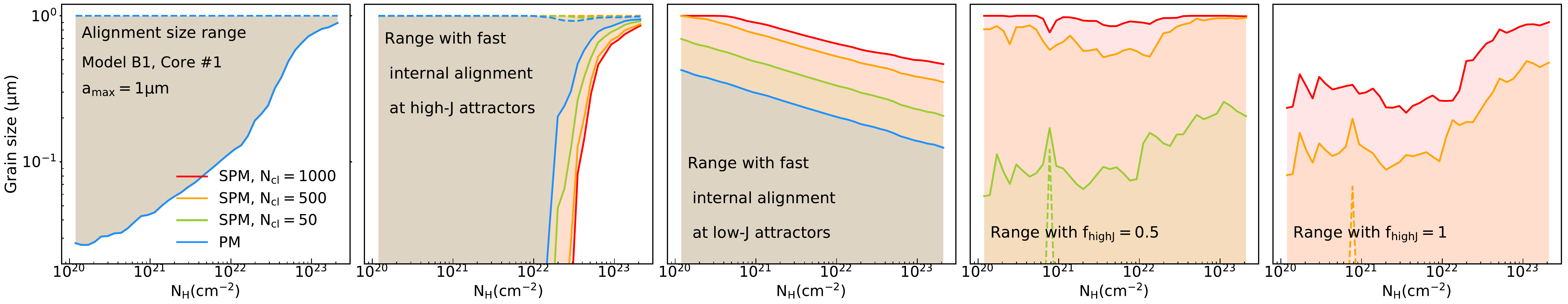}
    \caption{Effect of iron cluster sizes inside dust grains on the magnetic alignment state obtained within 1 pc of three starless cores identified in Figure \ref{fig:gas_density}. First column: variation of the alignment size range as a function of $N_{\rm H}$. Second and third columns: variation of size range with fast internal relaxation at high- and low-\textit{J} attractors. Grains beyond this area will have inefficient internal alignment by slow Barnett relaxation. Fourth and fifth columns: variation of the size range having $f_{\rm high-J} = 0.5$ and $f_{\rm high-J} = 1$ by MRAT alignment (colored area). Area from solid line to $a_{\rm max} = 1\mum$ in the fourth column indicates grains to be aligned with $\B$ by RATs with $f_{\rm high-J} = 0.25$. The alignment size range tends to be narrower while dust grains tend to be aligned with $\B$ at low-\textit{J} attractors, i.e., lower $f_{\rm high-J}$, with slow internal relaxation in the core center. The net alignment degree of dust grains in the inner core reduces significantly with decreasing grain magnetic susceptibility. This issue becomes worse inside starless cores formed from weakly magnetized clouds owing to the low local magnetic field strength and high gas concentration there.}
     \label{fig:grain_alignment}
\end{figure*}

\begin{figure*}
\centering      
\includegraphics[width=\textwidth,height=\textheight,keepaspectratio]{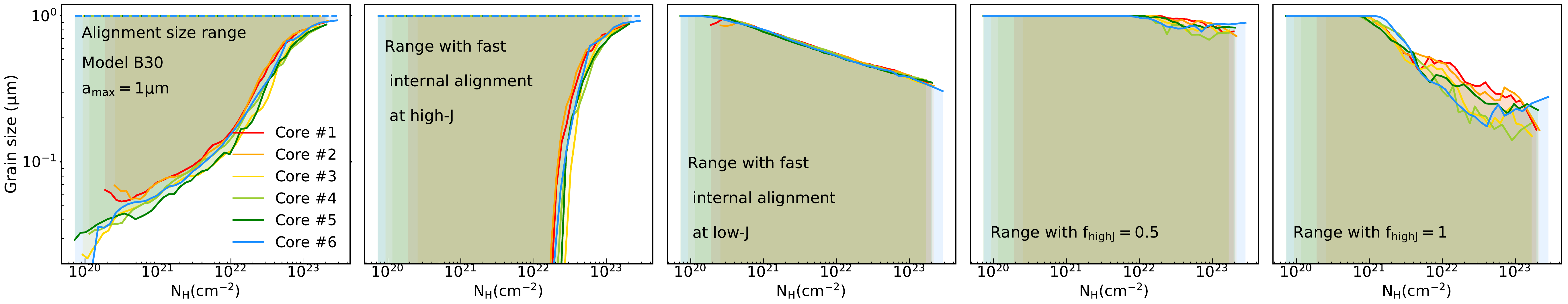}
\includegraphics[width=\textwidth,height=\textheight,keepaspectratio]{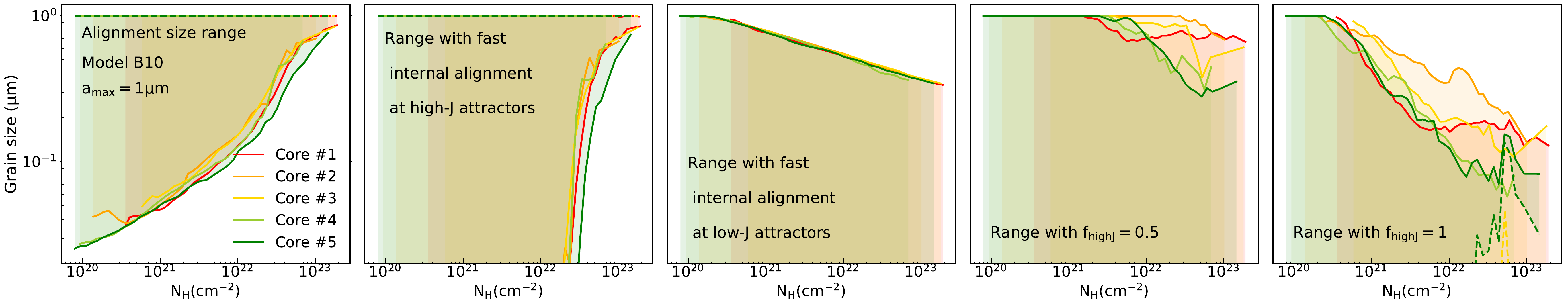}
\includegraphics[width=\textwidth,height=\textheight,keepaspectratio]{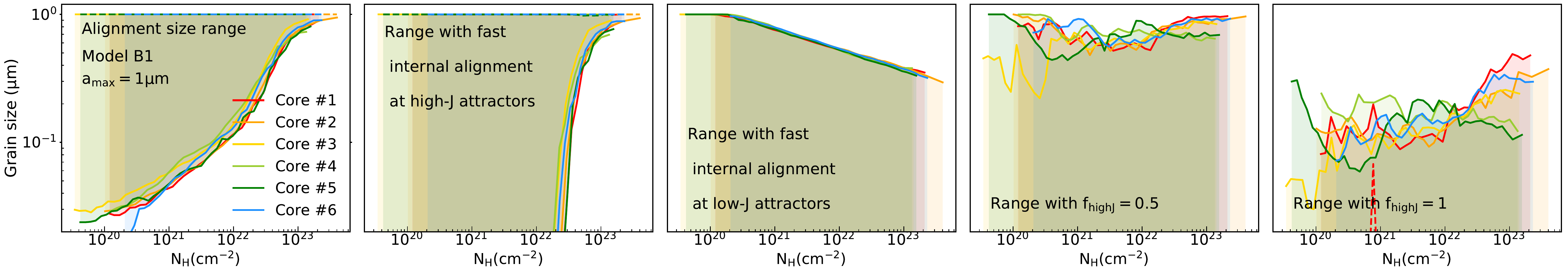}
    \caption{Similar results to Figure \ref{fig:grain_alignment} but for all cores found in the cloud model B30 (upper row), B10 (middle row), and B1 (lower row), assuming dust grains are SPM with $N_{\rm cl} = 500$ and $a_{\rm max} = 1\mum$. We found similar behaviors of aligned dust grains inside different cores belong to different cloud models }
     \label{fig:grain_alignment_all_core}
\end{figure*}

We show in the first column of Figure \ref{fig:grain_alignment} the mean variation of the alignment size range (colored area) within 1 pc of cores $\#2$, $\#4$, and $\#1$ identified in models B30, B10, and B1, from top to bottom (see Figure \ref{fig:gas_density}). The lower boundary of the alignment size range is determined by the suprathermal rotation condition $\langle a_{\rm align} \rangle$ (Figures \ref{fig:p_Slos_gamma_align_NH}), while the upper boundary is determined by the fast Larmor precession condition (Section \ref{sec:modelling_grain_alignment}). The two next columns of Figure \ref{fig:grain_alignment} show the mean variation of the size range with fast internal relaxation at high$-$ and low$-$\textit{J} attractors with column densities. And the fourth and fifth columns illustrate the variation of the size range having $f_{\rm high-J}=0.5$ and $f_{\rm high-J}=1$ by MRAT mechanism. The absence of lines in the fourth and fifth columns implies RAT is the major alignment mechanism of the considered dust model inside starless cores, with a typical $f_{\rm high-J}\approx0.25$. 

In general, the alignment size range will become narrower toward the core center due to the misalignment of sub-micron grains with $\B$ inside the core center by inefficient RATs. The variation of the alignment size range inside starless cores is independent of the grain magnetic properties and the magnetic field morphology. Accompanied by the reduction of the alignment size range at higher $N_{\rm H}$, the internal alignment degree and the effectiveness of MRAT alignment also reduce toward the core center owing to the efficient gaseous damping effect. However, how bad the net magnetic alignment degree can be will strongly depend on the size of iron clusters embedded inside dust grains. For example, almost all SPM grains with $N_{\rm cl} > 500$ (red and orange) will be aligned with $\B$ by MRAT mechanism, with $f_{\rm high-J} \sim 0.5-1$ even at $N_{\rm H} > 10^{22}\cm^{-2}$ (fourth and fifth columns). From the second column, grains at high-\textit{J} attractors still have fast internal relaxation in the core center. Therefore, SPM grains with $N_{\rm cl} > 500$ still have moderated magnetic alignment degree inside starless and prestellar cores. However, they never reach the perfect alignment state as assumed in model ideal$-$RAT because of the presence of $50\%$ of grains with slow internal relaxation at low-\textit{J} attractors in the core center (third column). That explains why model MRAT always shows stronger depolarization and lower $p(\%)$ in the core center compared to model ideal$-$RAT. In contrast to SPM grains with $N_{\rm cl} > 500$, PM and SPM grains with $N_{\rm cl} \leq 50$ (blue and green) are majorly aligned with $\B$ by RATs, with typical $f_{\rm high-J} = 0.25$ at $N_{\rm H} > 10^{22}\cm^{-2}$ (fourth and fifth columns). Their internal alignment efficiency inside the core is also poor, that explains why they can produce the prominent depolarization feature similar to the alignment loss case (Figures \ref{fig:p_NH_Ncl} and \ref{fig:p_NH_amax}), which happens when grains grow insignificantly with $a_{\rm max} \sim 0.25-0.5\mum$ in the core center. The dependence of the grain alignment state on the grain magnetic properties is quite consistent among different starless and prestellar cores identified in both intermediately and weakly magnetized clouds, as shown in Figure \ref{fig:grain_alignment_all_core}. It implies that the reduction of the grain alignment efficiency by gaseous damping must be considered carefully while interpreting dust polarization properties, especially within $0.2$ pc scale of starless and prestellar cores (where $N_{\rm H} > 10^{22}\cm^{-2}$, Section \ref{sec:discuss_grain_alignment}). Similar tests for cores formed inside strong feedback environments need to be carried out to generalize the above conclusion.

\section{Effect of grain aspect ratio and grain porosity}\label{sec:s_P_depolarization}
 \begin{figure}
\centering
    \includegraphics[width=0.45\textwidth,height=0.45\textheight,keepaspectratio]{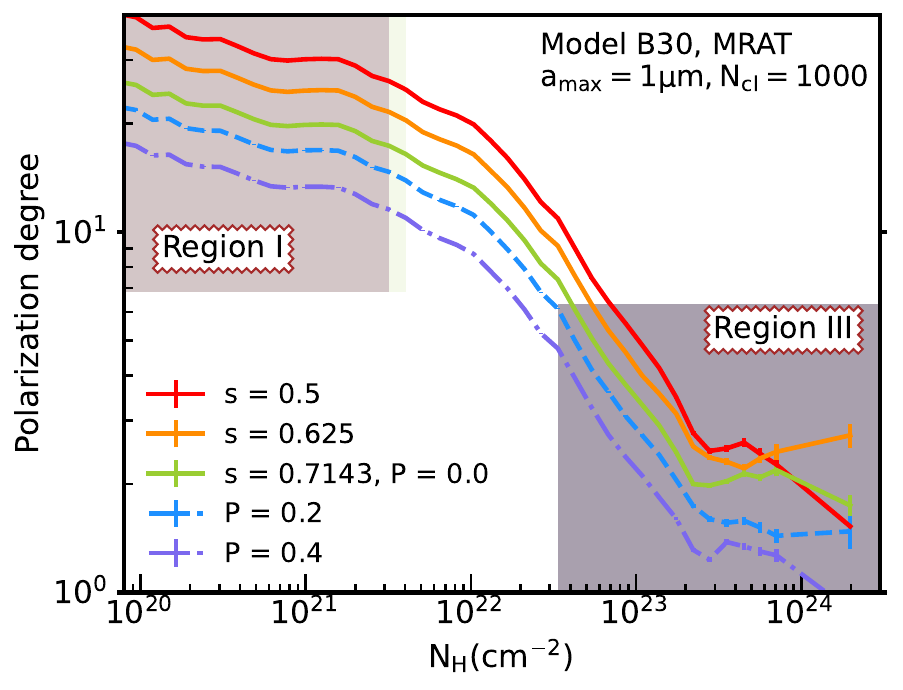}
    \caption{Effect of grain axial ratio $s$ and grain porosity $P$ on the $p–N_{\rm H}$ relation, assuming model MRAT for SPM grains with $N_{\rm cl} = 1000$ and $a_{\rm max} = 1\mum$. Each solid line shows the mean polarization trend averaged over all cores identified in model B30. Increasing the grain axial ratio (i.e., grains becoming more spherical) and increasing the grain porosity reduce the observed polarization degree. The values of $s$ and $P$ do not affect the boundary of region I and III, but $s$ and $P$ can affect the slope of $p-N_{\rm H}$ if grains change their structure inside starless cores.}
     \label{fig:p_NH_s_P}
\end{figure}

 \begin{figure*}
\centering \includegraphics[width=\textwidth,height=\textheight,keepaspectratio]{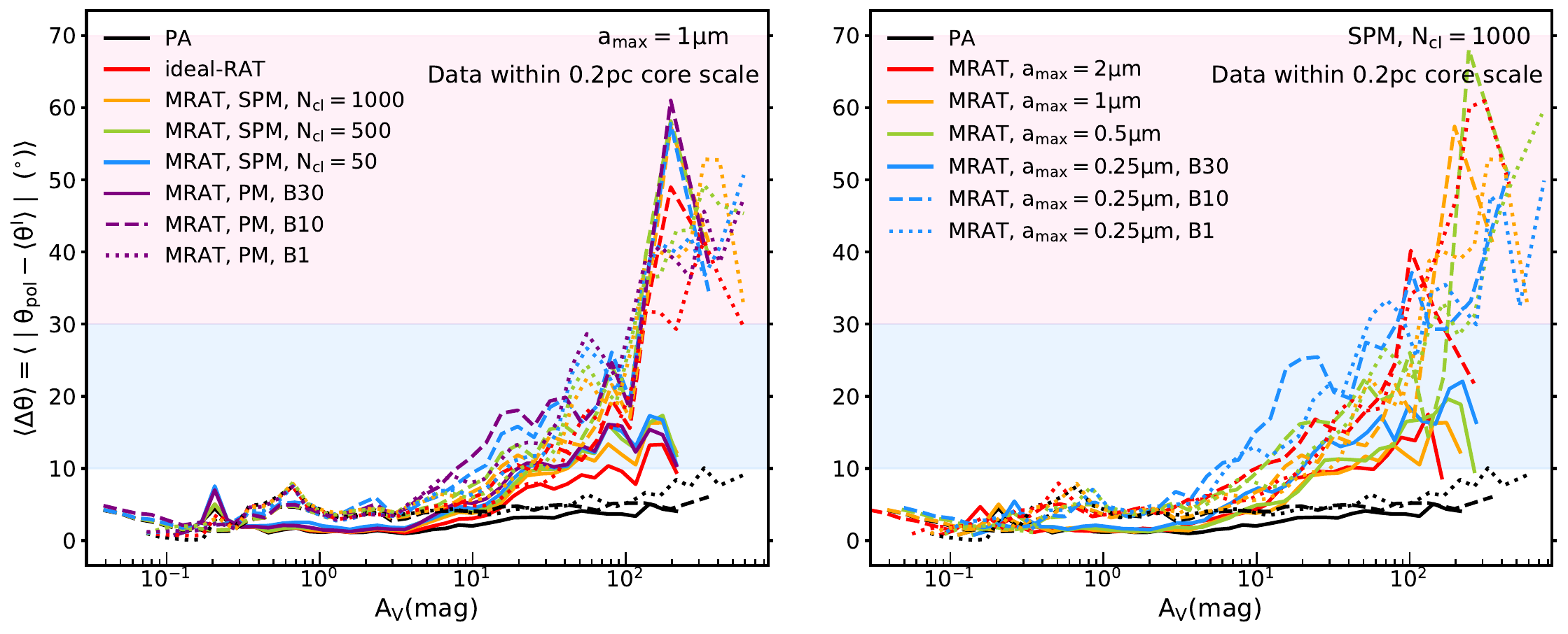}
    \caption{Similar to Figure \ref{fig:delta_phi_Av} but for the angle difference $\langle \Delta \theta \rangle$ seen between the magnetic field inferred from dust polarization and the intensity$-$weighted mean fields. The field inferred from dust polarization inside starless and prestellar cores tends to infer intensity-weighted field rather than the density-weighted fieldn density, with smaller $\delta \phi < 10^{\circ}$ in cases of model PA (black lines). However, similar to Figure \ref{fig:delta_phi_Av}, the angle deviation will increase systematically when considering the grain alignment effect into account, reaching $\langle \Delta \theta \rangle \geq 30^{\circ}$, regardless of grain properties. The angle deviation in the core center increases faster inside cores formed inside weakly-magnetized clouds.}
     \label{fig:delta_phi_Av_weight_intensity}
\centering \includegraphics[width=\textwidth,height=\textheight,keepaspectratio]{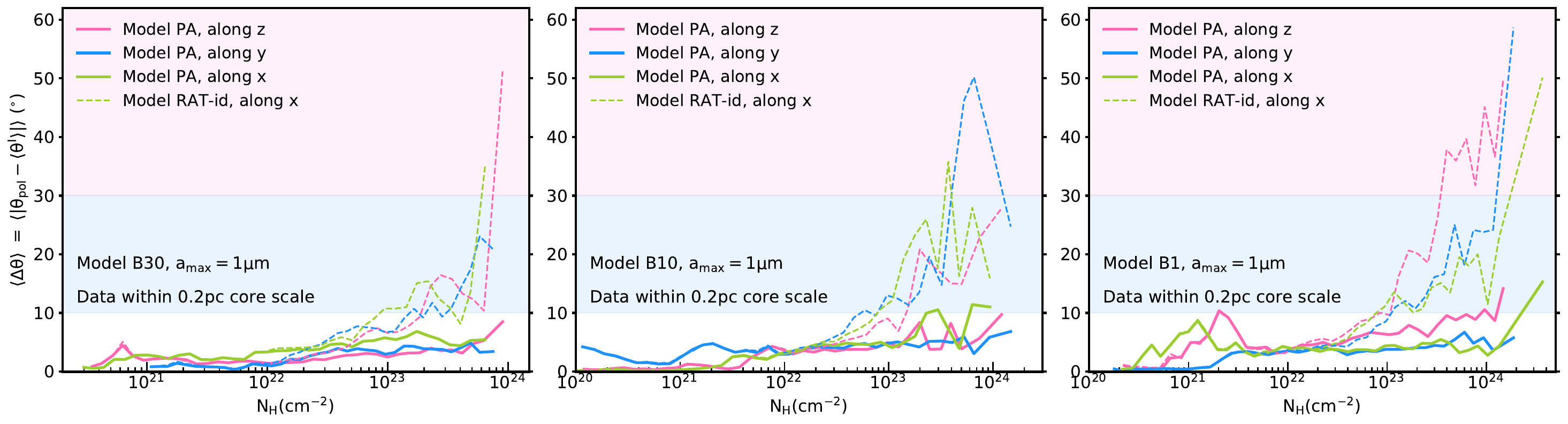}
    \caption{Similar to Figure \ref{fig:delta_phi_NH_weight_density_direction} but for the angle deviation seen between the inferred field and intensity$-$weighted magnetic fields. The viewing angle can slightly strengthen the angle deviation between two fields in the core center by $\sim 5-10^{\circ}$ (model PA, solid lines). But regardless of the viewing angle, the systematic increase of $\langle \Delta \theta \rangle > 30^{\circ}$ toward the core center due to the grain alignment effect is still seen in both intermediately and weakly magnetized clouds.}
     \label{fig:delta_phi_NH_weight_intensity_direction}
\end{figure*}

We show in Figure \ref{fig:p_NH_s_P} the effect of the grain axial ratio $s$ and grain porosity $P$ on the variation of $p(\%)$ across $N_{\rm H}$, considering model MRAT for SPM grains with $N_{\rm cl} = 1000$ and $a_{\rm max} = 1\mum$. We only show the results for the cloud model B30. The polarization degree obtained within 1 pc core scale declines with decreasing the axial ratio (i.e., grains become more spherical) and increasing the porosity owing to the reduced  polarization efficiency of dust grains. For highly elongated grains with $s = 0.5$ (red curve), one can get $p \sim 35\%$ in the outer core where $N_{\rm H} \sim 10^{20}\cm^{-2}$ and $p \sim 3\%$ in the core center. In contrast, for highly porous grains with $P = 0.4$ and $s = 0.7143$ (purple), $p(\%)$ is substantially lower, with $p \sim 13\%$ at $N_{\rm H} \sim 10^{20}\cm^{-2}$ and $p \sim 1\%$ in the innermost region of starless cores. 

In contrast to the effects of grain magnetic properties and grain growth, varying the grain axial ratio and porosity do not change the position of $N_{\rm H,trans}^{\rm I-II}$ and $N_{\rm H,trans}^{\rm II-III}$ because we still assume uniform values of $s$ and $P$ inside our clouds and cores. The grain alignment efficiency is found to weakly depend on $s$ and $P$ \footnote{\cite{Jager_2024} showed that radiative torques acting on ballistic aggregate grains are $\sim 10$–$100$ times weaker than those acting on compact grains in ranges $\lambda/a \sim 1$–$100$. However, because radiative torques are dominated by interactions between photons and grains that are comparable in size, the resulting maximum angular velocities gained by RATs differ only by a factor of $\sim 2$ between aggregate and compact grains. Consequently, grain geometry has a limited impact on alignment efficiency.}, which may induce insignificant effect on the change of depolarization mechanisms with $N_{\rm H}$. However, it may affect the steepness of $p-N_{\rm H}$ relation, as it can become steeper if grains become highly porous or more spherical toward the core center; or shallower if grains grow and become more elongated toward the core center. 

\section{Accuracy of magnetic fields inferred from dust polarization - compare with intensity$-$weighted mean field}\label{sec:appen_wI}
The intensity-weighted mean magnetic field is calculated using the same equation as for the density-weighted mean field (Equation \ref{eq:<theta>}), except that the weighting factor $n_{\rm H}$ is replaced by $n_{\rm H}B_{\lambda}(T_{\rm d})$, where $B_{\lambda}(T_{\rm d})$ is the Planck function with the dust temperature $T_{\rm d}$ obtained from the RT simulations (Section \ref{sec:MCRT}). We denote the intensity$-$weighted mean field angle as $\langle \theta ^{\rm I}\rangle$. Figure \ref{fig:delta_phi_Av_weight_intensity} presents the mean angle difference between the magnetic field inferred from dust polarization $\theta_{\rm pol}$ and $\langle \theta ^{\rm I}\rangle$ as a function of $N_{\rm H}$ obtained within 0.2 pc core scale, considering different dust models, dust properties, and cloud models, assuming $a_{\rm max}=1\mu m$. Similar results but for different viewing directions are shown in Figure \ref{fig:delta_phi_NH_weight_intensity_direction}. Comparing results to Sections \ref{sec:discuss_trace_B} and \ref{sec:view_angle}, it seems that dust polarization closely infers the intensity-weighted mean magnetic field rather than the density-weighted mean field, with the typical angle deviation of $\langle \Delta \theta \rangle < 10^{\circ}$ obtained in model PA in the core center. It happens because cold grains in the core center may not radiate strong emission at $850\mum$, that reduces the information of magnetic fields carried by this area in the observed dust polarization signal. That explains why comparing the field inferred from dust polarization with the density$-$weighted mean field (which emphasize the field morphology in the densest area) causes higher angle deviation. However, this effect is not significant compared to the impact of grain alignment. One still sees the systematic increase of $\langle \Delta \theta \rangle$ towards the maximum column density, reaching $\langle \Delta\theta \rangle \sim 30$-$60^{\circ}$ from model ideal$-$RAT in the core center, regardless of the cloud magnetization and viewing angle.

\section{Synthetic observed dust polarization and Fitting method}\label{sec:append_add_noise}

\subsection{Fitting method} \label{sec:fitting}  
For the power-law fitting, we first debias the observed polarized intensity $I_{\rm p}$ to correct for noise bias, following \cite{Wardle_1974},  \cite{Kwon_2018}:
\bea 
I_{\rm p,db} =  \sqrt{Q^{2} + U^{2} - \frac{1}{2}(\delta Q^{2} + \delta U^{2})},
\ena
which gives the debias polarization degree of:
\bea 
p_{\rm db} = \frac{I_{\rm p,db}}{I},
\ena 
where $\delta Q$ and $\delta U$ are the measurement errors of Stokes $Q$ and $U$, which are approximated as $\delta Q = \delta U = \sigma_{\rm QU}$ in our study (following the approach of \citealt{Kate_2019}). The uncertainty of the debiased polarization fraction, $\delta p$, is given by:
\bea 
\delta p_{\rm db} = \Bigg(\frac{Q^{2}\delta Q^{2} + U^{2}\delta U^{2}}{I^{2}(Q^{2}+U^{2})} + \frac{\delta I^{2}(Q^{2}+U^{2})}{I^{4}}\Bigg)^{\frac{1}{2}},
\ena
where $\delta I = \sigma_{\rm I}$ is the measurement error of Stokes $I$. We fit the high S/N polarization data that satisfies $I/\delta I > 10$ and $p_{\rm db}/\delta p \geq 3$ with the power-law function, from which we derive the power index $\alpha_{\rm power-law}$.

For the Ricean fitting, we use the non-debiased polarization fraction $p_{\rm ndb}$ and fit $p_{\rm ndb}-I$ with the mean of the Ricean distribution, which is given by \cite{Kate_2019}:
\bea  
p_{\rm ndb} = \sqrt{\frac{\pi}{2}} \left( \frac{I}{\sigma_{QU}} \right)^{-1} \, \mathcal{L}_{\frac{1}{2}} \left[ -\frac{p_{\sigma_{\rm QU}}^{2}}{2} \left( \frac{I}{\sigma_{QU}} \right)^{2(1-\alpha_{\rm ricean)}} \right],
\label{eq:p_ricean}
\ena
where $p_{\sigma_{\rm QU}}$ is the polarization fraction at the rms noise level of Stokes $Q$ and $U$, and $\alpha_{\rm ricean}$ is the power index of the $p-I$ relation recovered by Ricean fitting. Both $p_{\sigma_{\rm QU}}$ and $\alpha_{\rm ricean}$ are free parameters and will be determined through the fitting process. In above equation, $\mathcal{L}_{1/2}$ denotes the Laguerre polynomial of order $1/2$, defined as :

\bea 
    \mathcal{L}_{\frac{1}{2}}\!(x)
    = e^{\frac{x}{2}}
    \left[
        \left(1 - x\right)
        I_{0}\!\left(\frac{-x}{2}\right)
        - x I_{1}\!\left(\frac{-x }{2}\right)
    \right],
\ena
where $I_{0}$ and $I_{1}$ are the modified Bessel functions of order zero and one. An example of the Power-law and Ricean fitting to our synthetic data is presented in Appendix \ref{sec:append_add_noise}.

\begin{figure*}
\centering      
\includegraphics[width=\textwidth,height=\textheight,keepaspectratio]{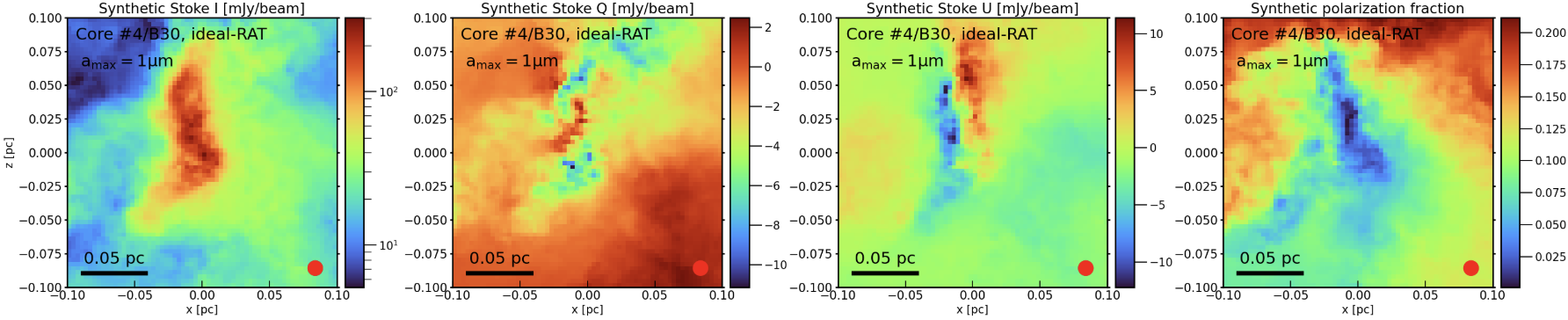}
\includegraphics[width=\textwidth,height=\textheight,keepaspectratio]{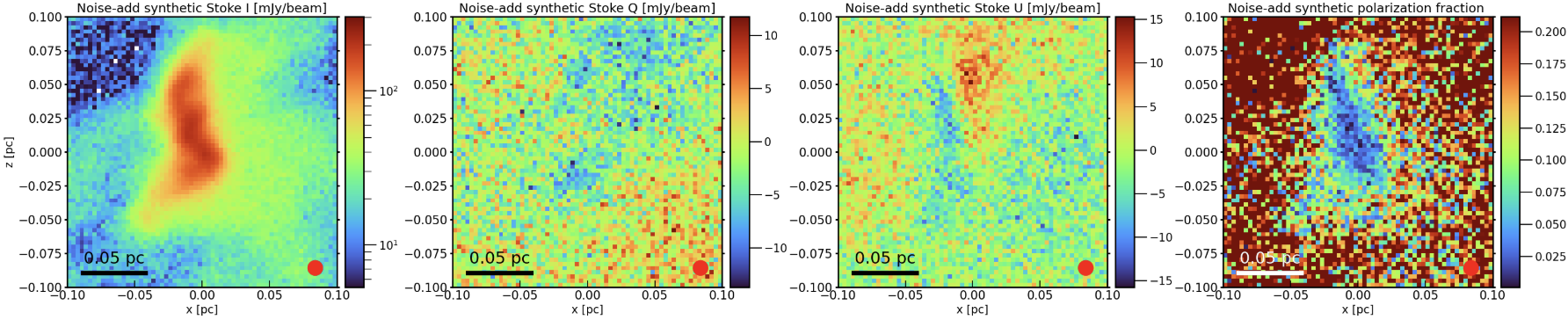}
\includegraphics[width=\textwidth,height=\textheight,keepaspectratio]{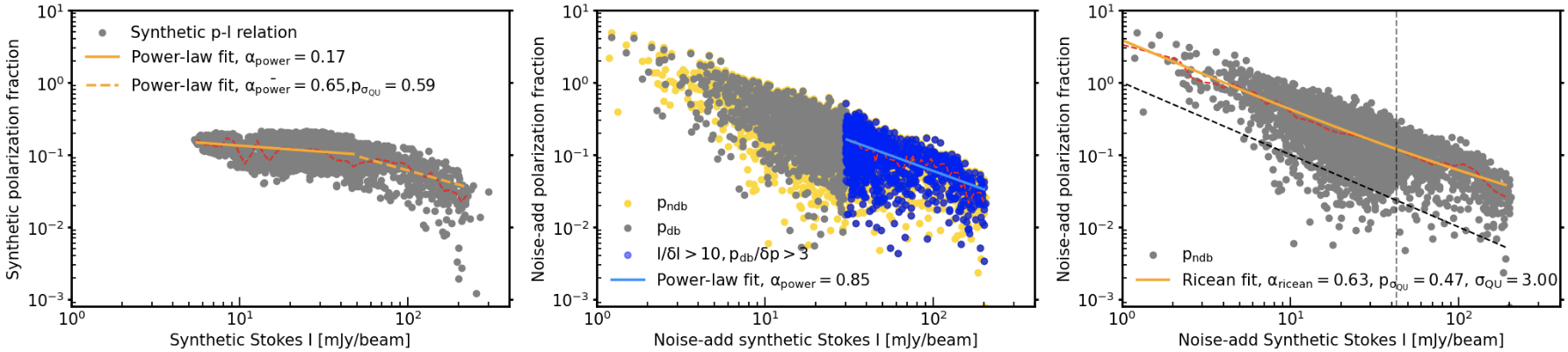}
    \caption{Synthetic Stokes $I,Q,U$, and polarization degree maps of core $\#4$ in cloud model B30 before (upper row) and after (middle row) adding noise, considering model ideal$-$RAT with $a_{\rm max} = 1\mum$. Synthetic map is rebin to 560 au (which corresponds to JCMT pixel size of $4 "$. Noise introduces artificially high polarization fractions ($p>1$) in the outermost, low-intensity regions. Lower row: $p-I$ fitting results. Left panel: power-law fitting to synthetic $p–I$ relation (data without noise), with red line presenting the mean variation of $p(\%)$ with $I$, dashed and solid yellow lines presenting the fitting power-law to outer and inner area. Middle panel: power-law fitting applied to debiased noisy polarization data satisfying $\rm I/\delta I \geq 10$ and $\rm p/\delta p \geq 3$. Right panel: Ricean fitting applied to non-debiased noisy polarization data. Ricean fitting can return the power index derived from synthetic $p-I$ relation $\alpha = 0.65$ more accurately than power$-$law fitting.}
     \label{fig:fitting_ideal_RAT}
\end{figure*}

We show in the upper and middle rows of Figure \ref{fig:fitting_ideal_RAT} an example of the synthetic Stokes $I$, $Q$, $U$, and polarization degree maps at $850\mum$ within 0.2 pc core scale, before and after adding noise (Section \ref{sec:adding_noise}). This example corresponds to core $\#4$ in model B30, assuming model ideal$-$RAT with $a_{\rm max}=1\mum$. The power-law fitting to the synthetic $p-I$ relation (without noise), and Ricean and power-law fitting to the noisy synthetic $p-I$ relation are shown in the lower row. The slope of the synthetic $p-I$ relation (lower left panel) changes from $\alpha\sim0.17$ in the outer core where $I<50$ mJy/beam to higher $\alpha\sim0.65$ in the inner core where $I>50$ mJy/beam due to the change in the depolarization mechanism (Figures \ref{fig:p_NH_Ncl} and \ref{fig:p_NH_amax}). The Ricean fitting to non-debiased polarization data (lower middle panel) can recover the second $p-I$ slope rather well, giving $\alpha_{\rm ricean}\sim0.6$. In contrast, the power-law fitting to debiased polarization data (lower right panel) slightly overestimates the $p-I$ slope with $\alpha_{\rm power-law}\sim0.77$ due to the residual noise contamination in debiased data.

\subsection{Noise effect on fitting}
\begin{figure*}
\centering      
\includegraphics[width=\textwidth,height=\textheight,keepaspectratio]{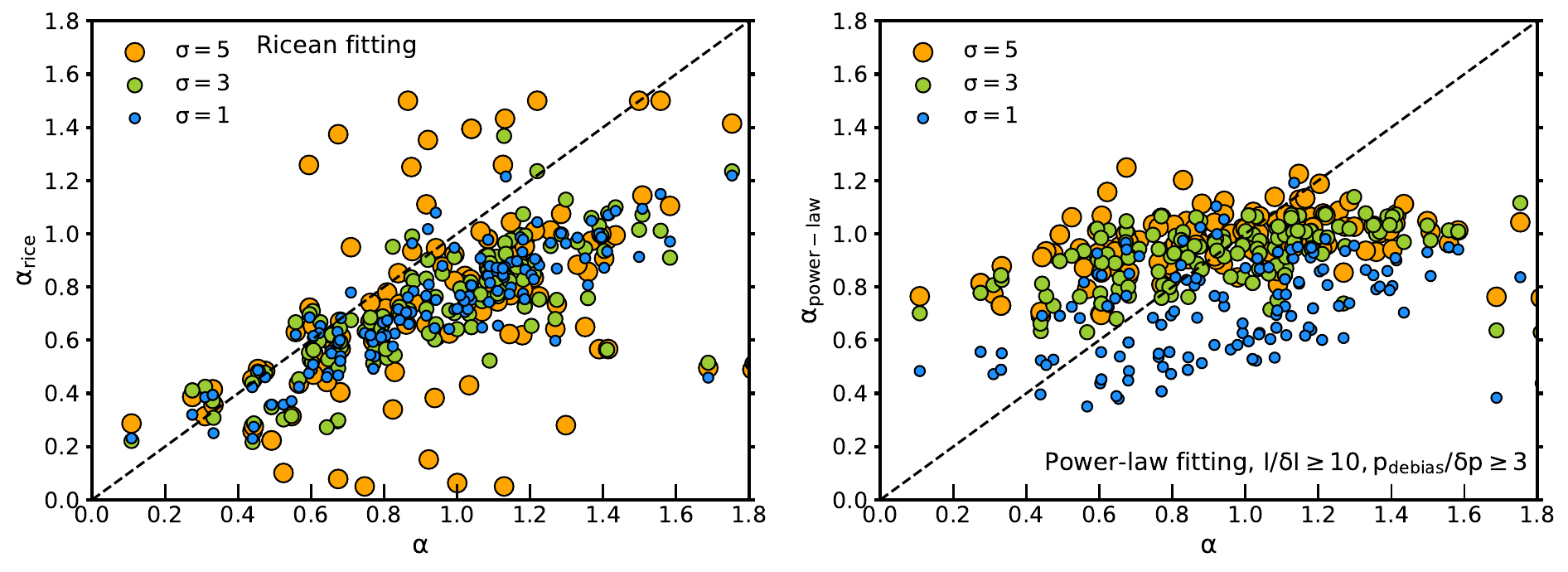}
    \caption{Ricean and Power-law fitting with different applied noise levels: strong noise $\sigma = 5$ (orange), moderate noise $\sigma = 3$ (green), small noise $\sigma = 1$ (blue). }
     \label{fig:fitting_sigma}
\end{figure*}

The impact of including noise on the $\alpha$ fitting is illustrated in Figure \ref{fig:fitting_sigma}. Strong levels of noise ($\sigma = 5$, orange points) cause a large uncertainty in Ricean fitting while $\alpha > 0.4$, and systematically overestimate $\alpha$ derived from Power-law fitting regardless of the real slope $\alpha$ from objects. In contrast, weak noise ($\sigma = 1$, blue points) allows Power-law fitting to return values closer to $\alpha$ in case $\alpha <0.6$, but it also induces larger uncertainty of $\alpha_{\rm power-law}$ in case $\alpha > 0.6$.

\label{lastpage}
\end{document}